\documentclass[prb,showpacs,twocolumn]{revtex4-1}
\usepackage{graphicx,amsmath}
\usepackage{float}
\usepackage{subfigure}
\usepackage{mathrsfs}
\usepackage{graphics}
\usepackage{epstopdf}
\usepackage{amssymb}
\usepackage{hyperref}
\usepackage{color}
\usepackage{braket}
\usepackage{multirow}
\graphicspath{{/media/kallol/7224D84124D809CD/corona/PhD_work/RMO/Draft/Draft0/figures/}}
\begin{document}
\author{Kallol Mondal and Charudatt Kadolkar}
\affiliation{Department of Physics, Indian Institute of Technology Guwahati, Guwahati, Assam 781039, India}
\date{\today}
\title{Regular magnetic orders in triangular and kagome lattices}
\begin{abstract}
We investigate the possible \textit{regular magnetic order}(RMO) for the spin models with global $O(3)$ spin rotation, based on a group theoretical approach for triangular and kagome lattices. The main reason to study these RMOs is that they are good variational candidates for the ground states  of many specific models. In this work, we followed the prescription introduced by Messio et al. (L. Messio, C. Lhuillier, and G. Misguich, Phys. Rev. B, 2011, 83, 184401) for the $p6m$ group and extended their work for different subgroups of $p6m$, i.e., $p6$, $p3$, $p3m1$, and $p31m$. We have listed all the possible regular magnetic orders for kagome and triangular lattices, which fall into the category of these groups. We calculate the energy and the spin structure factors for each of these states.
\end{abstract}
\maketitle
\section{\label{sec:intro}Introduction}
Finding the ground state of different frustrated magnets is a long-sought goal in condensed matter physics. The interest is triggered by the possibility of disordered state and their relationship to high-temperature superconductivity~\cite{anderson1987resonating}. Each of those disordered phases is interesting in its own way and appears with different types of order and excitations associated with unusual quantum numbers. However, it turns out that the magnetic structure of the ordered phases of such magnets becomes non-trivial. For example, the classical ground of many such frustrated spin systems has ``accidental degeneracies'' which are lifted by quantum fluctuations. In those cases, quantum fluctuations induce magnetic order in the spin system, known as ``order from disorder''~\cite{villain1974quantum,rastelli1987order,henley1989ordering}.

In general, finding the classical ground state of such a spin system is a challenging problem to solve. For instance, there exists no general method to figure out the ground state spin configuration of a non-Bravais lattice described by simple Heisenberg Hamiltonian with $O(3)$ symmetry is given by
\begin{equation}
H= \sum_{ij} J(|\vec{\mathbf{R}}_i - \vec{\mathbf{R}}_j|)~ \vec{\mathbf{S}}_i \cdot \vec{\mathbf{S}}_j
\end{equation}
where $J$ is the exchange integral. $\vec{\mathbf{S}}_i$ and $\vec{\mathbf{S}}_j$ are the classical spin vectors at $i$-th and $j$-th site, whose position vectors are given by $\mathbf{R}_i$ and $\mathbf{R}_j$ respectively.

A family of spin configurations that respect all the lattice symmetries of a given lattice modulo global spin transformations is termed as the \textit{regular magnetic order}(RMO)~\cite{messio2011lattice}. The approach of finding the RMOs is analogous to Wen's classification of quantum spin liquids based on the concept projective symmetry group~\cite{wen2002quantum}. The most simple example is the Neel state in one dimension. In that case, any lattice transformation can be compensated by appropriate spin rotations. By construction, the family of RMOs only depends on the symmetries of the model, i.e., lattice symmetries and spin transformations(rotations and spin flips), and does not rely on the strength of the couplings present in the model. These RMOs become very interesting in the case of frustrated magnets, as these are the good variational candidates for many spin models. For example, for Heisenberg spins on kagome lattice, the ground state is found to possess non-coplanar spin structures with first, second, and third neighbor exchange interactions~\cite{PhysRevB.72.024433, PhysRevLett.101.106403, Janson_2009}. Apart from that, these RMOs can provide useful insights into the experimental data of magnetic materials where the information about the lattice is known, but the values of the couplings present in the material are unknown. In that case, the magnetic correlation can be directly compared to these states. If these two correlation matches, we can extract some useful information about the couplings present in the material.

In this work, we mainly focus on the Heisenberg model, and we treat each spin as a three-dimensional unit vector. We use a simple group theoretical approach to investigate the possible regular magnetic order in triangular and kagome lattices. So, we have considered the wallpaper groups, which include the triangular and kagome geometries. The triangular Bravais lattice is characterized by (i) three or six-fold rotation symmetry (ii) unit cell is a rhombus with angles $60^o$ and $120^o$. There is a total of seventeen wallpaper groups in 2-dimension\citep{schwarzenberger197417}. There are five wallpaper groups $p6, p3, p3m1, p31m$, and $p6m$ (in Hermann-Mauguin notation) that include the triangular and kagome geometry. The wallpaper group $p6m$ is already done by Messio et al. ~\cite{messio2011lattice}. We have extended their work and mainly focused on the rest of the four wallpaper groups $p6, p3, p3m1$, and $p31m$. We will consider each of them separately and list all the possible regular magnetic orders for each of those groups.

The layout of this paper is as follows in sec. II, we introduce the mathematical formulation and a brief of how to find the RMOs. In sec. III, sec. IV, sec. V. and sec. VI, we discuss the construction of RMOs for the wallpaper groups $p6, p3, p3m1$, and $p31m$ and list all the possible regular magnetic orders for each of those groups. In sec. VII, we briefly discuss the regular magnetic order in the case of the $p1$ group. In sec. VIII, we discuss all the obtained RMOs along with their energies and the spin structure factors. In sec. IX, we make the closing remarks.
\section{Mathematical Formulation}
Let us denote the group of transformation  $ \mathcal{G}_{\mathcal{L}}$ on a lattice $\mathcal{L}$ as $\mathcal{G}_{\mathcal{L}} = \{ \sigma : \mathcal{L} \rightarrow \mathcal{L} \}$. Let $\mathcal{H}$ be the spin configuration space that is $\mathcal{H} = \{ \phi | \phi : \mathcal{L} \rightarrow \mathbb{S}^2\}$. There are two types of transformation on the $\mathcal{H}$.
 
 \vspace{0.25cm}
\noindent(i) \textbf{Lattice transformations:} For each $\sigma \in \mathcal{G}_{\mathcal{L}}$, there is  $O_\sigma : \mathcal{H}\rightarrow \mathcal{H}$ such that $O_\sigma \phi(i) = \phi (\sigma^{-1}(i))$. Since this group is isomorphic to $\mathcal{G}_\mathcal{L}$, we will refer to this group as $\mathcal{G}_\mathcal{L}$ hoping that there is no confusion.

 \vspace{0.25cm}
\noindent(ii) \textbf{Spin rotations:} There can be two types of spin rotations; local and global. Fora global spin rotation $g \in O(3)$, we have $R_g : \mathcal{H}\rightarrow \mathcal{H}$ such that $(R_g \phi)(i) = g \phi(i)$. We will refer to this group of global spin rotations by the same symbol $O(3)$.
 
  \vspace{0.25cm}
\noindent The group of transformations $\mathcal{G}$ on $\mathcal{H}$, is a direct product of $~\mathcal{G}_{\mathcal{L}} $ and $  O(3)$ i.e. $ \mathcal{G} = \mathcal{G}_{\mathcal{L}}\times  O(3)$ 

\vspace{0.25cm}
\noindent \textbf{Stabilizer group of spin transformation:}
 Let $\phi$ be a spin configuration. The set of all transformations that leaves the spin configuration $\phi$ invariant is denoted by $G_\phi$. So, $G_\phi = \{ g \in \mathcal{G} ~|~ g \phi = \phi \}$. Then $G_\phi$ is a subgroup of $\mathcal{G}$ and is called the \textit{stabilizer} group of the spin configuration $\phi$. Let $G^S_\phi = O(3) \cap G_\phi$ that is the $G^S_\phi$ is a subgroup of $O(3)$ that leaves the spin configuration $\phi$ invariant and we call it the invariant spin symmetry group(ISG).

\vspace{0.25cm}
\noindent \textbf{Regular structures: }
A spin configuration $\phi \in \mathcal{H}$ is called \textit{regular} if, for every $\sigma \in \mathcal{G}_\mathcal{L}$, there is $g_\sigma \in O(3)$ such that $g_\sigma (\sigma (\phi))= \phi$.  For a regular structure, we can prove the following theorem which states that \textit{if a spin configuration $\phi$ is regular, then the group $~G_\phi/ G_S~$ is isomorphic to $~\mathcal{G}_{\mathcal{L}}$}.  The proof is given in the appendix. This theorem is central to the idea of the algebraic symmetry group presented in the next section. It is not easy to list out all the RMOs of a system. However, we can find all RMOs by using a systematic procedure based on algebraic symmetry groups.

\vspace{0.25cm}
\noindent \textbf{Algebraic symmetry groups:}
\noindent A subgroup of $~O(3) \times \mathcal{G}_{\mathcal{L}}~$ that is homomorphic to $\mathcal{G}_{\mathcal{L}}$ is called an \textit{algebraic symmetry group}$(G)$. If $\phi$ is a regular structure, then it has one of the algebraic symmetry groups. The idea is to find the entire list of such groups. However, if $\phi$ is regular, then for any $g \in O(3)$, $~ g \phi$ is also regular and $G_{g \phi} = g G_{\phi}g^{-1}$. Then $G_{g \phi}/ G_{g \phi}^S$ is isomorphic to $G_\phi/G^S_\phi$. We only need to list the equivalence classes of algebraic symmetry groups, and loosely we will call the equivalence classes also by the same name.

Let $G \subset \mathcal{G}$ be an ASG. Then, there is a homomorphism $\xi : \mathcal{G}_{\mathcal{L}} \rightarrow G$. Let us consider three elements $\sigma_1,\sigma_2$ and $\sigma_3$ of $\mathcal{G}_{\mathcal{L}}$ with the algebraic relation $\sigma_1 \sigma_2 = \sigma_3$. Let, $G_{\sigma_1}, G_{\sigma_2}$ and $ G_{\sigma_3}$ be the images of the symmetry elements  $\sigma_1,\sigma_2$ and $ \sigma_3$ respectively under the homomorphism$(\xi)$. Also, under $\xi$, identity is mapped to ISG$(G^S_\phi)$. The relation $\sigma_1\sigma_2 \sigma_3^{-1} = I$ imposes an algebraic constraint on the  images $G_{\sigma_1}, G_{\sigma_2}$ and $ G_{\sigma_3}$ which is given by
\begin{equation}
G_{\sigma_1}\sigma_1G_{\sigma_2}\sigma_2 \sigma_3^{-1}G_{\sigma_3}\in G^S_\phi, \hspace{0.5cm} \forall ~\sigma_1,\sigma_2 ~\in ~\mathcal{G}_{\mathcal{L}}
\label{eqn:ch4-ssg}
\end{equation}
Since the elements of $~\mathcal{G}_{\mathcal{L}}$ and $g_\sigma$ commutes, we have
\begin{eqnarray}
G_{\sigma_1}G_{\sigma_2}\sigma_1\sigma_2 \sigma_3^{-1}G_{\sigma_3}^{-1} & \in & G^S_\phi \nonumber\\
G_{\sigma_1}G_{\sigma_2}G_{\sigma_3}^{-1} & \in & G^S_\phi \nonumber
\end{eqnarray}
So, we have started with the algebraic relation between the elements of $\mathcal{G}_{\mathcal{L}}$, and we end up with a relation in pure spin transformations.

We must mention that the algebraic symmetry group only depends on the ISG, and also on the algebraic relations between the generators of the $\mathcal{G}_{\mathcal{L}}$, but there is no direct dependence on the lattice $\mathcal{L}$. The spin rotation group for Heisenberg spins is $O(3)$ group. Now the possible ISG can only be isomorphic to one of the groups $\{I\}, \mathbb{Z}_2$ or $O(2)$, which leads to the non-coplanar, coplanar, and collinear spin configurations, respectively. The first case, $G^S_\phi = \{I\}$ is the most interesting one and we will consider it for each of the wallpaper groups $p6,p3,p31m$ and $p3m1$.
\subsection{Construction of regular magnetic order}
An algebraic symmetry group $G$ is compatible with a spin configuration $\phi$ if $g  \phi = \phi$ for each $g \in G$. The next step is to find all the compatible states corresponding to each of the algebraic symmetry groups. This step is explicitly dependent on the lattice $\mathcal{L}$. To construct the compatible states with a given algebraic symmetry group, we first fix the spin direction at site $i$. Then we apply all the elements of lattice symmetry group $\mathcal{G}_{\mathcal{L}}$ to extract the spin arrangement of all other sites. Here we must mention that two elements of lattice symmetry group $X,Y \in \mathcal{G}_{\mathcal{L}}$ leads to the same site $i$ i.e. $X(i) = Y(i)$, then we must have $G_X(i) = G_{Y}(i)$. This will result in a constraint on the direction of the spin at site $i$ or indicate no compatible states for the given mapping.
  
To construct the regular magnetic orders, we have the following steps. First, we fix the ISG. Then we look for the algebraic relation between the generators of the $\mathcal{G}_{\mathcal{L}}$. These algebraic relations impose constraint on $g_\sigma \in O(3)$ for each $\sigma \in \mathcal{G}_{\mathcal{L}}$. In the second step, we have to determine the spin configuration compatible with each of the algebraic symmetry groups if there exists.
\begin{figure*}[ht!]
	\includegraphics[width=0.4\textwidth]{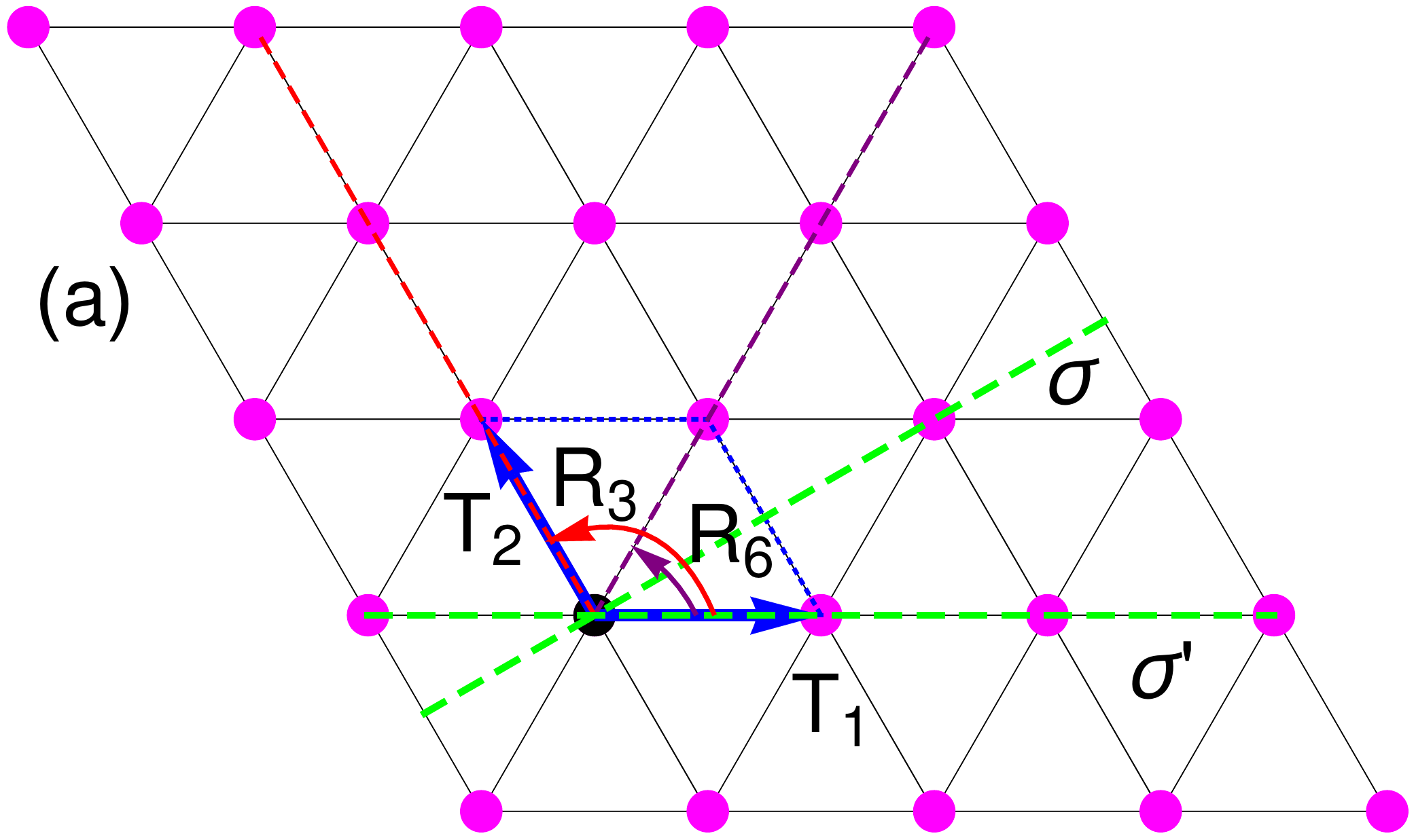}\hspace{0.25cm}
	\includegraphics[width=0.4\textwidth]{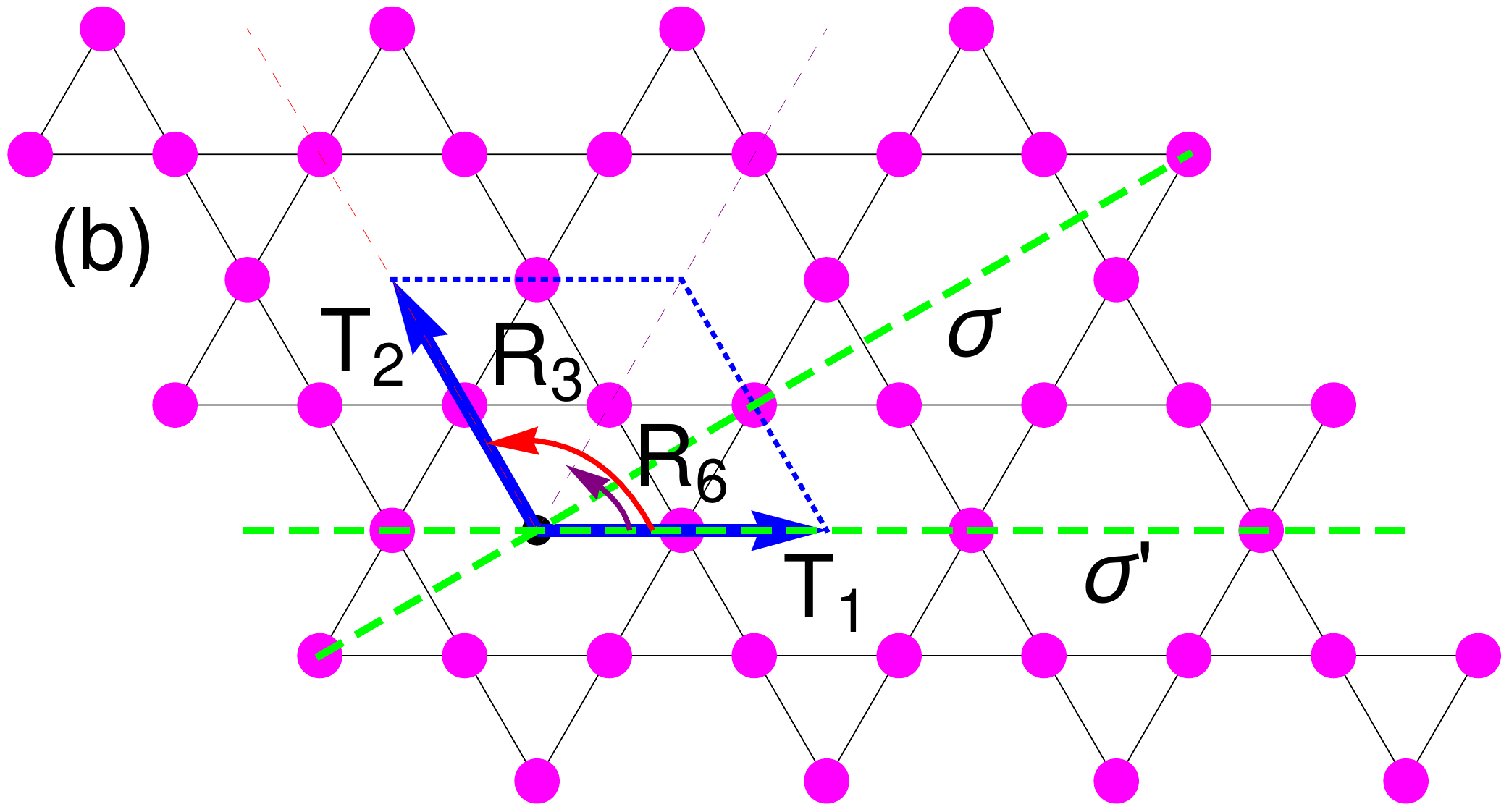}
 	\caption{Generators of all the wallpaper groups $p6, p3, p3m1$, and $p3m1$ in case of (a) triangular lattice, and  (b) kagome lattice. Blue rectangles indicate the unit cell of the respective lattices. Here $R_6$ and $R_3$ represents six-fold and three-fold rotations about the origin, respectively. $T_1$ and $T_2$ be the translation along the horizontal and the vertical direction, respectively. $\sigma^\prime$ be the reflection about a plane parallel to $T_1$ and $\sigma$ be the reflection about a plane at angle $30^o$ with respect to $T_1$.}
	\label{fig:generator}
\end{figure*}
\subsection{Application to the symmetry group of triangular lattice}
 There are a total of five wallpaper groups $p6, p3, p3m1, p31m$, and $p6m$, that support the triangular geometry. 
Let us denote the lattice by $\mathcal{L}$. Let, $(r_1,r_2)$ be the oblique co-ordinates with respect to the basis $\vec{a} = (1,0)$ and $\vec{b} = (\frac{1}{2},\frac{-\sqrt{3}}{2})$. Then the generators for these groups are  combinations of the lattice transformations $T_1, T_2,R_6,R_3, \sigma,$ and $ \sigma^\prime$ where, $T_1, T_2$ are two translations, $R_6, R_3$ are six fold and three fold rotations about the origin respectively, $\sigma$ and $\sigma^\prime$ are the reflections about a plane at an angle $30^o$ and parallel to the x-axis as shown in Fig.~\ref{fig:generator}. For example, $p6$ is generated by $T_1,T_2$ and $R_6$, and $p3m1$ is generated by $T_1,T_2,R_3$ and $\sigma$. The action of each of the generators on the lattice sites are given by
\begin{subequations}
\begin{eqnarray}
T_1 : (r_1,r_2) & \rightarrow & (r_1 + 1 ,r_2) \\
T_2 : (r_1,r_2) & \rightarrow & (r_1 ,r_2 +1) \\
R_6 : (r_1,r_2) & \rightarrow & (r_1 - r_2 ,r_1)\\
R_3 : (r_1,r_2) & \rightarrow & (- r_2 ,r_1-r_2) \\
\sigma : (r_1,r_2) & \rightarrow & (r_1 ,r_1-r_2)  \\
\sigma^\prime : (r_1,r_2) & \rightarrow & (r_1- r_2, -r_2)
\end{eqnarray}
\end{subequations}
Now, in the following section, we will consider each of the groups separately.
 \section{$p6$ group}
For the $p6$ group, the lattice symmetry group $\mathcal{G}_{\mathcal{L}}$ is $p6$. Every element of $p6$ can be  written in the  form $R_6^r T_1^{t_1}T_2^{t_2}$ where $r=0,1,2,3,4,5$ and $t_1, t_2 \in \mathbb{Z}$.  Let us consider the mapping $G$ from $\mathcal{G}_{\mathcal{L}}$ to $\mathcal{G}_{\mathcal{L}} \times O(3)$. The mapping $G$ can be constructed from the images of the generators of $\mathcal{G}_{\mathcal{L}}$ using Eq.~\ref{eqn:ch4-ssg}. Now, the relation between the generators of $p6$ group are following
\begin{subequations}
\begin{eqnarray}
T_1 T_2 & = & T_2 T_1 \\
T_1 R_6 T_2 & = & R_6 \\
R_6 T_1 T_2 & = & T_2 R_6 \\
R_6^6 & = & I
\end{eqnarray}
\end{subequations}
Let $G_{T_1}, G_{T_2}$, and $G_{R_6}$ be the image of the generators $T_1,T_2$, and $R_6$. Then, we have the following equations.
\begin{subequations}
\begin{eqnarray}
G_{T_1} G_{T_2} & = & G_{T_2} G_{T_1} \\
G_{T_1} G_{R_6} G_{T_2} & = & G_{R_6} \\
G_{T_2} G_{R_6} & = & G_{R_6} G_{T_1} G_{T_2} \\
G_{R_6}^6 & = & I
\end{eqnarray}
\end{subequations}
Each element $G_X$, corresponding to an elements $X \in \mathcal{G}_{\mathcal{L}}$ is characterized by its determinant $\epsilon_X = \pm 1$ and a rotation $R_{\hat{n}_X \theta_X}$ by an angle $\theta _X \in [0, \pi]$ about an axis $\hat{n}_X$ such that $G_X = \epsilon_X R_{\hat{n}_X \theta_X}$. For the $p6$ group, we get total twelve solutions as given below
\begin{subequations}
\begin{eqnarray}
&\text{(i)} &G_{T_1} = I, G_{T_2} = I, \hspace{0.3cm} \text{and} \hspace{0.3cm} G_{R_6} = \epsilon_{R_6} I \\
&\text{(ii)} &G_{T_1} = I, G_{T_2} = I, \hspace{0.3cm} \text{and} \hspace{0.3cm}G_{R_6} = \epsilon_{R_6} R(\hat{n}, \frac{\pi}{3}) \\
& \text{(iii)} ~& G_{T_1} = I, G_{T_2} = I, \hspace{0.3cm} \text{and} \hspace{0.3cm}G_{R_6} = \epsilon_{R_6} R(\hat{n}, \frac{2\pi}{3}) \\
& \text{(iv)} ~& G_{T_1} = I, G_{T_2} = I, \hspace{0.3cm} \text{and} \hspace{0.3cm}G_{R_6} = \epsilon_{R_6} R(\hat{n}, \pi) \\
&\text{(v)}~ &G_{T_1} = G_{T_2} = R(\hat{n}, \frac{2\pi}{3}),\hspace{0.1cm}\text{and}\hspace{0.1cm}G_{R_6} = \epsilon_{R_6} R(\hat{n}_6,\pi)  \nonumber  \\
&~~ &\text{where} ~~\hat{n}_6 \perp \hat{n}\\
&\text{(vi)} &G_{T_1} = R(\hat{x}, \pi), G_{T_2} = R(\hat{y}, \pi) ~~ \text{and}\hspace{0.1cm} \nonumber \\
& ~~&G_{R_6} = \epsilon_{R_6} R(\hat{n}, \frac{2\pi}{3}) ~~\text{with}~ \hat{n} = (1,1,1)
\end{eqnarray}
\label{eqn:ch4-p6-mapping}
\end{subequations}
where $\epsilon_{R_6}$ can take values $\pm1$.
\subsection{RMOs in triangular lattice}
The position vector for any arbitrary point is given by $\vec{r} = m \vec{a} + n \vec{b}$ where $m,n \in \mathbb{Z}$. Let us consider that the spin configuration at starting point is $\phi(0,0)$. So, the spin configuration at any arbitrary site is given by
\begin{equation}
\phi(m,n) = G_{T_1}^m G_{T_2}^n \phi(0,0)
\end{equation}
Under $6$-fold rotation, we must have,
\begin{equation}
\phi(m,n) = G_{R_6} \phi(n,n-m)
\end{equation}
After some manipulation we get,
\begin{equation}
G_{T_1}^m G_{T_2}^n \phi(0,0) = G_{T_1}^m G_{T_2}^n G_{R_6}\phi(0,0)
\end{equation}
This implies that $\phi(0,0)$ must be an eigenvector of $G_{R_6}$ with an eigenvalue $+1$. Now we can list all the possible RMOs corresponding to each of the mapping given in Eq,~\ref{eqn:ch4-p6-mapping} for the triangular lattice.

\vspace{0.25cm}
\noindent For (i)-(iii), we can set $\hat{n}$ to be in the $(1,1,1)$ direction. For $\epsilon_{R_6} =+1$ case, $(1,1,1)$ is the eigenvector of $ G_{R_6}$ with positive eigenvalue. So, $\phi(0,0)$ can be taken as $(1,1,1)$. The resulting RMO is well-known ferromagnetic state, as shown in Fig.~\ref{Ch4:fig:ferro-state}. But, For $\epsilon_{R_6} = -1$, all the eigenvalues of $G_{R_6}$ are negative. So, there is no compatibles RMOs possible for these mappings.

\vspace{0.25cm}
\noindent (iv) We choose $\hat{n} = (1,1,1)$ and the resulting RMO is a ferromagnet for $\epsilon_{R_6} = 1$. The spins are aligned ferromagnetically along $(1,1,1)$. But, for $\epsilon_{R_6} = -1$, we also get ferromagnetic state where the spins are not aligned in the $(1,1,1)$ directions, rather it lies in a plane perpendicular to the $(1,1,1)$ directions.

\vspace{0.25cm}
\noindent (v) Since in this case $\hat{n}_6 \perp \hat{n}$, we choose $\hat{n} = (1,1,1)$ and $\hat{n}_6 = (1,-1,0)$. For $\epsilon_{R_6} = +1$ case, $(-1,1,0)$ is the eigenvector of $ G_{R_6}$ with positive eigenvalue. So, the $\phi(0,0)$ can be taken as $(-1,1,0)$ and the resulting RMO is a planar structure, containing three sub-lattices. This structure is known as coplanar state, as shown in the Fig.~\ref{Ch4:fig:Q0-planar-state}.

For $\epsilon_{R_6} = -1$ case, we have two eigenvalues of $ G_{R_6}$ which are positive and the associated eigenvectors are $(1,1,0)$ and $(0,0,1)$. So, any linear combination of these two vectors can be taken as $\phi(0,0)$. In this case, we get a non-coplanar structure with three sub-lattices, as shown in Fig.~\ref{Ch4:fig:Q0-umbrella-state}.

\vspace{0.25cm}
\noindent (vi) Here, $\hat{n}$ is taken in the $(-1,-1,-1)$ direction. For $\epsilon_{R_6} = +1$, $(1,1,1)$ is the eigenvector of $ G_{R_6}$ with positive eigenvalue and hence $\phi(0,0)$ can be taken as $(1,1,1)$. In this case the RMO obtained is tetrahedral which has four sub-lattices as shown in the Fig.~\ref{Ch4:fig:Tetrahedral-state}.

For $\epsilon_{R_6} = -1$, all the eigenvalues of $ G_{R_6}$ is negative. So, there will be no RMO corresponding to this algebraic symmetry group.
\begin{table*}[ht!]
\begin{center}
\begin{tabular}{|c|c|c|c|c|c|c|}
 \hline
 \hline
  No.                     &    $G_{T_1}$                                               &     $G_{T_2} $                                           &    $G_{R_6}$                                           & $\hat{n}$                              &  Triangular   &    Kagome  \\ 
 \hline
\multirow{2}*{1} & \multirow{2}*{I}                                  & \multirow{2}*{I}                                  &       I                                                   &    Any                         &         F         &  F     \\  \cline{4-7}
                            &                                                              &                                                             &     -I                                                   &    Any                         &         $\times$       &   $\times$      \\
\hline
\multirow{2}*{2} & \multirow{2}*{I}                                  & \multirow{2}*{I}                                 &       $R(\hat{n}$,$\frac{ \pi}{3})$          &    Any                        &         F        &  F    \\  \cline{4-7}
                            &                                                              &                                                             &     -$R(\hat{n}$,$\frac{ \pi}{3})$       &     Any                         &      $\times$      & $Q=0$ (P)  \\
\hline
\multirow{2}*{3} & \multirow{2}*{I}                                  & \multirow{2}*{I}                                  &       $R(\hat{n}$,$\frac{2 \pi}{3})$      &    Any                            &       F          & $Q=0$(U), F     \\  \cline{4-7}
                            &                                                              &                                                            & -$R(\hat{n}$,$\frac{2 \pi}{3})$&     Any                                       &     $\times$         &    $\times$ \\
\hline                  
\multirow{2}*{4} &            \multirow{2}*{I}                      & \multirow{2}*{I}                                  &       $R(\hat{n},\pi)$           &   Any                                              &          F       & F     \\  \cline{4-7}
                            &                                                            &                                                              &     -$R(\hat{n},\pi)$          &     Any                                              &         F         &   F     \\
\hline
\multirow{2}*{5} & \multirow{2}*{$R(\hat{n},\frac{2 \pi}{3}) $} & \multirow{2}*{$R(\hat{n},\frac{2 \pi}{3}) $}  &       $R(\hat{n}_6,\pi)$      &   $\hat{n}_6 \perp \hat{n}$                     &        P        &  $\sqrt{3} \times \sqrt{3}$  \\  \cline{4-7}
                            &                                                            &                                                              &     -$R(\hat{n}_6,\pi)$      &     $\hat{n}_6 \perp \hat{n}$                   &         U       &  $\sqrt{3} \times \sqrt{3}$ (U),F   \\
\hline
\multirow{2}*{6} & \multirow{2}*{$R(\hat{x}, \pi$)}       & \multirow{2}*{$R(\hat{y}, \pi$) }        &       $R(\hat{n},\pi)$           &  $\frac{-1}{\sqrt{3}}(1,1,1)$   &         T        &                   O    \\  \cline{4-7}
                           &                                                            &                                                              &       -$R(\hat{n},\pi)$         &  $ \frac{-1}{\sqrt{3}}(1,1,1)$   &        $\times$          &  O, $C_1$, $I_1$, $C_2$, $I_2$ \\
\hline
\hline
\end{tabular}
\caption{List of regular magnetic orders in  $p6$ wallpaper group. }
\label{tbl:p6group}
\end{center}
\end{table*}

\subsection{RMOs in kagome lattice}
Since, the lattice symmetry group $\mathcal{G}_{\mathcal{L}}$ of kagome lattice is isomorphic to the lattice symmetry group of triangular lattice, the algebraic solutions remains same for both the lattice. Let the three sub-lattices A, B and C are located at $\frac{1}{2}\vec{a}, \hspace{0.1cm}\frac{1}{2}\vec{b}$ and $-\frac{1}{2} ( \vec{a} + \vec{b})$. Consider a lattice point of sub-lattice index $\mu$ in a unit cell with left bottom corner at $m \vec{a} + n \vec{b} $. Clearly
\begin{equation}
\phi(m,n,B) = G_{R_6}^2 \phi(n-m,-m,A)
\end{equation}
After few steps of algebra we get,
\begin{equation}
G_{T_1}^m G_{T_2}^n \phi(0,0,B) = G_{T_1}^{m} G_{T_2}^n G_{R_6}^2 \phi(0,0,A)
\end{equation}
Thus we get, the relation between the three sub-lattices,
\begin{equation}
\phi(0,0,A) = G_{R_6}^2 \phi(0,0,C) = G_{R_6}^4 \phi(0,0,B) = G_{R_6}^6 \phi(0,0,A) \nonumber
\end{equation}
But we also have the condition that $\phi(0,0,A) = G_1 G_6^3 \phi(0,0,A)$. For each possibilities of RMOs , spins at each site must follow the above conditions. We can choose $\phi(0,0,A)$ as the eigenvector of $ G_1 G_{R_6}^3$ corresponding to the positive eigenvalue.
\vspace{0.25cm}

\noindent (i) This is the trivial case where all the symmetry operations are identity. For $\epsilon_{R_6} = +1$, all the eigenvalues of $ G_1 G_{R_6}^3$ are positive. Any linear combination of the eigenvectors can be taken as $\phi(0,0,A)$. The resulting RMO is the well known ferromagnetic state, as shown in Fig.~\ref{Ch4:fig:ferro-kagome-state}.

For $\epsilon_{R_6} = -1$, all the eigenvalues are negative. So, in this case, there will not be any compatible RMO.

\vspace{0.25cm}
\noindent (ii) This case is very similar to the previous one. We choose $\hat{n}$ to be along $(1,1,1)$ direction. For $\epsilon_{R_6} = +1$ case, $(1,1,1)$ is eigenvector of $ G_1 G_{R_6}^3$ with positive eigenvalue and the resulting RMO is ferromagnetic state.

For $\epsilon_{R_6} = -1$, there are two eigenvalues of $ G_1 G_{R_6}^3$ which are positive and the corresponding eigenvectors are $(-1,0,1)$ and $(-1,1,0)$. So we can take any linear combination of these two vectors as $\phi(0,0, A)$. In this case, the resulting RMO is $Q=0$ planar state, which is shown in the Fig.~\ref{Ch4:fig:Q0-planar-kagome-state}.

\vspace{0.25cm}
\noindent (iii) Here too, we choose $\hat{n}$ to be along $(1,1,1)$ direction. For $\epsilon_{R_6} =+1$ case, all the eigenvalue of $ G_1 G_{R_6}^3$ are positive. So, any linear combination of these eigenvectors can be taken as $\phi(0,0, A)$, and the resulting RMO is a $Q=0$ umbrella state as shown in Fig.~\ref{Ch4:fig:Q0-umbrella-kagome-state}.

For $\epsilon_{R_6} = -1$ case, all the eigenvalue of $ G_1 G_{R_6}^3$ are negative, so there will not be any compatible RMO for this mapping.

\vspace{0.25cm}
\noindent (iv) In this case, we choose $\hat{n}$ to be along $(1,1,1)$ direction. By looking at the positive eigenvalues of $ G_1 G_{R_6}^3$ we choose $\phi(0,0,A)$. Here we get ferromagnetic state for both $\epsilon_{R_6} = \pm 1$.

\vspace{0.25cm}
\noindent (v) Since in this case $\hat{n}_6 \perp \hat{n}$, we choose $\hat{n} = (1,1,1)$ and $\hat{n}_6 = (1,-1,0)$. For $\epsilon_{R_6} =+ 1$ case, $\phi(0,0,A)$. can be taken as $(-1,0,1)$ and we get $\sqrt{3}\times \sqrt{3}$ planar state with three sub-lattices as shown in Fig.~\ref{Ch4:fig:root3Xroot3-planar-kagome-state}.

For $\epsilon_{R_6} = -1$ case, $(1,0,1)$ and $(0,1,0)$ are the eigenvectors corresponding to the positive eigenvalues of $ G_1 G_{R_6}^3$. So we can take linear combination of these two vectors as $\phi(0,0,A)$. Here we get, non-coplanar $ \sqrt{3}\times \sqrt{3}$ umbrella state with three sub-lattices as shown in Fig.~\ref{Ch4:fig:root3Xroot3-planar-kagome-state}.

\vspace{0.25cm}
\noindent (vi) In this case, $\hat{n}$ is chosen to be along $(1,1,1)$. For $\epsilon_{R_6} = 1$ case, $\phi(0,0,A)$ can be taken as $(1,0,0)$ and the resulting RMO is the Octahedral state which has six sub-lattices and the spins are pointing towards the corner of an octahedron as shown in Fig.~\ref{Ch4:fig:Octahedral-kagome-state}. The magnetic unit-cell contains 12 sites.

Now, $\epsilon_{R_6} = -1$ case is the most interesting one. Here, $(0,0,1)$ and $(0,1,0)$ are the eigenvectors of $ G_1 G_{R_6}^3$ with positive eigenvalues. So, any linear combination of these two vectors can be chosen as $\phi(0,0,A)$. Here, we choose $\phi(0,0,A) = (0, \cos(\pi t), \sin(\pi t))$ where $t$ is a continuous parameter. As we increase the value of $t$, starting from $0$ to $+1$, we get icosahedron states with a series of regular structures as summarized in the tabular form given below.
\begin{table}[h!]
\begin{center}
\begin{tabular}{ |c| c| c| }
\hline
\hline
No.  & Parameter(t)& RMO \\ 
 \hline
1 & 0 & Octahedral~\ref{Ch4:fig:Octahedral-kagome-state} \\  
 \hline
 2 &$\tan^{-1}(1/4)$ & Cuboc1~\ref{Ch4:fig:cuboc1-kagome-state}    \\ 
  \hline
 3 & $(\tan^{-1} \Phi)/\pi $&  Regular icosahedron$(I_1)$~\ref{Ch4:fig:Icosahedron1-kagome-state}  \\ 
  \hline
 4 &$\tan^{-1}(3/4)$ & Cuboc2~\ref{Ch4:fig:cuboc2-kagome-state}   \\ 
  \hline
 5 &1/2 +$(\tan^{-1} \Phi)/\pi $&  Regular icosahedron$(I_2)$~\ref{Ch4:fig:Icosahedron2-kagome-state}  \\ 
 \hline
 \hline
\end{tabular}
\caption{List of regular magnetic orders for the solution-(vi) compatible with the $p6$ wallpaper group. (Here, the symbol $\Phi$ indicates the golden ratio)}
\label{tbl:p6groupsolution}
\end{center}
\end{table}

We list all the possible RMOs corresponding to each of the algebraic symmetry groups in the case of triangular and kagome lattices for the wallpaper group $p6$. Details of these states are given in the Discussion section. (`$\times$' mark implies there is no RMO corresponding to the algebraic symmetry group)
\begin{table*}[ht!]
\begin{center}
\begin{tabular}{|c|c|c|c|c|c|c|}
 \hline
 \hline
No.		&    $G_{T_1}$            &     $G_{T_2} $             &    $G_{R_3}$              &  Direction                 &  Triangular   &    Kagome  \\ 
\hline
1		&    I                             &     I                               &    I                               &  -                               &           F       &    F  \\ 
\hline
2              &    I                            &     I           &   $R(\hat{n}, \frac{2\pi}{3})$            &  Any                 &  F   &    F,U  \\ 
\hline
3              &$R(\hat{n}, \frac{2\pi}{3})$&   $R(\hat{n}, \frac{2\pi}{3})$         &   I             &  Any                 &  U   &$\sqrt{3} \times \sqrt{3}$(U)   \\ 
\hline
4              &$R(\hat{n}, \frac{2\pi}{3})$&  $R(\hat{n}, \frac{2\pi}{3})$   &   $R(\hat{n}, \frac{2\pi}{3})$     &  Any   &  F   &  $U_1$  \\ 
\hline
5              &$R(\hat{n}, \frac{2\pi}{3})$&   $R(\hat{n}, \frac{2\pi}{3})$   &  $R(-\hat{n}, \frac{2\pi}{3})$ &  Any &  F    &    $U_2$   \\ 
\hline
6              &$ R(\hat{x},\pi)$&  $ R(\hat{y},\pi)$ &   $R(\hat{w}, \frac{2\pi}{3})$           &  $\hat{w}=(1,1,1)$                 &  T   & $T,O,C_1,C_2,I_1,I_2$   \\ 
\hline
\hline
\end{tabular}
\caption{List of regular magnetic orders in $p3$ wallpaper group}
\label{tbl:ch4:p3-rmo-tbl}
\end{center}
 \end{table*}
 \section{$p3$ group}
 For this wallpaper group, the generators are $T_1, T_2$ and  $R_3$. Now, we can write the product of generators in the following form $R_3^r T_1^{t_1}T_2^{t_2}$ where  $r=0,1,2$ and $t_1, t_2 \in \mathbb{Z}$. Let $G_{T_1}, G_{T_2}$, and $G_{R_3}$ be the image of the generators $T_1,T_2$, and $R_6$. Then, we have the following equations
 \begin{subequations}
\begin{eqnarray}
G_{T_1}G_{T_2} & = & G_{T_2}G_{T_1}  \\
G_{R_3}G_{T_1} & = & G_{T_2}G_{R_3}  \\
G_{R_3} G_{T_2} & = & G_{T_2}^{-1} G_{T_1}^{-1} G_{R_3}  \\
G_{R_3}^3 & = & I 
\end{eqnarray}
\label{ch3:eqn-p3-mapping}
\end{subequations}
Here $\epsilon_{R_3}$ can only be $+1$. 
\subsection{RMOs in triangular lattice}
Let us consider that the spin configuration at the starting point is $\phi(0,0)$. So, the spin configuration at any arbitrary site with oblique coordinate $(m,n)$ is given by
\begin{equation}
\phi(m,n) = G_{T_1}^m G_{T_2}^n \phi(0,0)
\end{equation}
Under $R_3$ rotation, we must have,
\begin{equation}
\phi(m,n) = G_{R_3} \phi(n-m,-m)
\end{equation}
After some algebra we get the following condition
\begin{equation}
G_{T_1}^m G_{T_2}^n \phi(0,0) = G_{T_1}^m G_{T_2}^n G_{R_3} \phi(0,0)
\end{equation}
In the last step, we have used the fact that $G_{R_3} G_{T_1} G_{R_3}^{-1} = G_{T_2}$ and $G_{R_3} G_{R_2} G_{R_3}^{_1} = G_{T_1}^{-1}G_{T_2^{-1}}$. Thus $\phi(0,0)$ becomes the eigenvectors of $G_{R_3}$ with eigenvalue $+1$. Now, we can list the possible RMOs for the $p3$ group in case triangular lattice. The compatible RMOs are ferromagnetic state, umbrella state, which includes the coplanar state as well as ferromagnetic state and tetrahedral state.
\subsection{RMOs in kagome lattice}
Consider an arbitrary lattice point with sublattice index $\mu$ in a unit cell with left bottom corner at $m \vec{a} + n \vec{b} $. Clearly
\begin{eqnarray}
\phi(m,n,B) & =& G_{R_3} \phi(n-m,-m,A) \nonumber \\
& = & G_{R_3}G_{T_1}^{n-m}G_{T_2}^{-m} \phi(0,0,A)
\end{eqnarray}
After doing some algebra we get,
\begin{equation}
G_{T_2}^{n}G_{T_1}^m \phi(0,0,B) = G_{T_2}^n G_{T_1}^m G_{R_3} \phi(0,0,A)
\end{equation}
Thus we get,
\begin{eqnarray}
\phi(0,0,A) & = & G_{R_3} \phi(0,0,C) \nonumber \\
& = & G_{R_3}^2 \phi(0,0,B) = G_{R_3}^3 \phi(0,0,A)
\end{eqnarray}
This implies that $\phi(0,0, A)$ must be an eigenvector of $G_{R_3}^3$, which is the identity in all cases. Therefore there is no restriction on $\phi(0,0,\mu)$. Using this, we can calculate the spin arrangement for the entire lattice. The first five solutions lead to the ferromagnetic, umbrella, and $\sqrt{3} \times \sqrt{3}$ umbrella states,  umbrella1($U_1$) state, and  umbrella2($U_2$) state. Each of these umbrella states interpolates from planar structures to the planar structures. The last solution is quite interesting; much like the $p6$ group, we get icosahedron states with a series of regular structures. Finally, we list all the possible RMOs compatible with wallpaper group $p3$ in Table.~\ref{tbl:ch4:p3-rmo-tbl}.
\begin{table*}[ht!]
\begin{center}
\small
\begin{tabular}{|c|c|c|c|c|c|c|c|}
 \hline
 \hline
  No.                     &    $G_{T_1}$                                          &     $G_{T_2} $      &     $G_{R_3} $                                           &    $G_\sigma$                                           & $\hat{n}$                              &  Triangular   &    Kagome  \\ 
 \hline
\multirow{2}*{1} & \multirow{2}*{I}                             &     \multirow{2}*{I}      & \multirow{2}*{I}                                  &       I                                                   &    -                         &         F         &  F     \\  \cline{5-8}
                            &                                                         &          &                                                             &     -I                                                   &    -                         &         $\times$       &   $\times$      \\
\hline                  
\multirow{2}*{2} &            \multirow{2}*{I}              &     \multirow{2}*{I}           & \multirow{2}*{I}                                  &       $R(\hat{n},\pi)$           &   Any                                              &          F       & F     \\   \cline{5-8}
                            &                                                      &           &                                                              &     -$R(\hat{n},\pi)$          &     Any                                              &         F         &   F     \\
\hline
\multirow{2}*{3} & \multirow{2}*{$R(\hat{n},\frac{2 \pi}{3}) $} &    \multirow{2}*{$R(\hat{n}, \frac{2\pi}{3})$}   & \multirow{2}*{I}  &       $R(\hat{n}_\sigma,\pi)$      &   $\hat{n}_\sigma \perp \hat{n}$                     &        P        &  $\sqrt{3} \times \sqrt{3}$(p)  \\   \cline{5-8}
                            &                                                           &      &                                                              &     -$R(\hat{n}_\sigma,\pi)$      &     $\hat{n}_\sigma \perp \hat{n}$                   &         U       &  $\sqrt{3} \times \sqrt{3}$ (U),F   \\
\hline
\multirow{2}*{4} & \multirow{2}*{$R(\hat{x}, \pi$)}      & \multirow{2}*{$R(\hat{y}, \pi$) }             & \multirow{2}*{$R(\hat{w}, \frac{2\pi}{3})$}   &       $R(\hat{n},\pi)$           &  (-1,0,1)   &         $\times$        &                   $C_1$    \\   \cline{5-8}
                           &                                                            &      &                                                              &       -$R(\hat{n},\pi)$         &  (-1,0,1)   &        T          &$O,I_1,T_1,I_2,T_2,C_2$\\
\hline
\hline
\end{tabular}
\caption{List of regular magnetic orders in  $p3m1$ wallpaper group($\hat{w} = (1,1,1)$)}
\label{tbl:p3m1group}
\end{center}
\end{table*}
 \section{$p3 m1$ group}
 In this wallpaper group, the lattice symmetry group $\mathcal{G}_\mathcal{L}$ is generated by four generators $T_1, T_2, R_3$ and $\sigma$. Now, we can write product of generators in the following form $\sigma^s R_3^r T_1^{t_1}T_2^{t_2}$ where  $r=0,1,2$, $s =0,1$ and $t_1, t_2 \in \mathbb{Z}$. Let $G_{T_1}, G_{T_2}$, $G_{R_3}$, and $G_\sigma$ be the image of the generators $T_1,T_2$, $R_3$, and $\sigma$. Then, we have the following equations.
\begin{subequations}
\begin{eqnarray}
G_{T_1} G_{T_2}  & = & G_{T_2} G_{T_1} \\
G_{T_1} G_{R_3} & =  & G_{R_3} G_{T_1}^{-1} G_{T_2}^{-1} \\
G_{T_2} G_{R_3} &= & G_{R_3}G_{T_1} \\
G_{T_2} G_\sigma & = & G_\sigma G_{T_2}^{-1} \\
G_{T_1} G_{\sigma} & = & G_\sigma G_{T_1}G_{T_2}   \\
G_{R_3} G_\sigma & = &  G_\sigma G_{R_3}^2 \\
G_{R_3}^3 & = & I \\
G_\sigma^2 & = & I  
\end{eqnarray}
\end{subequations}
Here, we see that $\epsilon_\sigma $ can take values $\pm 1$. 
\subsection{RMOs in triangular lattice}
Like previous cases, we can show that $\phi(0,0)$ must be an eigenvector of not only $G_{R_3}$ but also $G_\sigma$ with a positive of eigenvalue. If the eigenvector does not match, then there will not be any RMO. In this case, we obtained ferromagnetic state, coplanar state, umbrella state, and tetrahedral state for different algebraic symmetry groups as shown in Table.~\ref{tbl:p3m1group}.
\subsection{RMOs in kagome lattice} In case of kagome lattice we can show that $\phi(0,0,A)$ must be an eigenvector of $G_{T_1} G_\sigma G_{R_3}^2$ associated with positive eigenvalue. First, three solutions with $\epsilon_\sigma = \pm 1$ leads to the ferromagnetic and $\sqrt{3} \times \sqrt{3}$ umbrella states which includes planar states and also ferromagnetic states. However, for the fourth solution with $\epsilon_\sigma = 1$ leads to cuboc1 state and $\epsilon_\sigma = -1$ case is the most interesting one. In this case, we get icosahedron states with a series of regular structures. Finally, we list all the possible RMOs compatible with wallpaper group $p3m1$ in Table.~\ref{tbl:p3m1group}.
\section{$p31m$ group}
For this wallpaper group, the generators are $T_1,T_2, R_3$ and $\sigma^\prime$. The basic difference between the symmetry groups, $p3m1$, and $p31m$ is the choice of the reflection axis. The $p3m1$ wallpaper group corresponds to the reflection lines, which make angles of $30^o, 90^o$ and $150^o$ with one of the translation vectors, whereas $p31m$ wallpaper group corresponds to the reflection lines at an angle $0^o, 60^o$ and $120^o$ with respect to a translation vector. Another notable difference is that if one draws a basic hexagon, then the reflection lines of the $p3m1$ group never pass through corners of the hexagon, while the reflection lines for $p31m$ always pass through the corners. Now, we can write product of generators in the following form $\sigma^{\prime s} R_3^r T_1^{t_1}T_2^{t_2}$ where $r=0,1,2$, $s =0,1$ and $t_1, t_2 \in \mathbb{Z}$. Let $G_{T_1}, G_{T_2}$, $G_{R_3}$, and $G_{\sigma^\prime}$ be the image of the generators $T_1,T_2$, $R_3$, and $\sigma^\prime$. Then, we have the following equations.
\begin{subequations}
\begin{eqnarray}
G_{T_1} G_{T_2} & = & G_{T_2} G_{T_1} \\
G_{T_1} G_{R_3} &= & G_{R_3} G_{T_1}^{-1}G_{T_2}^{-1}\\
G_{T_2} G_{R_3} & = & G_{R_3} G_{T_1} \\
G_{T_2} G_{\sigma^\prime} & = &G_{\sigma^\prime} G_{T_1}^{-1}G_{T_2}^{-1}\\
G_{T_1} G_{\sigma^\prime} & = & G_{\sigma^\prime} G_{T_1} \\
G_{R_3} G_{\sigma^\prime} & = & G_{\sigma^\prime} G_{R_3}^2 \\
G_{R_3} ^3& = & I \\
G_{\sigma^\prime}^2 & = & I
\end{eqnarray}
\end{subequations}
Here, we see that $\epsilon_{\sigma^\prime} $ can take values $\pm 1$.
\subsection{RMOs in triangular lattice}
Like previous cases, we can show that $\phi(0,0)$ must be an eigenvector or linear combination of the eigenvectors of not only $G_{R_3}$ but also $G_{\sigma^\prime}$ with positive eigenvalues. If there are no common eigenvectors between them, then there will not be any RMO. For this $p3m1$ group, we obtained ferromagnetic state, umbrella state, which includes planar structure as well as the ferromagnetic and tetrahedral state for different algebraic symmetry groups as shown in Table.~\ref{tbl:p31mgroup}.
\subsection{RMOs in kagome lattice}
For kagome lattice we can show that $\phi(0,0,A)$ can be chosen as the eigenvector of $G_{R_3}^2 G_\sigma G_{R_3}^{-1}$ corresponding to the positive eigenvalue. First, four solutions with $\epsilon_{\sigma^\prime} = \pm 1$ leads to the ferromagnetic and $\sqrt{3} \times \sqrt{3}$ umbrella states which includes planar states and also ferromagnetic states. However, for the fifth solution with $\epsilon_{\sigma^\prime} = 1$ leads to cuboc2 state and $\epsilon_{\sigma^\prime} = -1$ case is the most interesting one. For $\epsilon_{\sigma^\prime} = -1$, depending upon the initial vector $\phi(0,0,A)$ we get icosahedron states which contains several regular structure like octahedral, cuboc1, regular icosahedron1, tetrahedral and regular icosahedron2. Finally, we list all the possible RMOs compatible with wallpaper group $p31m$ in Table.~\ref{tbl:p31mgroup}.
\begin{table*}[ht!]
\begin{center}
\small
\begin{tabular}{|c|c|c|c|c|c|c|c|}
 \hline
 \hline
  No.                     &    $G_{T_1}$                                          &     $G_{T_2 }$      &     $G_{R_3} $                                           &    $G_{\sigma^\prime}$                                           & $\hat{n}$                              &  Triangular   &    Kagome  \\ 
 \hline
\multirow{2}*{1} & \multirow{2}*{I}                             &     \multirow{2}*{I}      & \multirow{2}*{I}                                  &       I                                                   &    -                         &         F         &  F     \\  \cline{5-8}
                            &                                                         &          &                                                             &     -I                                                   &    -                         &         $\times$       &   $\times$      \\
\hline                  
\multirow{2}*{2} &            \multirow{2}*{I}              &     \multirow{2}*{I}           & \multirow{2}*{I}                                  &       $R(\hat{n},\pi)$           &   Any                                              &          F       & F     \\   \cline{5-8}
                            &                                                      &           &                                                              &     -$R(\hat{n},\pi)$          &     Any                                              &         F         &   F     \\
\hline
\multirow{2}*{3} & \multirow{2}*{$R(\hat{n},\frac{2 \pi}{3}) $} &    \multirow{2}*{$R(\hat{n}, \frac{2\pi}{3})$}   & \multirow{2}*{I}  &      I      &   Any                     &        U        &  $\sqrt{3} \times \sqrt{3}$ (U),F \\   \cline{5-8}
                            &                                                           &      &                                                              &     -I     &     Any                &        $\times$      &  $\times$  \\
                            \hline
\multirow{2}*{4} & \multirow{2}*{$R(\hat{n},\frac{2 \pi}{3}) $} &    \multirow{2}*{$R(\hat{n}, \frac{2\pi}{3})$}   & \multirow{2}*{I}  &       $R(\hat{n}_\sigma,\pi)$      &   $\hat{n}_\sigma = \hat{n}$                     &        $\times$       & F    \\   \cline{5-8}
                            &                                                           &      &                                                              &     -$R(\hat{n}_\sigma,\pi)$      &     $\hat{n}_\sigma = \hat{n}$                   &         U       &  $\sqrt{3} \times \sqrt{3}$ (U),F   \\
\hline
\multirow{2}*{5} & \multirow{2}*{$R(\hat{x}, \pi$)}      & \multirow{2}*{$R(\hat{y}, \pi$) }             & \multirow{2}*{$R(\hat{w}, \frac{2\pi}{3})$}   &       $R(\hat{n},\pi)$           &  $(0,-1,1)$   &         $\times$        &             $C_2$    \\   \cline{5-8}
                           &                                                            &      &                                                              &       -$R(\hat{n},\pi)$         &  $(0,-1,1)$   &        T          &  O, $C_1$, $I_1$, $T$, $I_2$ \\
\hline
\hline
\end{tabular}
\caption{List of regular magnetic orders in  $p31m$ wallpaper group ($\hat{w} = (1,1,1)$)}
\label{tbl:p31mgroup}
\end{center}
\end{table*}
\section{Regular magnetic order with only translations ($p1$ group)}
The construction of RMOs using only translations is already discussed by Messio et al. ~\cite{messio2011lattice}. So, we will briefly discuss this here for the sake of completeness. In this case, the generators of the lattice symmetry group become commutative. So, the images by $G$ of the generators of the lattice symmetry group must satisfy the same algebraic constraints. Hence $G_{T_i}$ will commute among themselves. We can choose a reference unit cell with arbitrary spin directions. Then we choose an $O(3)$ rotation $\mathbf{R}(\hat{n},\alpha_i)$, associated with $G_{T_i}$ along the direction of the translation $T_i$. For fixed spin length, each spin characterized by two angles $\theta$ and $\phi$. Without losing generality, we can take the same axis $\hat{n}$ and unconstrained angles. We minimize the classical energy with respect to these eight parameters( six of them will come from the three sub-lattices and the other two from the $G_{T_i}$). These states correspond to the spin spiral(SS) states. Since the choices of the spins in the reference unit cell are arbitrary, the resulting spin spiral states not necessarily be planar. Another set of spin spiral states can be found by combining $G_{T_i}$ with $\pi$ rotations about some orthogonal spin direction.

\section{Discussions}

In this section, we will discuss each of the states found in all of the wallpaper groups as presented earlier. Messio et al. ~\cite{messio2011lattice} has shown that under some general conditions, these RMOs are a stationary point for the energy regardless of the Hamiltonian as long as the Hamiltonian commutes with the lattice symmetries. They have also argued that these RMOs that do not belong to a continuum are ``energetically stationary'' with respect to small spin deviations. Thus they are good candidates to be global energy minima. We calculate the energies and the equal time spin structure factors of each of the states. The spin structure factor(SSF) is defined as
\begin{equation}
S(\mathbf{Q}) = \frac{1}{N}\sum_{ij} e^{i~\mathbf{Q}\cdot (\mathbf{R}_i - \mathbf{R}_j)}  ~~\mathbf{S}_i \cdot \mathbf{S}_j
\label{eqn:sf}
\end{equation}

where $\mathbf{R}_i$ and $\mathbf{R}_j$ is the site index of $i$-th and $j$-th spins, respectively. $N$ denotes the total number of unit cells. The equal time structure factor $S(\mathbf{Q})$ is zero everywhere in the Brillouin zone except for a finite number of $\mathbf{Q}$ where there are sharp Bragg peaks. Magnetic long-range order is indicated by the sharp Bragg peaks in the hexagonal Brillouin zone.
We denote first, second, and third neighbor coupling by $J_1, J_2$ and $ J_3$ respectively( In the case of kagome lattice, there are two kinds of third neighbors, one is denoted by $J_3$, and the other one is denoted by $J_{3h}$). In all of the diagrams presented below, the unit cell is indicated by a dashed blue line. Each of the sub-lattices is shown by different colors in the diagram. The spin orientation is shown in the middle columns of the diagrams. The positions and weights of the Bragg peaks(shown by green dots) in the hexagonal Brillouin zone(BZ) and extended Brillouin zone(EBZ) is shown by the extreme right column for each of the diagrams. For better clarity, we have shown only one value of SSF in each of the Brillouin zones.
\begin{figure}[ht!]
\centering
	\includegraphics[width=0.2375\textwidth]{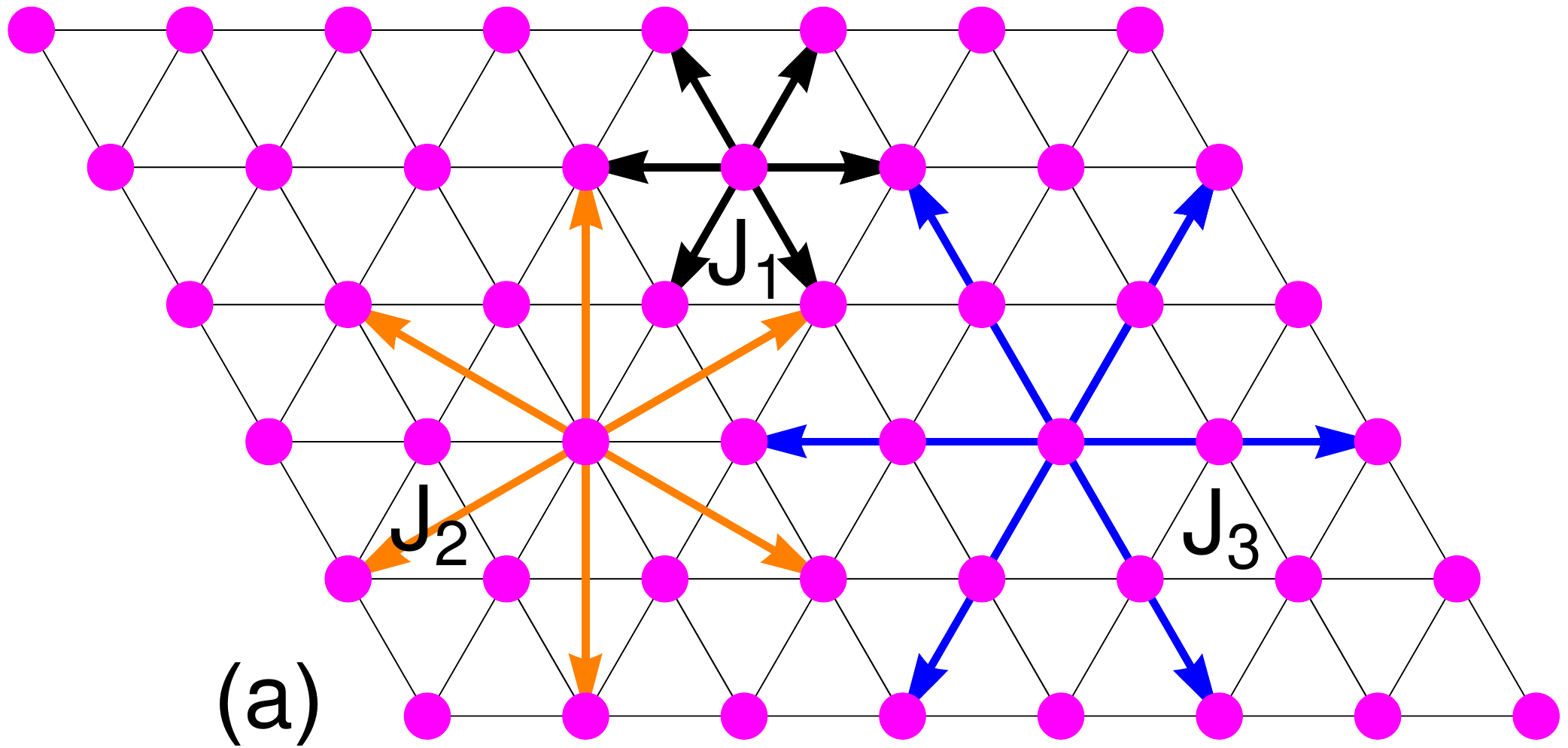}
	\includegraphics[width=0.2375\textwidth]{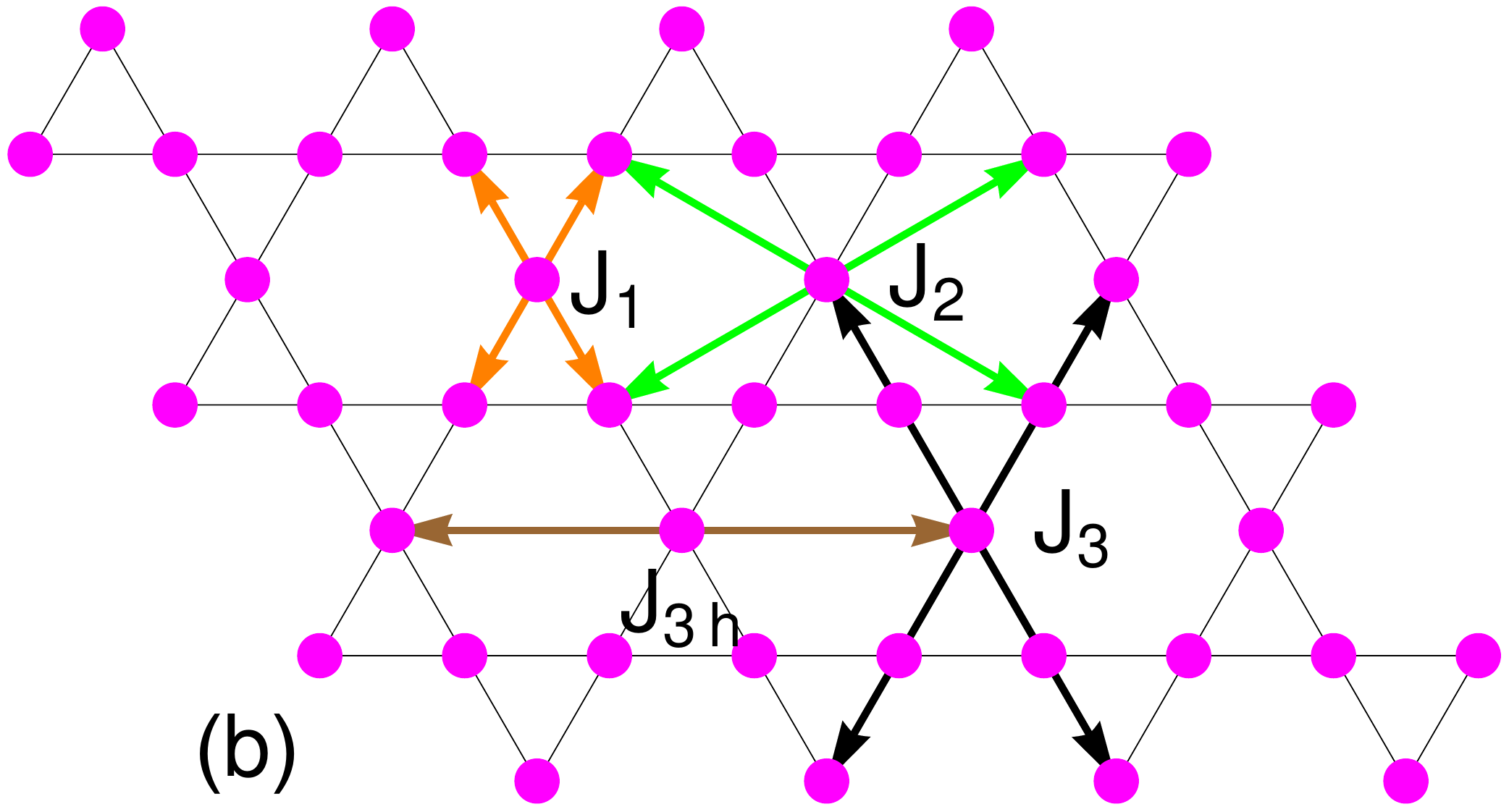}
	\caption{Pictorial description of exchange interaction in (a) triangular lattice (b) kagome lattice where $J_n$ be the exchange interaction between n-th neighbors}
	\label{fig:neighbors}
\end{figure}

\subsection{Triangular lattice } 
The triangular lattice consists of edge-sharing triangles and single-site unit cells. We have found the following RMOs for the triangular lattice.

\begin{figure}[ht!]
\begin{center}
\includegraphics[height=2.5cm,width=4.25cm]{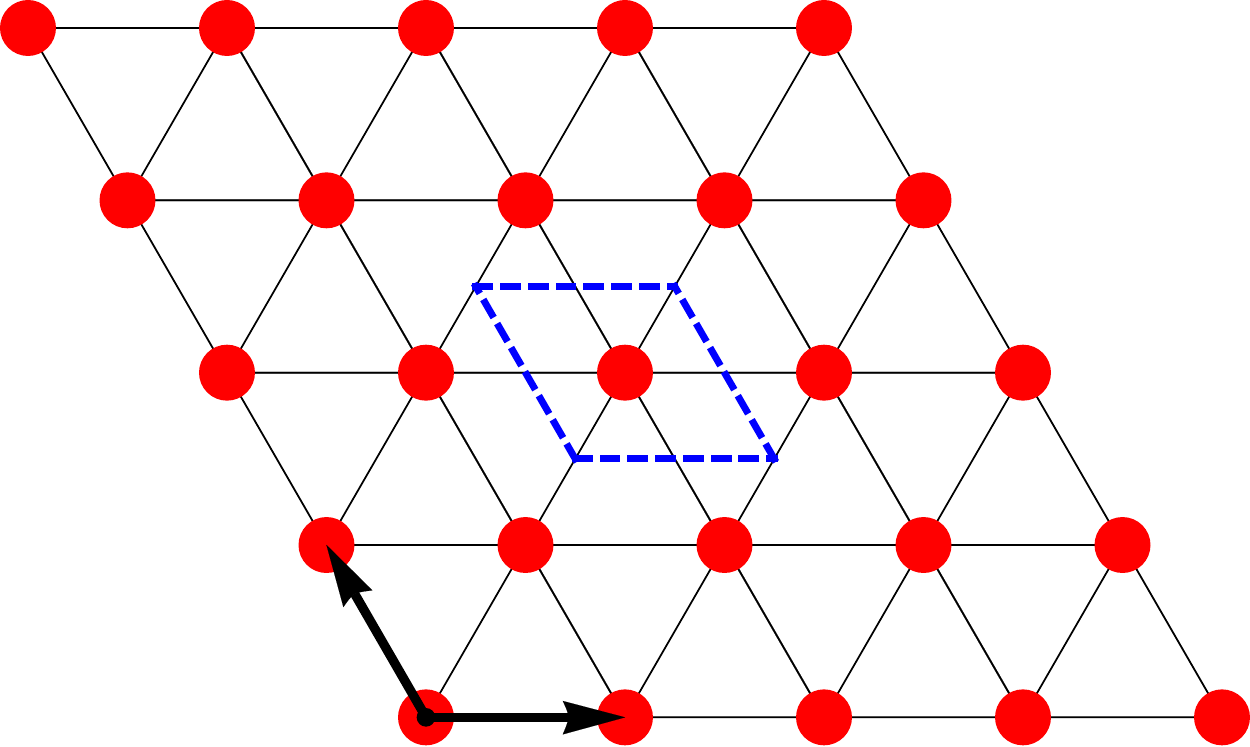}
\includegraphics[height=1.5cm,width=1.5cm]{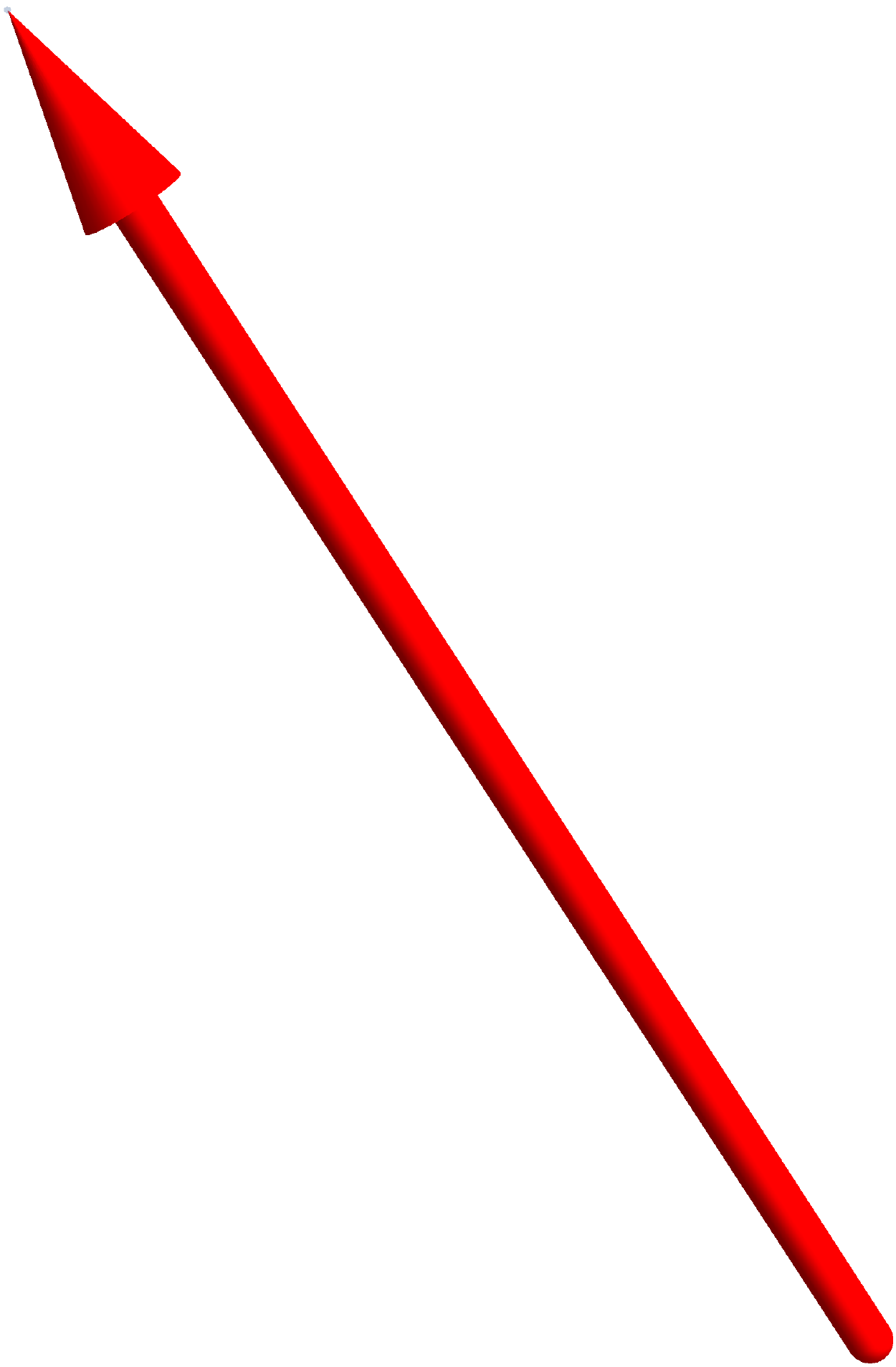}
\includegraphics[height=2.5cm,width=2.5cm]{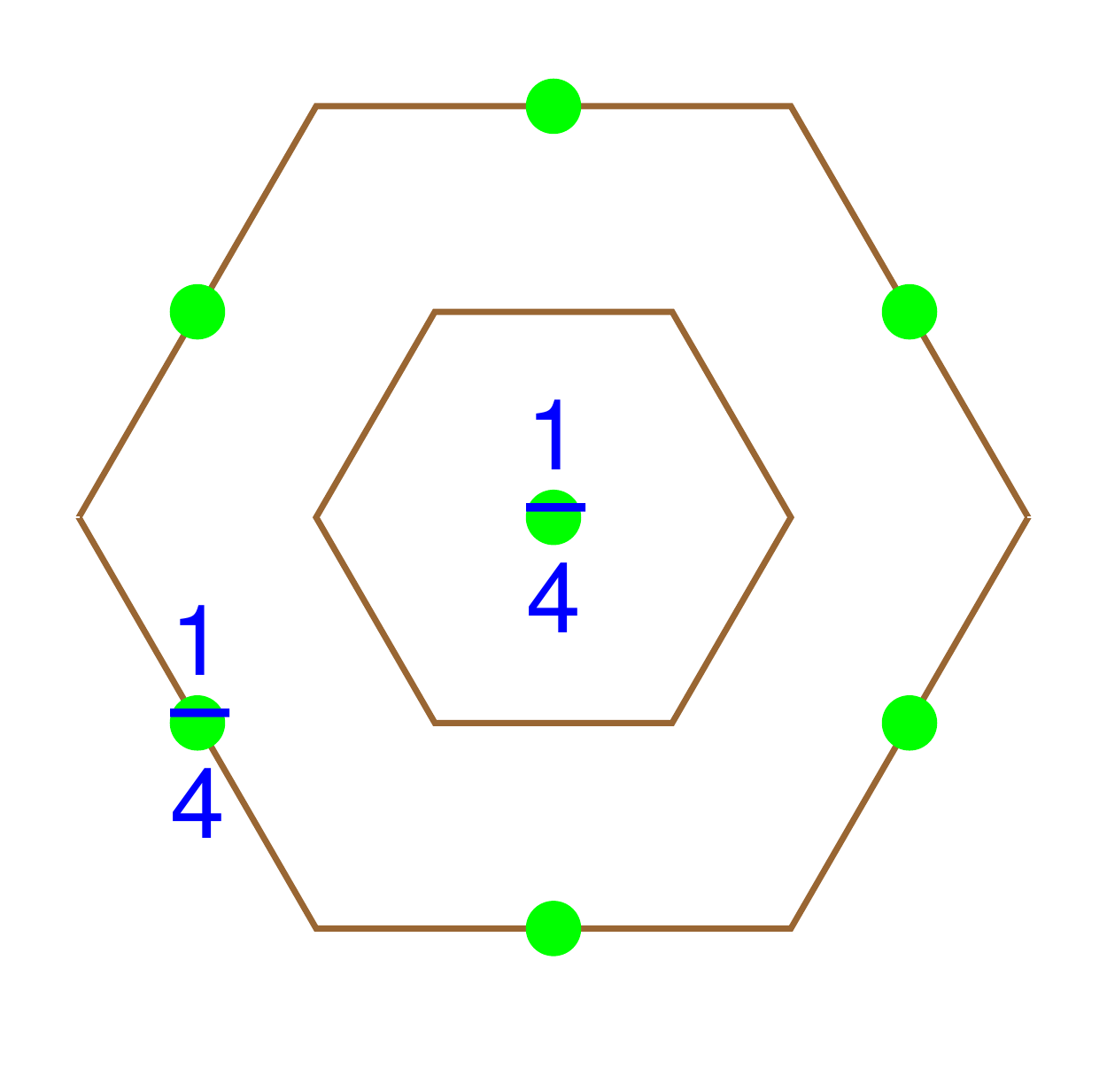}
\caption{Ferromagnetic(F) state}
\label{Ch4:fig:ferro-state}
\end{center}
\end{figure}

\vspace{0.2cm}
\noindent \textbf{(i) Ferromagnetic(F) state: } In the ferromagnetic state, all the spins are pointing in the same direction as shown in Fig.~\ref{Ch4:fig:ferro-state}. The unit cell contains one site. Energy per site is given by $E =6(J_1 +J_2 + J_3) $. 

\begin{figure}[ht!]
\begin{center}
\includegraphics[height=2.5cm,width=4.25cm]{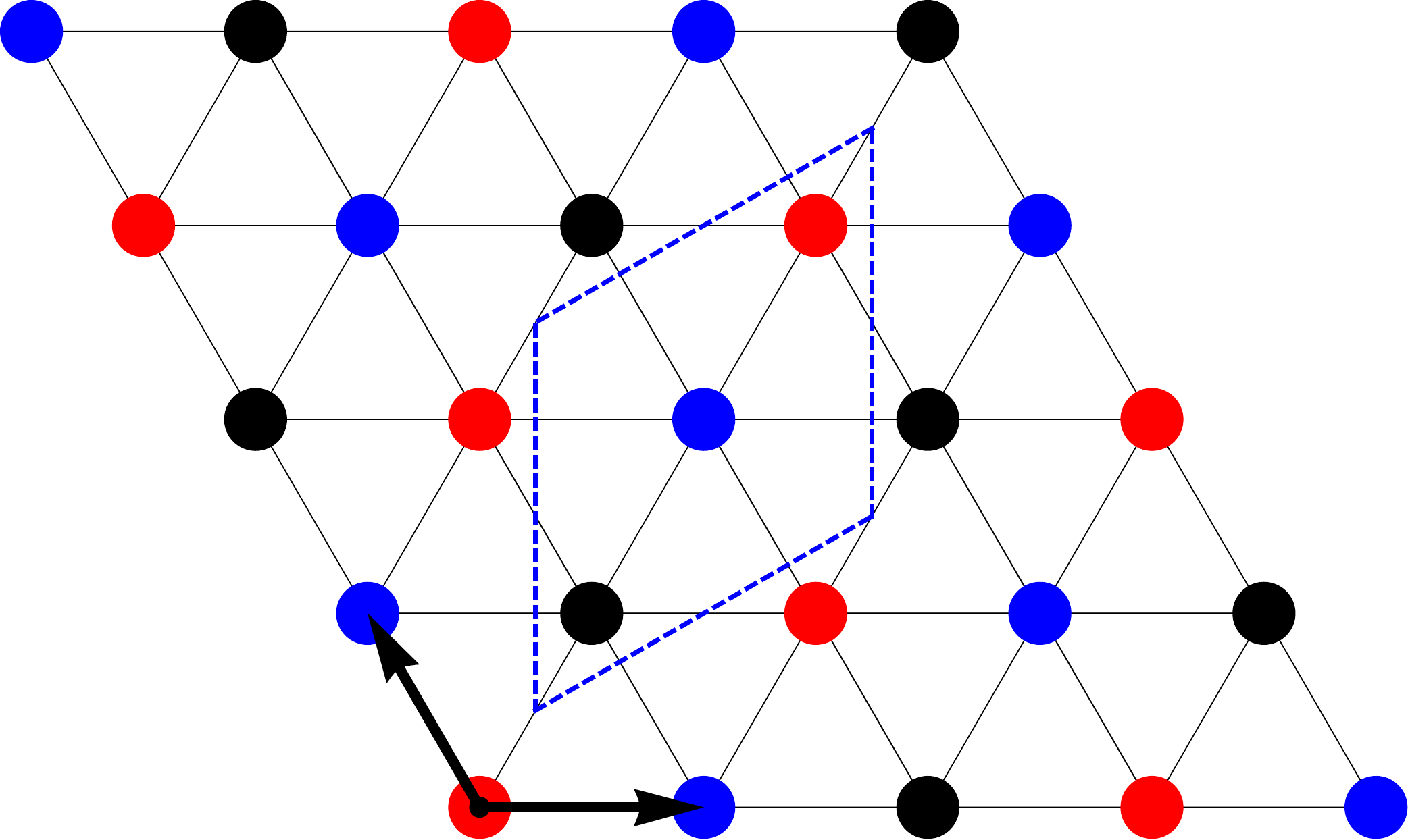}\hspace{-0.1cm}
\includegraphics[height=1.5cm,width=1.5cm]{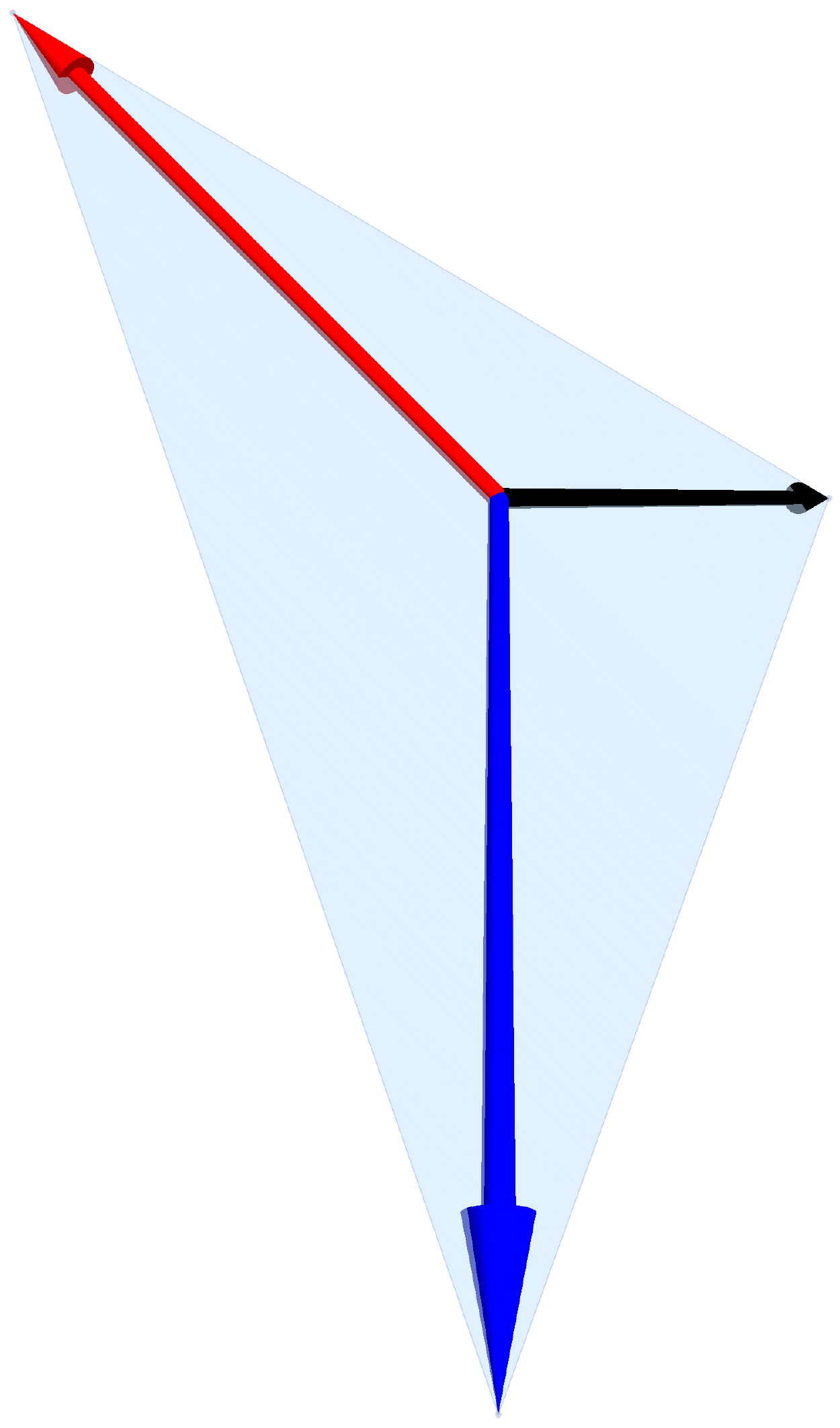}
\includegraphics[height=2.5cm,width=2.5cm]{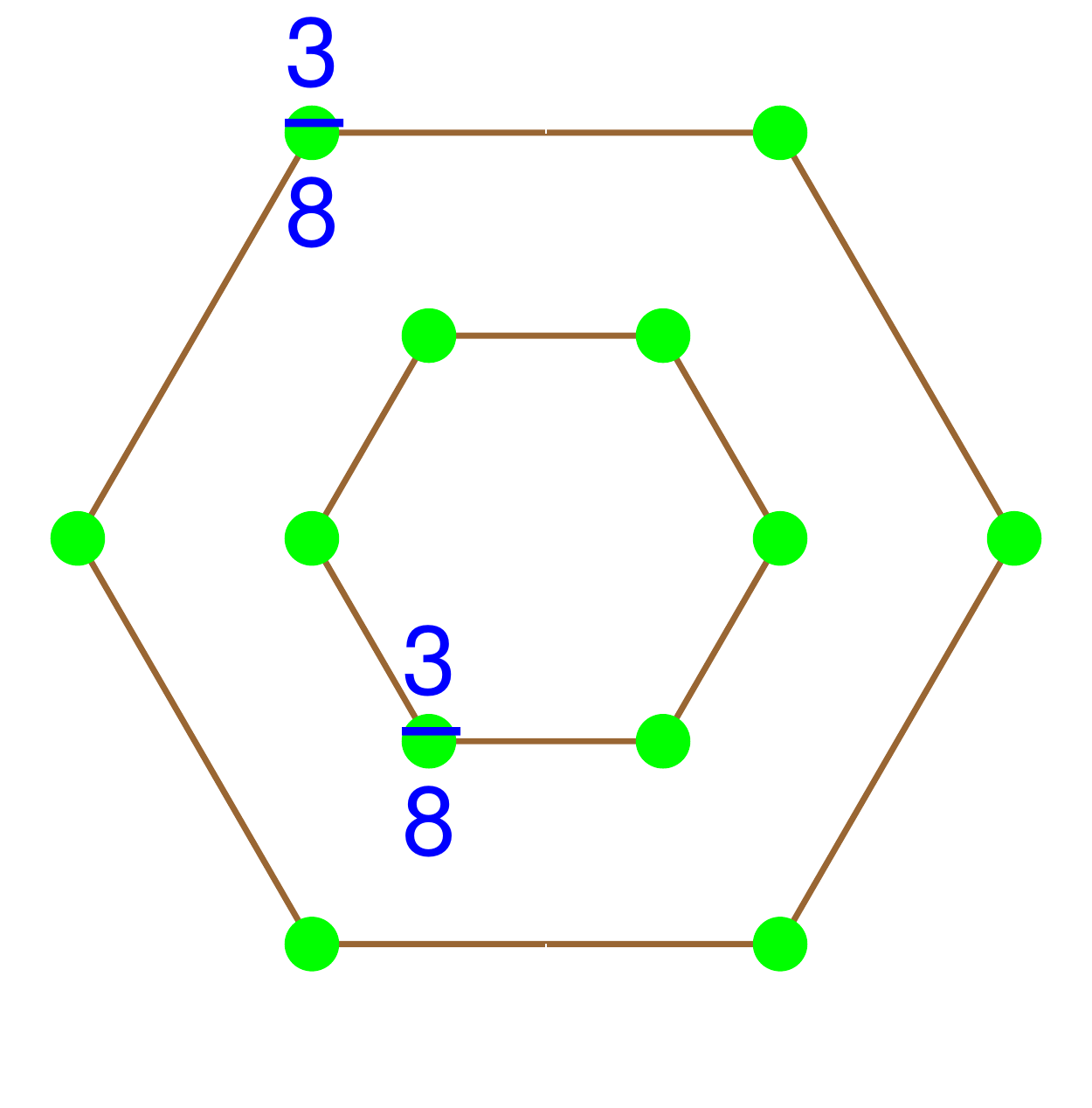}
\caption{Coplanar state.}
\label{Ch4:fig:Q0-planar-state}
\end{center}
\end{figure}

\vspace{0.2cm}
\noindent \textbf{(ii) Coplanar state: } This state contains three sub-lattices, and this structure is known as Coplanar structure, where in each triangle, the spins are at an angel $120^o$ to each other lying in a plane as shown in the Fig.~\ref{Ch4:fig:Q0-planar-state}. The unit cell contains three sites and the energy per site is given by $E = -3 J_1 + 6 J_2 - 3 J_3$.

\begin{figure}[ht!]
\begin{center}
\includegraphics[height=2.5cm,width=4.25cm]{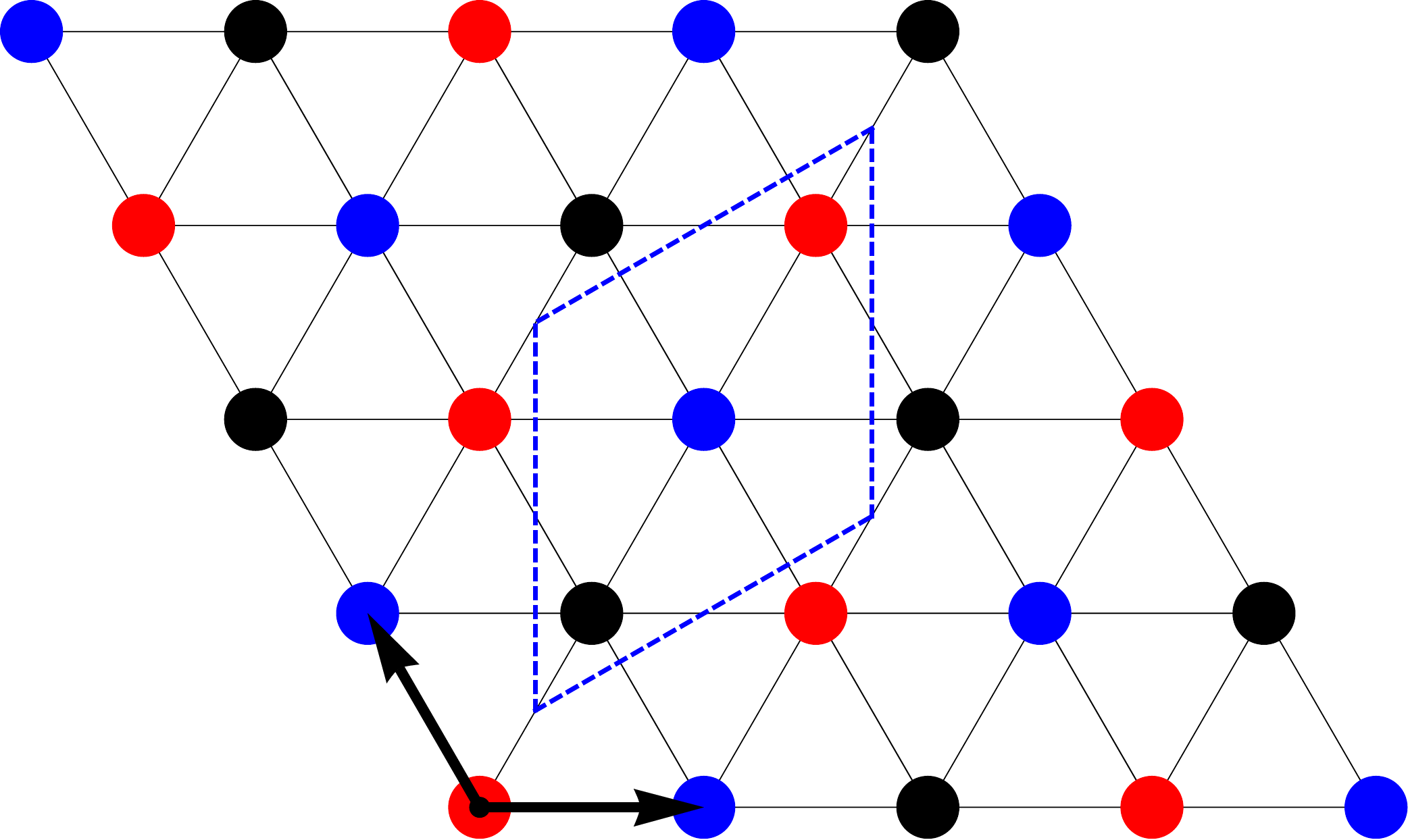}\hspace{0.75cm}
\includegraphics[height=2cm,width=2cm]{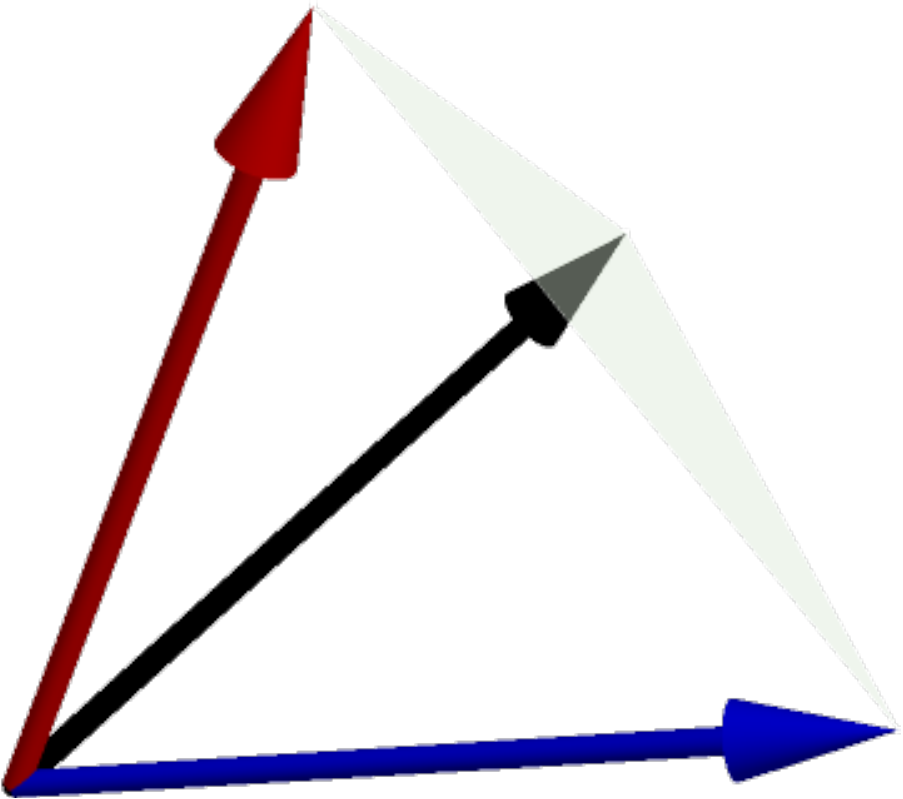}
\caption{F umbrella state.}
\label{Ch4:fig:Q0-umbrella-state}
\end{center}
\end{figure}

\vspace{0.2cm}
\noindent \textbf{(iii) F umbrella state: }
This is a non-coplanar structure with three sub-lattices, where the relative angle between the spins are the same and also $\leq 120^o$. In this case, the spin arrangements interpolate between the ferromagnetic state and the coplanar state. Some intermediate state is shown in Fig.~\ref{Ch4:fig:Q0-umbrella-state}. In case of umbrella states, the energy will always depend on the angle between the spins. The energy of a continuum can not be lower than the two extreme states, between which it interpolates. 

\begin{figure}[ht!]
\begin{center}
\includegraphics[height=2.5cm,width=4.25cm]{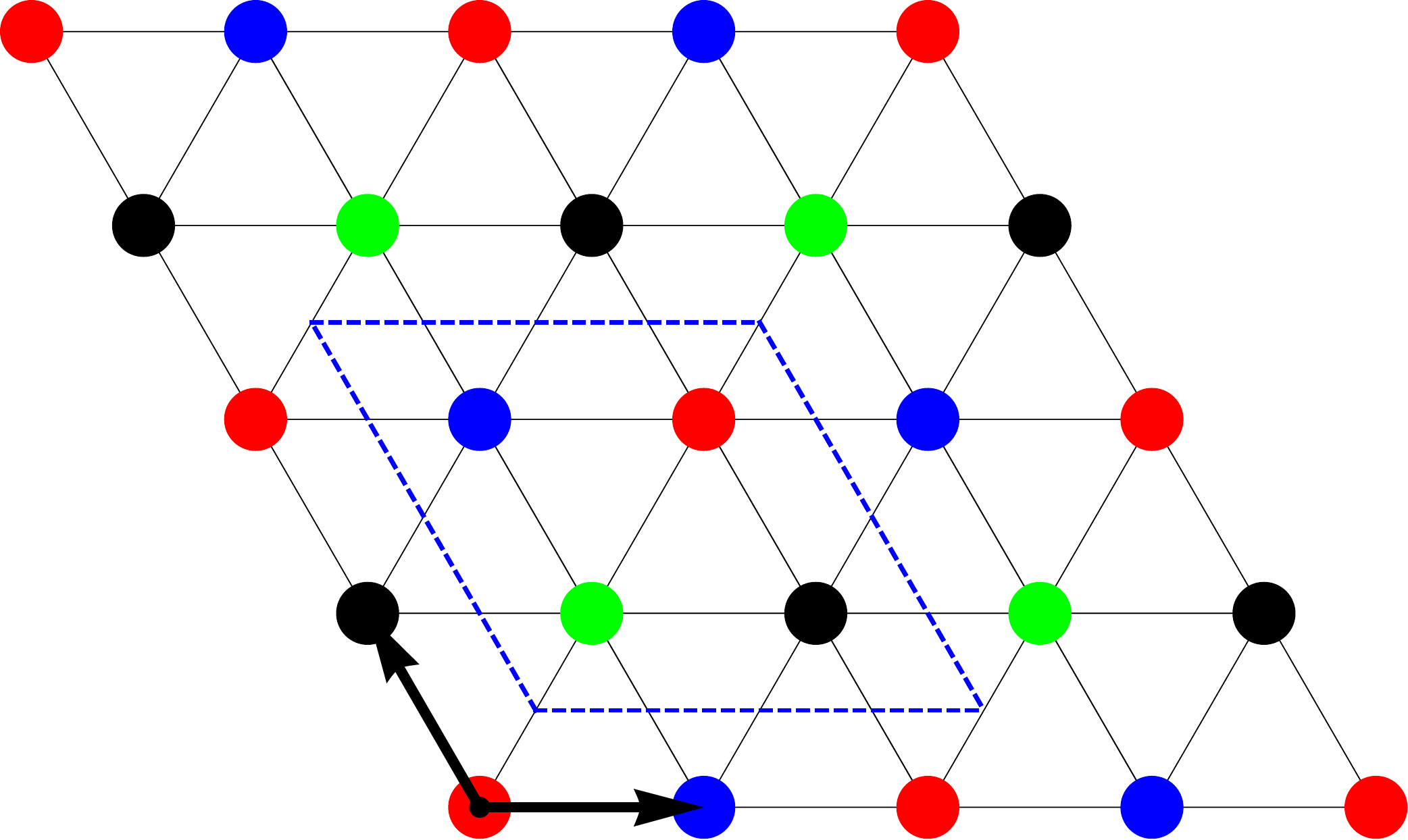}
\includegraphics[height=1.5cm,width=1.5cm]{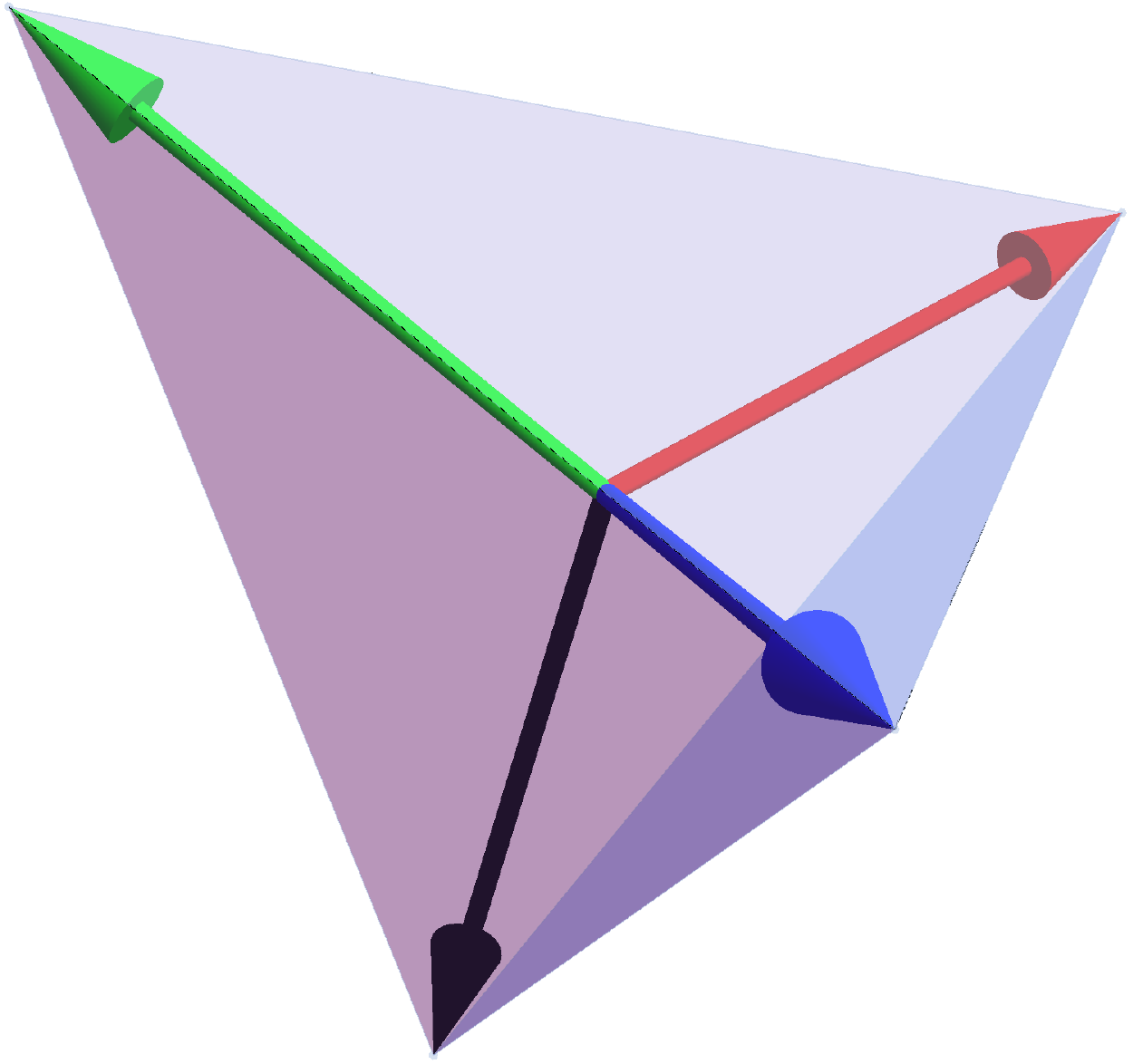}
\includegraphics[height=2.5cm,width=2.5cm]{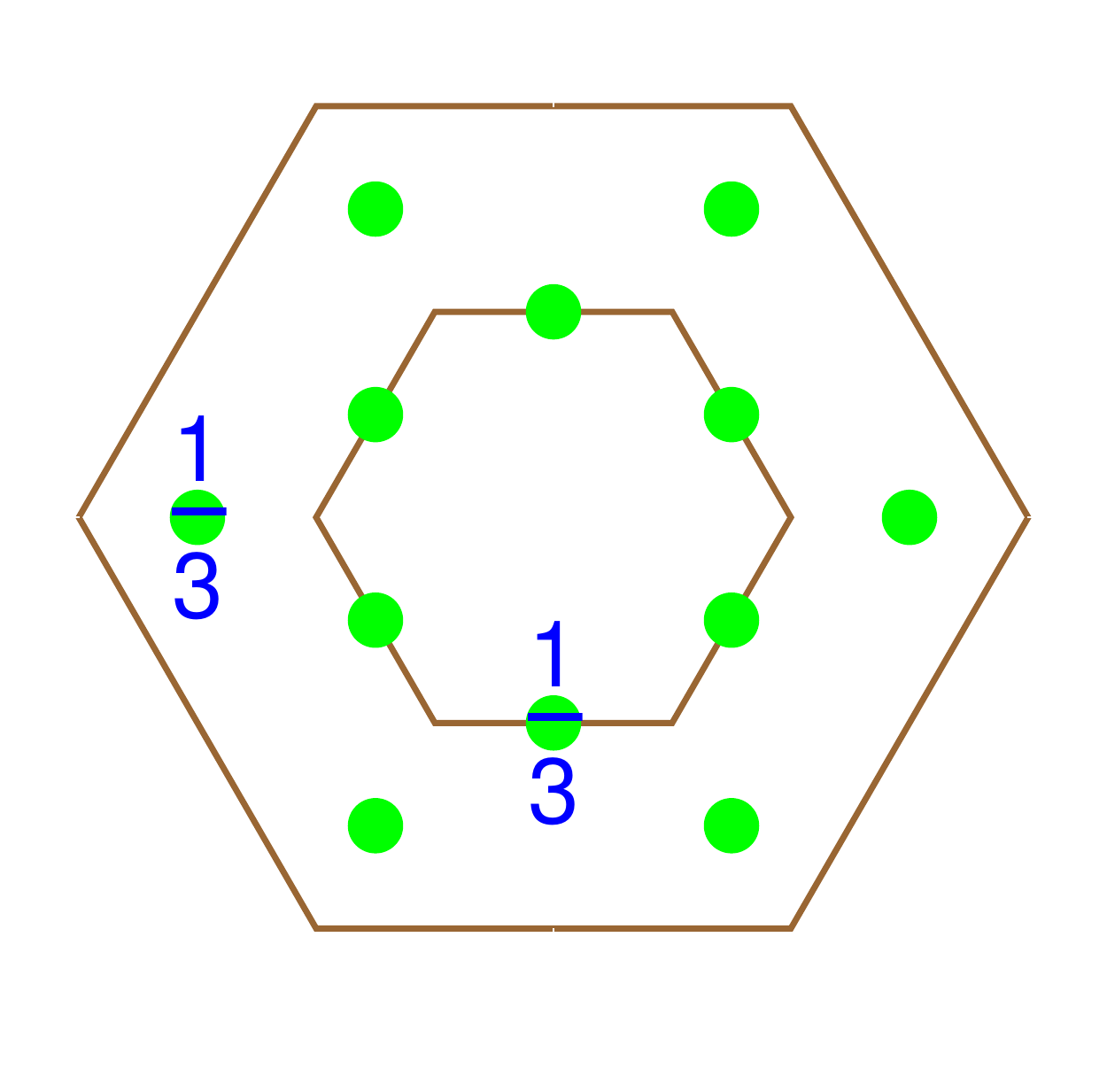}
\caption{Tetrahedral state.}
\label{Ch4:fig:Tetrahedral-state}
\end{center}
\end{figure}

\vspace{0.2cm}
\noindent \textbf{(iv) Tetrahedral(T) state: } The tetrahedral has four sub-lattices, pointing towards the corner of a tetrahedron, and the sub-lattices are shown by four different colors in Fig.~\ref{Ch4:fig:Tetrahedral-state}. The sign of $\phi(0,0)=\pm(1,1,1)$ determines the chirality of the spin configuration. The energy per site is given by $E = -2 J_1 -2 J_2 +6 J_3$.

\subsection{ Kagome lattice} 
The kagome lattice is a non-Bravais lattice consists of vertex sharing triangles. It has three sites per unit cell. For kagome lattice, we have got the following states.

\begin{figure}[ht!]
\begin{center}
\includegraphics[height=2.75cm,width=4.25cm]{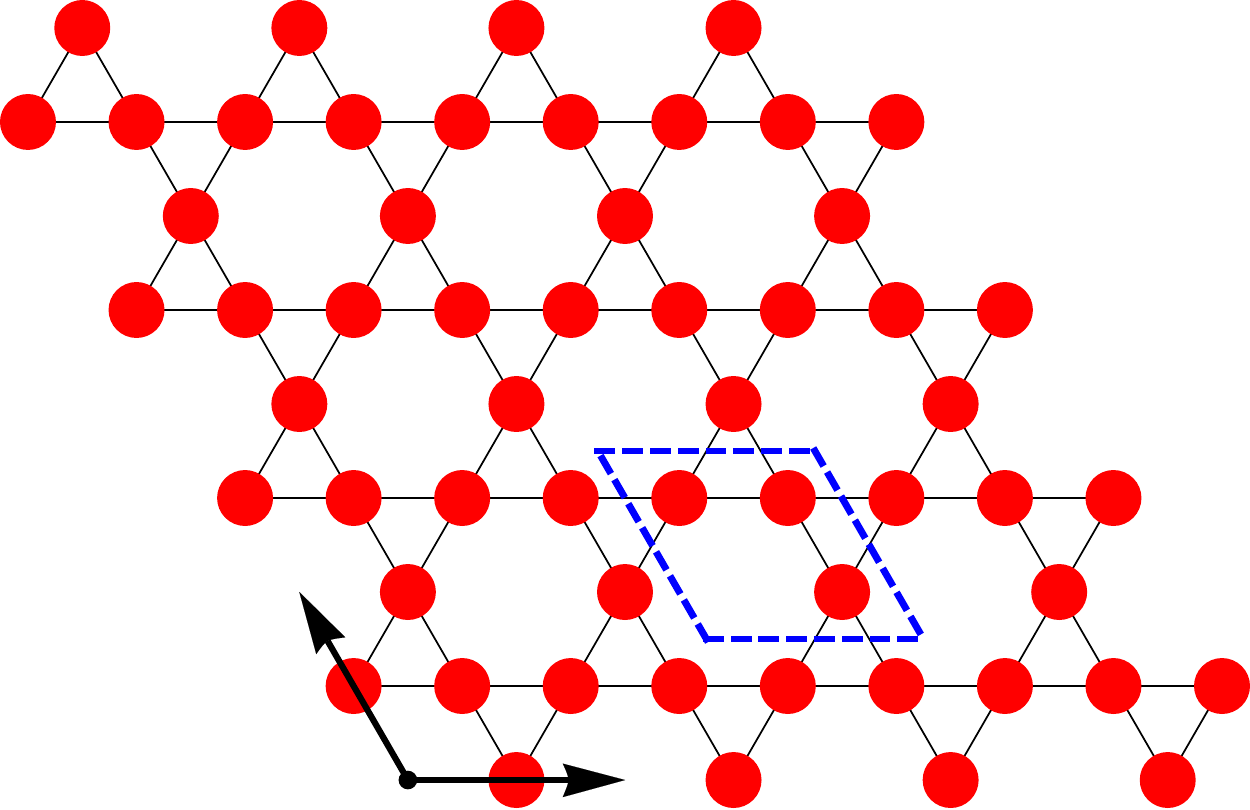}
\includegraphics[height=1.75cm,width=1.5cm]{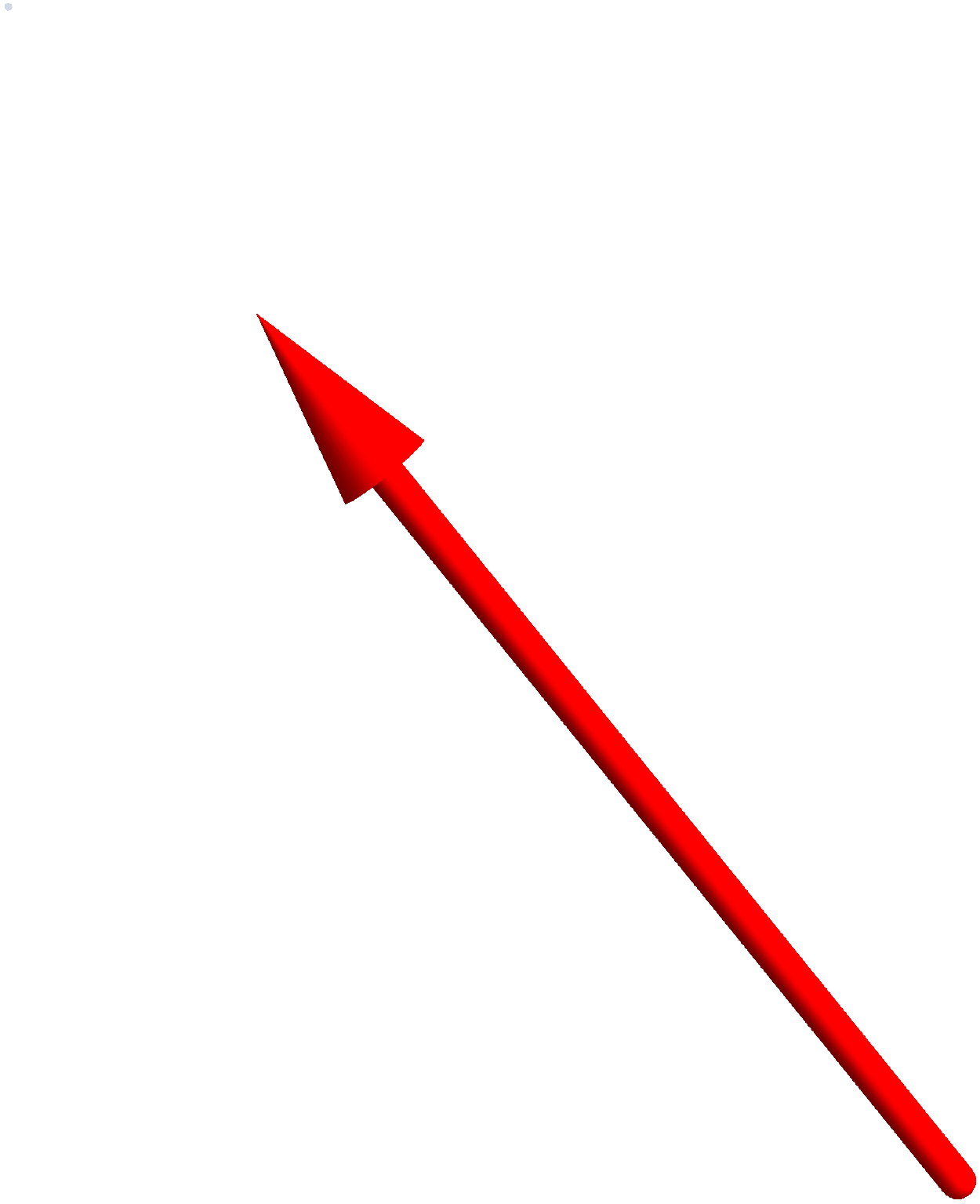}
\includegraphics[height=2.65cm,width=2.5cm]{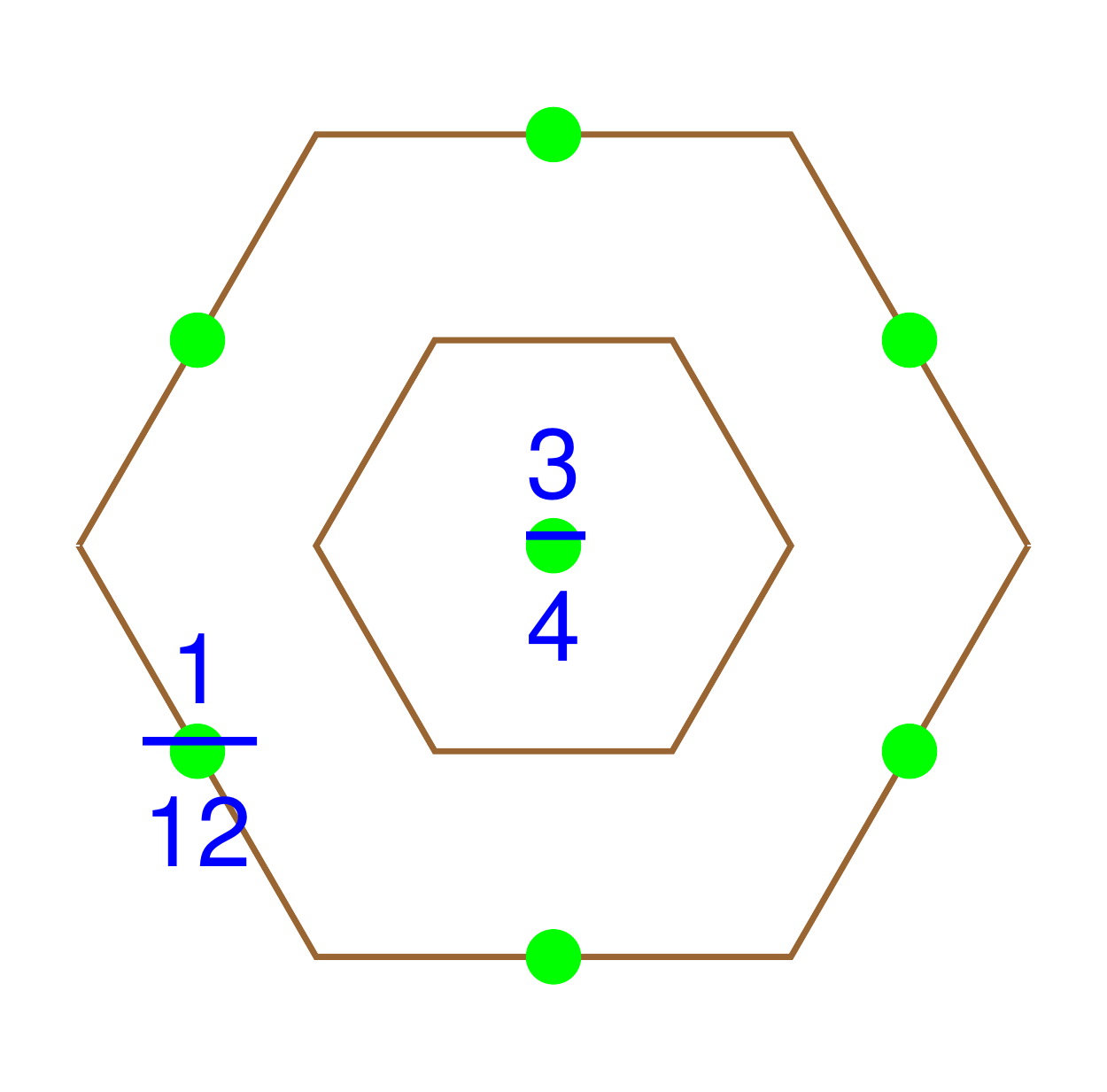}
\caption{Ferromagnetic(F) state}
\label{Ch4:fig:ferro-kagome-state}
\end{center}
\end{figure}

\vspace{0.2cm}
\noindent \textbf{(i) Ferromagnetic(F) state: } In ferromagnetic state, all the spins are pointing in the same direction as shown in Fig.~\ref{Ch4:fig:ferro-kagome-state}. The unit cell contains three sites and the energy per site is given by $E = 4 J_1 + 4 J_2 + 4 J_3 + 2 J_{3h}$.

\begin{figure}[ht!]
\begin{center}
\includegraphics[height=2.75cm,width=4.25cm]{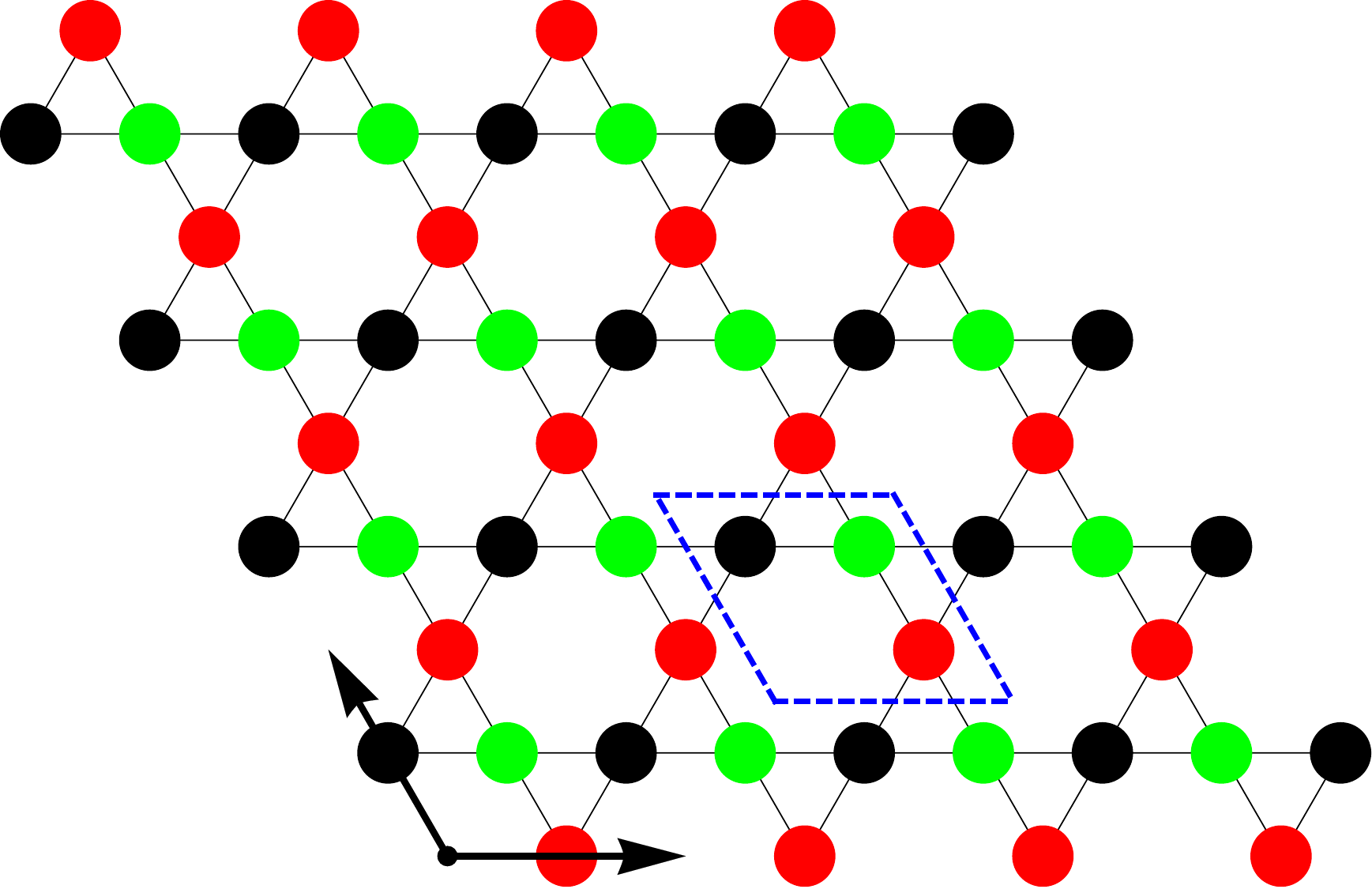}
\includegraphics[height=1.75cm,width=1.5cm]{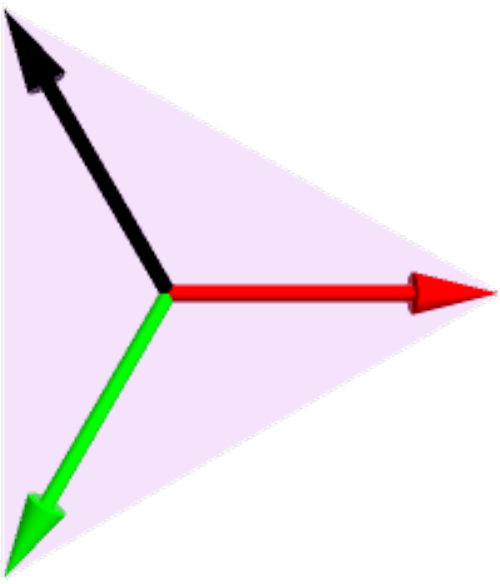}
\includegraphics[height=2.5cm,width=2.5cm]{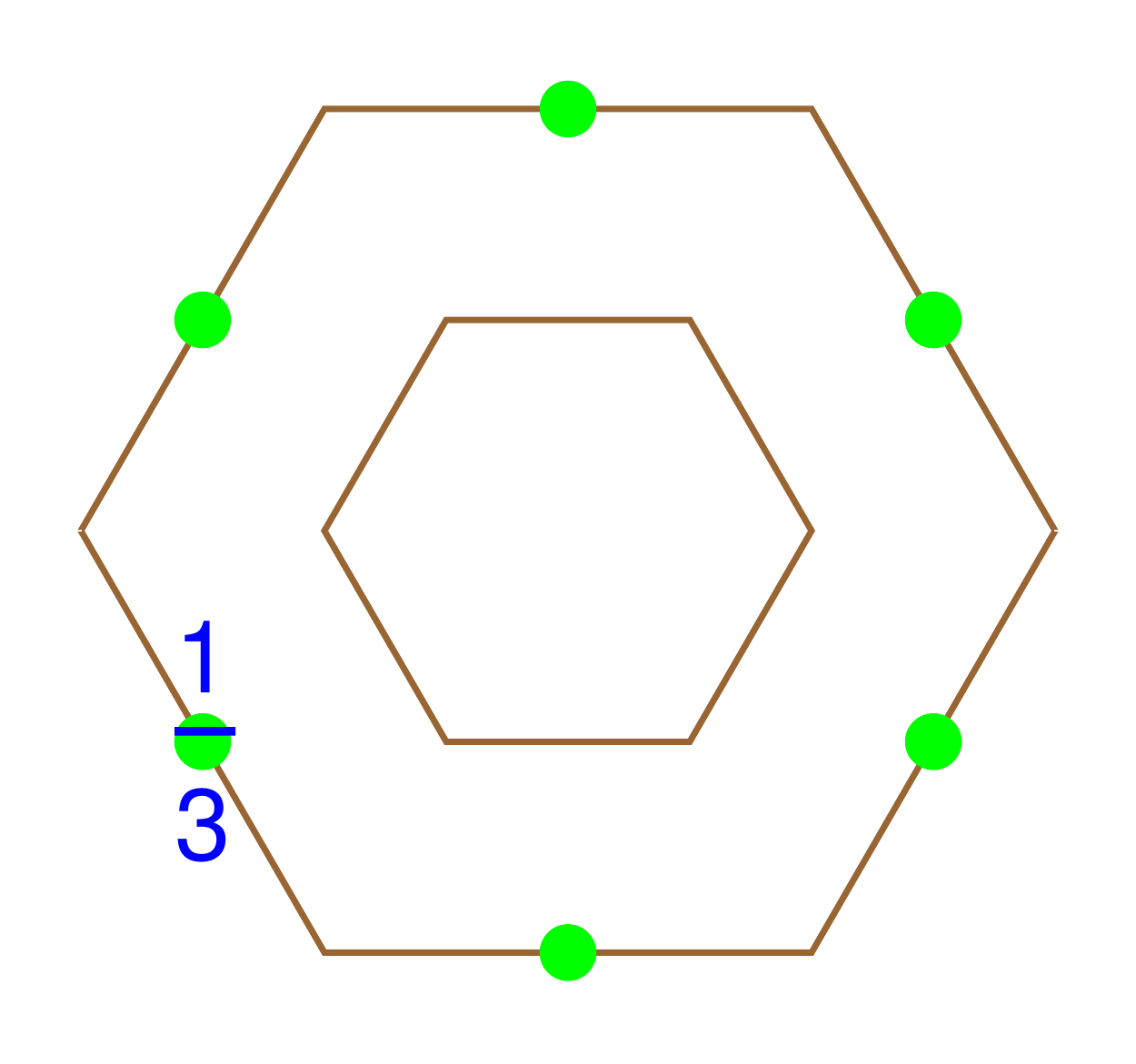}
\caption{$Q=0$ planar state}
\label{Ch4:fig:Q0-planar-kagome-state}
\end{center}
\end{figure}

\vspace{0.2cm}
\noindent \textbf{(ii) $\mathbf{Q = 0}$(P) state: } $\mathbf{Q=0}$ planar state has three sub-lattices, and the spins are at an angle $2 \pi/3$ to each other as shown in Fig.~\ref{Ch4:fig:Q0-planar-kagome-state}. The unit-cell contains three sites, and the spins lie in the kagome plane. For this state, the energy per site is given by $E = -2 J_1 - 2 J_2 + 4 J_3+2 J_{3h}$.

\begin{figure}[ht!]
\begin{center}
\includegraphics[height=2.75cm,width=4cm]{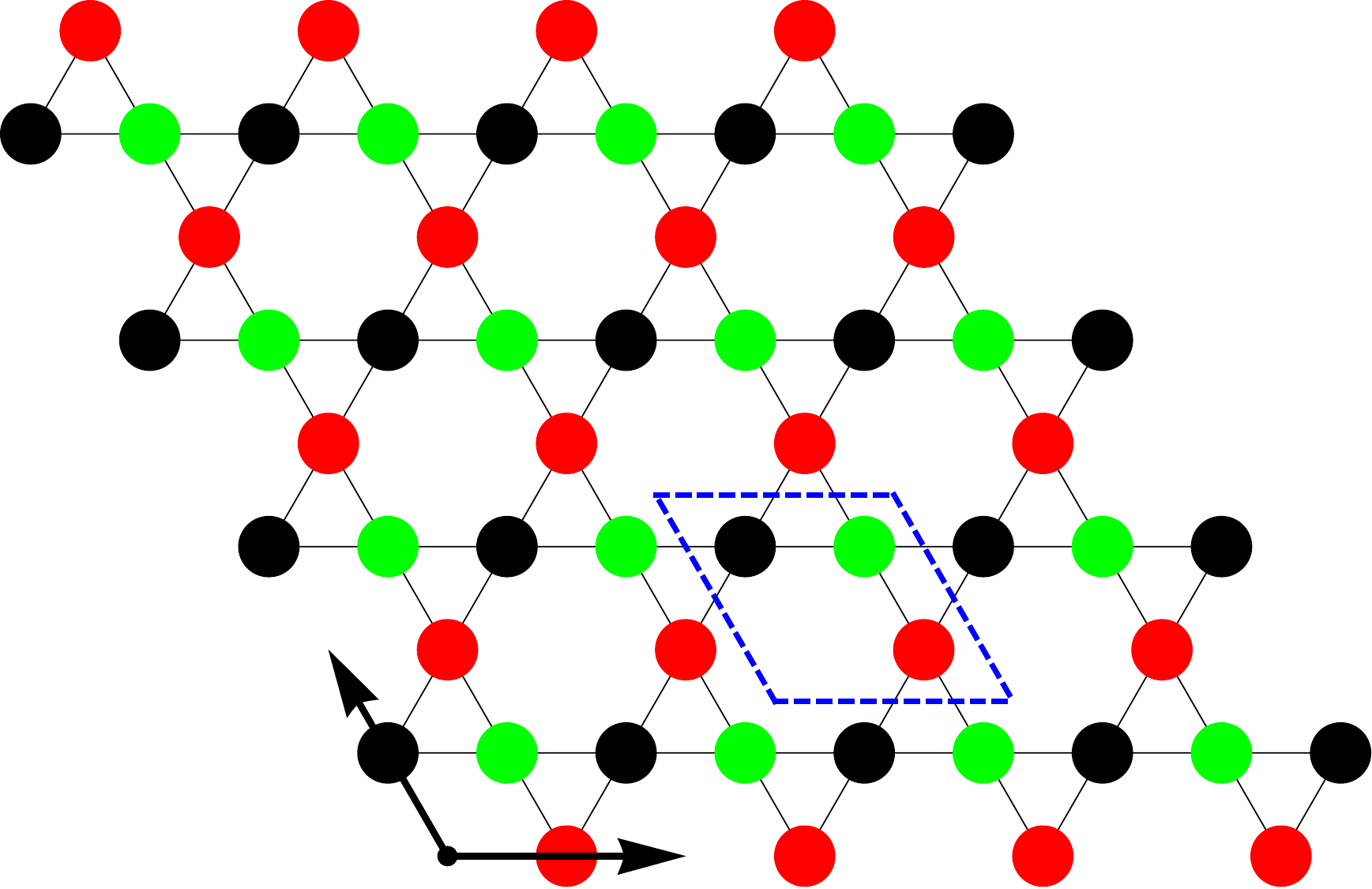}
\includegraphics[height=1.25cm,width=1.25cm]{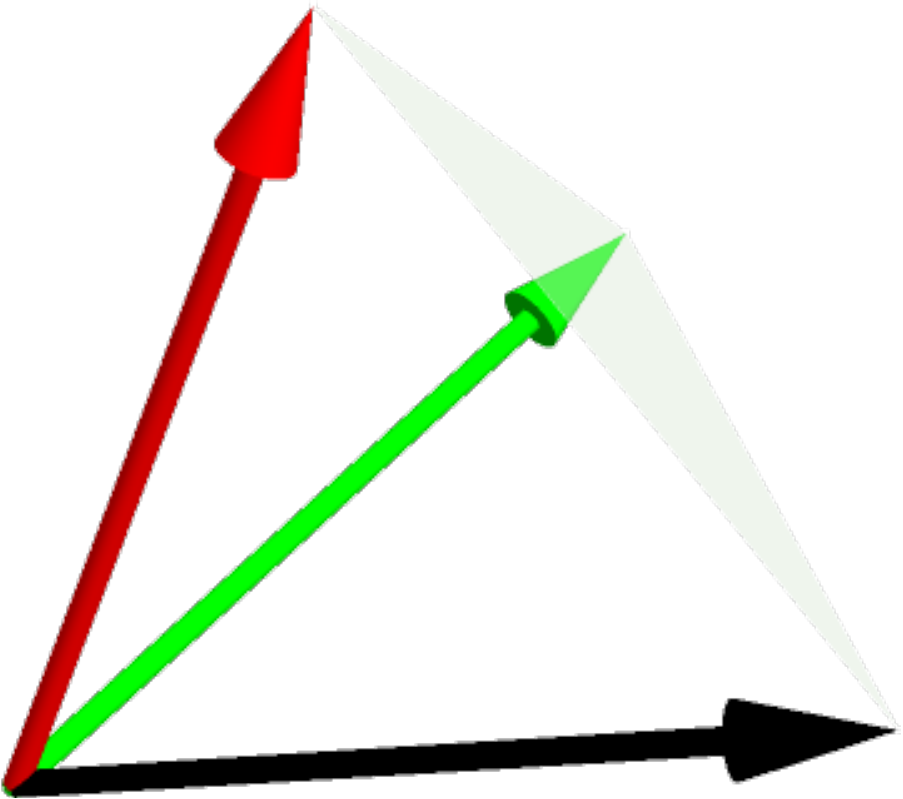}\hspace{-0.85cm}
\includegraphics[height=2.5cm,width=4cm]{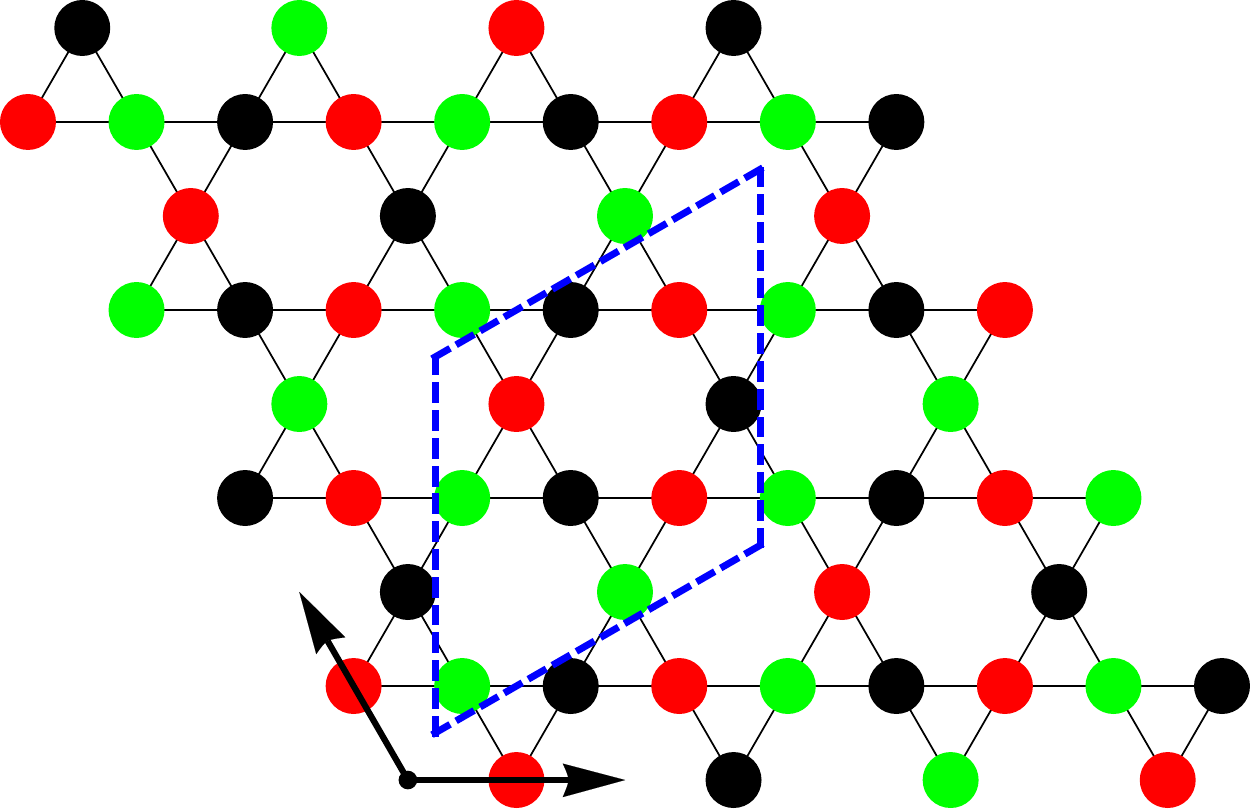}
\caption{$Q=0$ umbrella state and $\sqrt{3} \times \sqrt{3}$ umbrella state}
\label{Ch4:fig:Q0-umbrella-kagome-state}
\end{center}
\end{figure}

\vspace{0.2cm}
\noindent \textbf{(iii) $\mathbf{Q = 0}$(U) state: } $\mathbf{Q=0}$ umbrella state has three sub-lattices, and the relative angle between spins is identical. The angle varies from $0$ to $2 \pi/3$. If the relative angle between the spins is zero, it becomes ferromagnetic, whereas it becomes $\mathbf{Q=0}$ planar when the angle becomes $2 \pi/3$. Some intermediate state is shown in Fig.~\ref{Ch4:fig:Q0-umbrella-kagome-state}(left). In the case of umbrella states, the energy will always depend on the angle between the spins. The energy of a continuum can not be lower than the two extreme states, between which it interpolates. So, the $\mathbf{Q=0}$ umbrella state must lie between the energies of $\mathbf{Q=0}$ planar state and the ferromagnetic state.

\begin{figure}[ht!]
\begin{center}
\includegraphics[height=2.75cm,width=4.25cm]{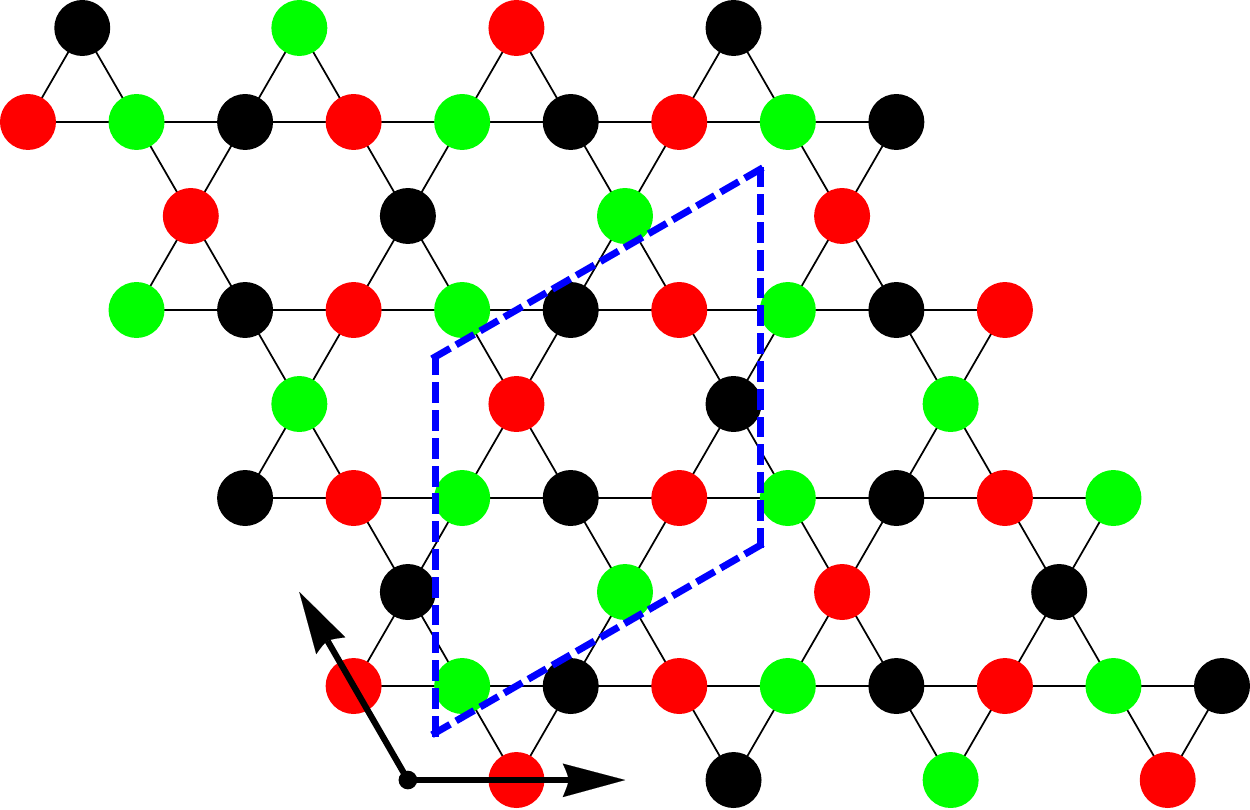}\hspace{-0.25cm}
\includegraphics[height=1.75cm,width=1.5cm]{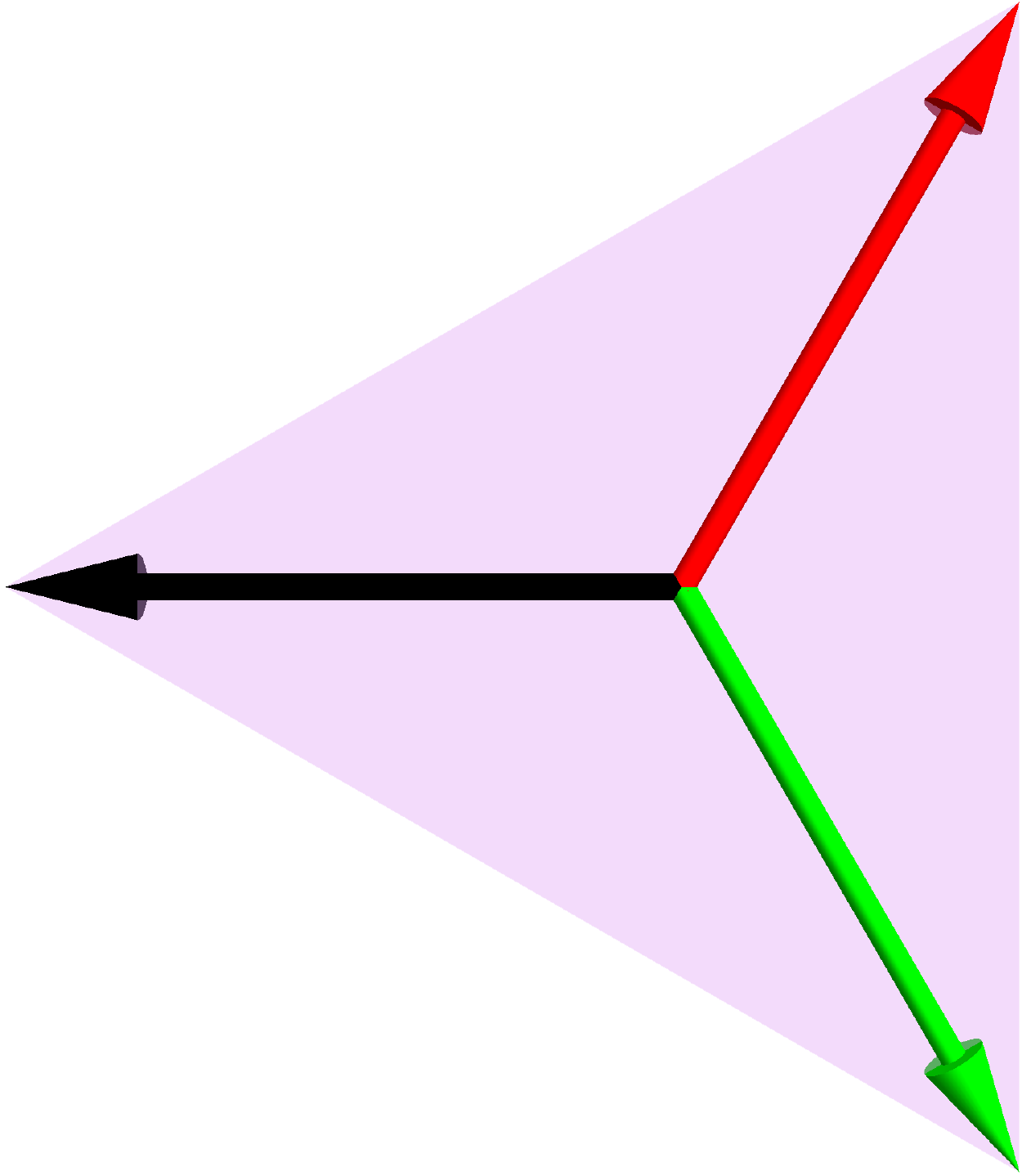} \hspace{0.2cm}
\includegraphics[height=2.5cm,width=2.5cm]{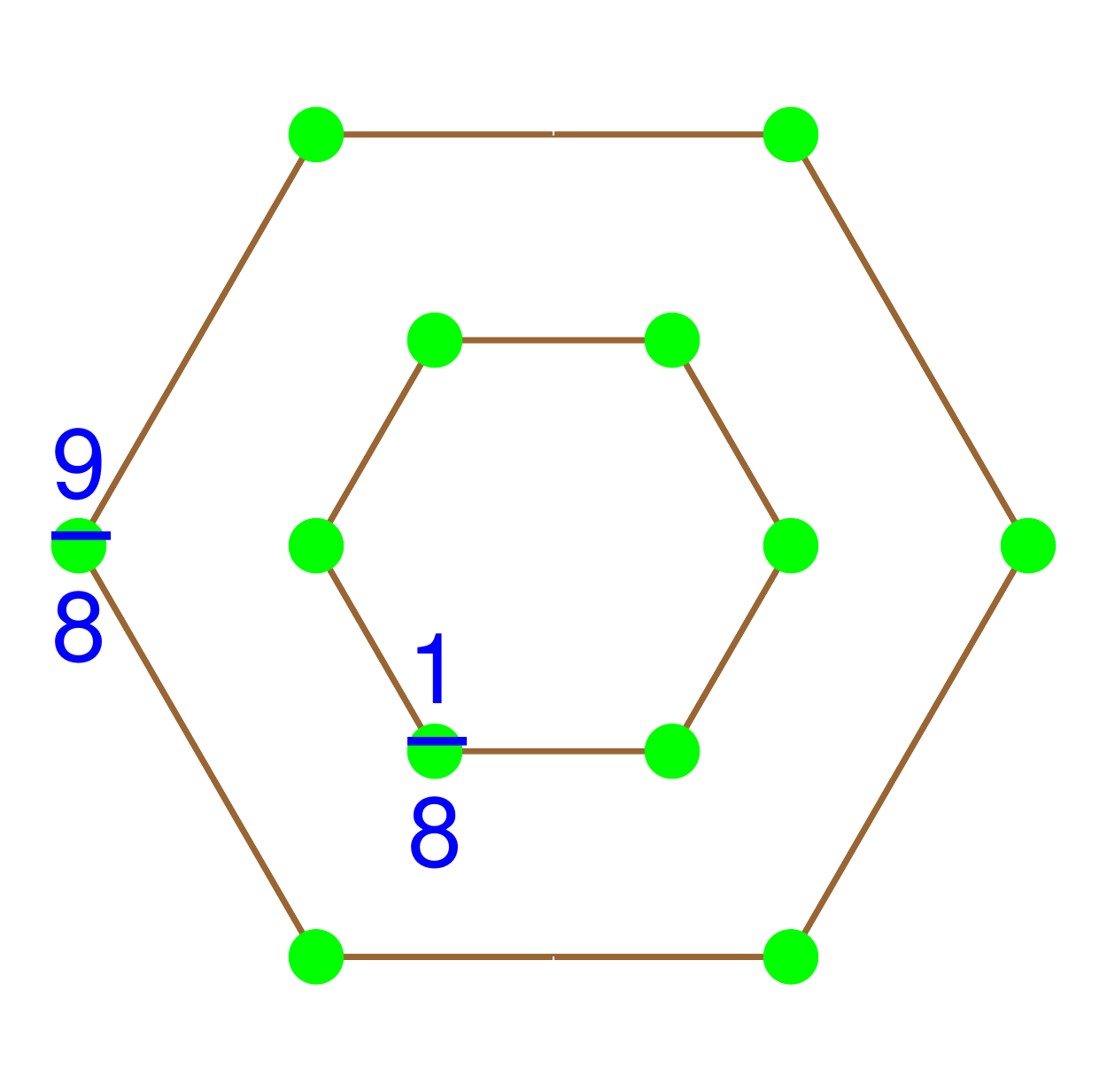}
\caption{$\sqrt{3}\times \sqrt{3}$ planar state}
\label{Ch4:fig:root3Xroot3-planar-kagome-state}
\end{center}
\end{figure}

\vspace{0.2cm}
\noindent \textbf{(iv) $\sqrt{3} \times \sqrt{3}$ (P) state: } $\sqrt{3}\times \sqrt{3}$ planar state has three sub-lattices and the relative angle between the spins is $2 \pi/3$. The unit cell contains nine sites as shown in Fig.~\ref{Ch4:fig:root3Xroot3-planar-kagome-state}. The energy per site is given by $E = - 2 J_1 + 4 J_2 - 2 J_3- J_{3h}$.

\vspace{0.2cm}
\noindent \textbf{(v) $\sqrt{3} \times \sqrt{3}$(U) state: } This state is a non-coplanar. It has three sub-lattices. The unit cell contains nine sites. In this case, the spin arrangements interpolate between the ferromagnetic state and the co-planar $ \sqrt{3}\times \sqrt{3}$ states. Some intermediate state is shown in Fig.~\ref{Ch4:fig:Q0-umbrella-kagome-state}(right).

\begin{figure}[ht!]
\begin{center}
\includegraphics[height=2.75cm,width=4.25cm]{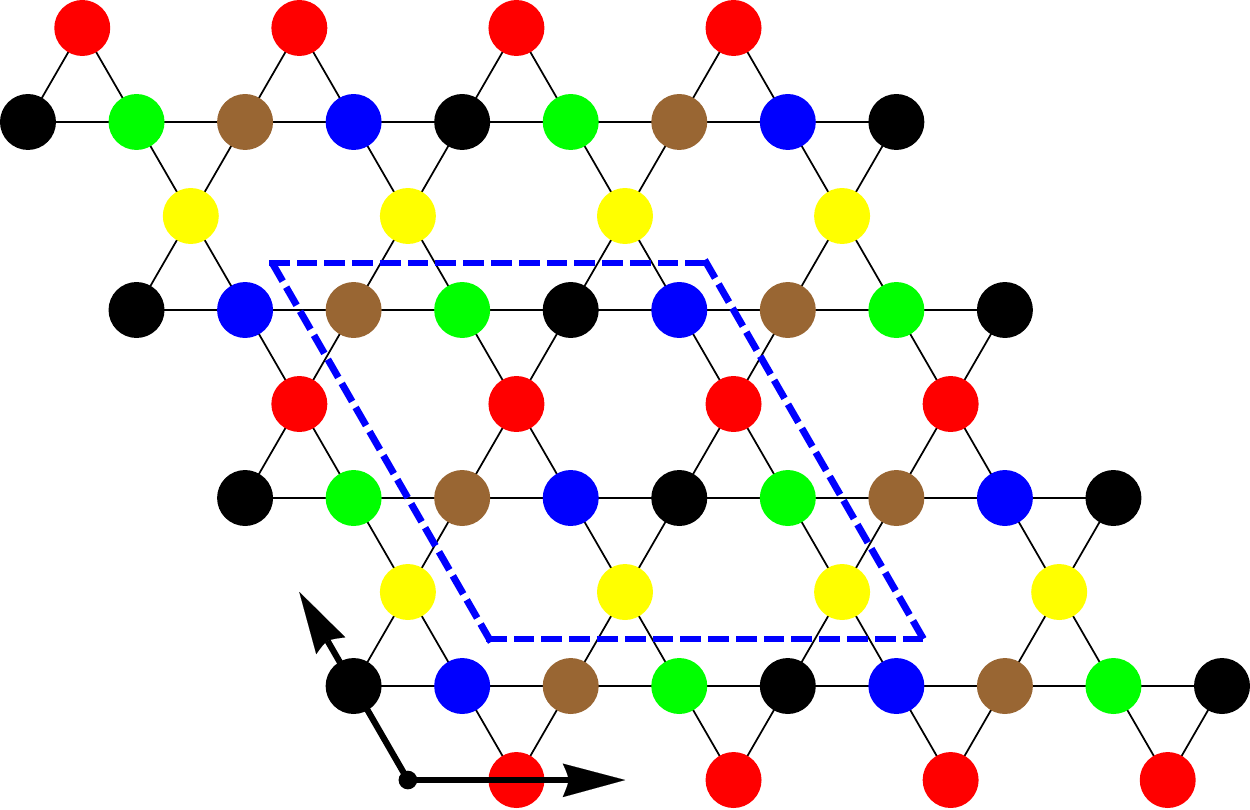}
\includegraphics[height=1.75cm,width=1.5cm]{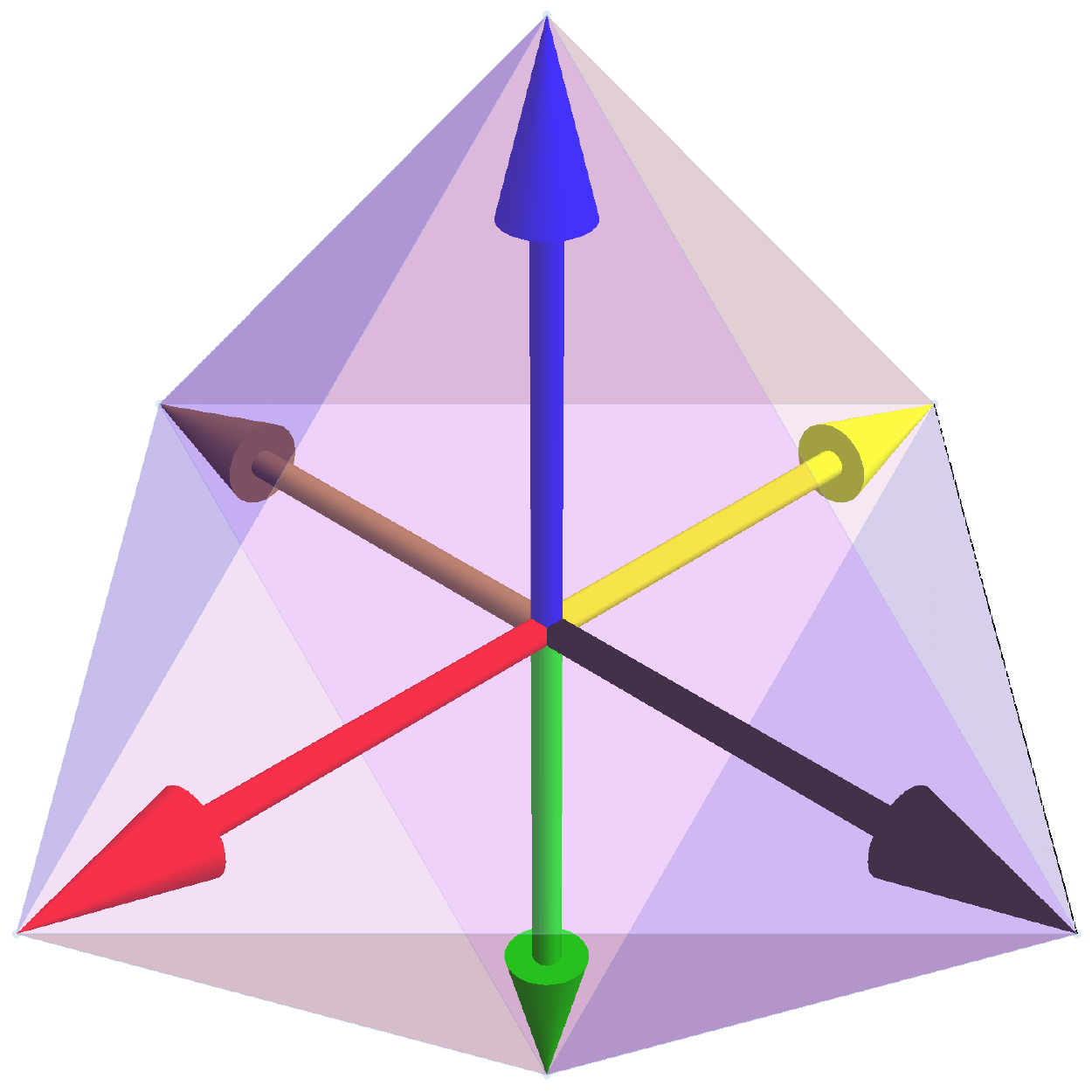}
\includegraphics[height=2.5cm,width=2.5cm]{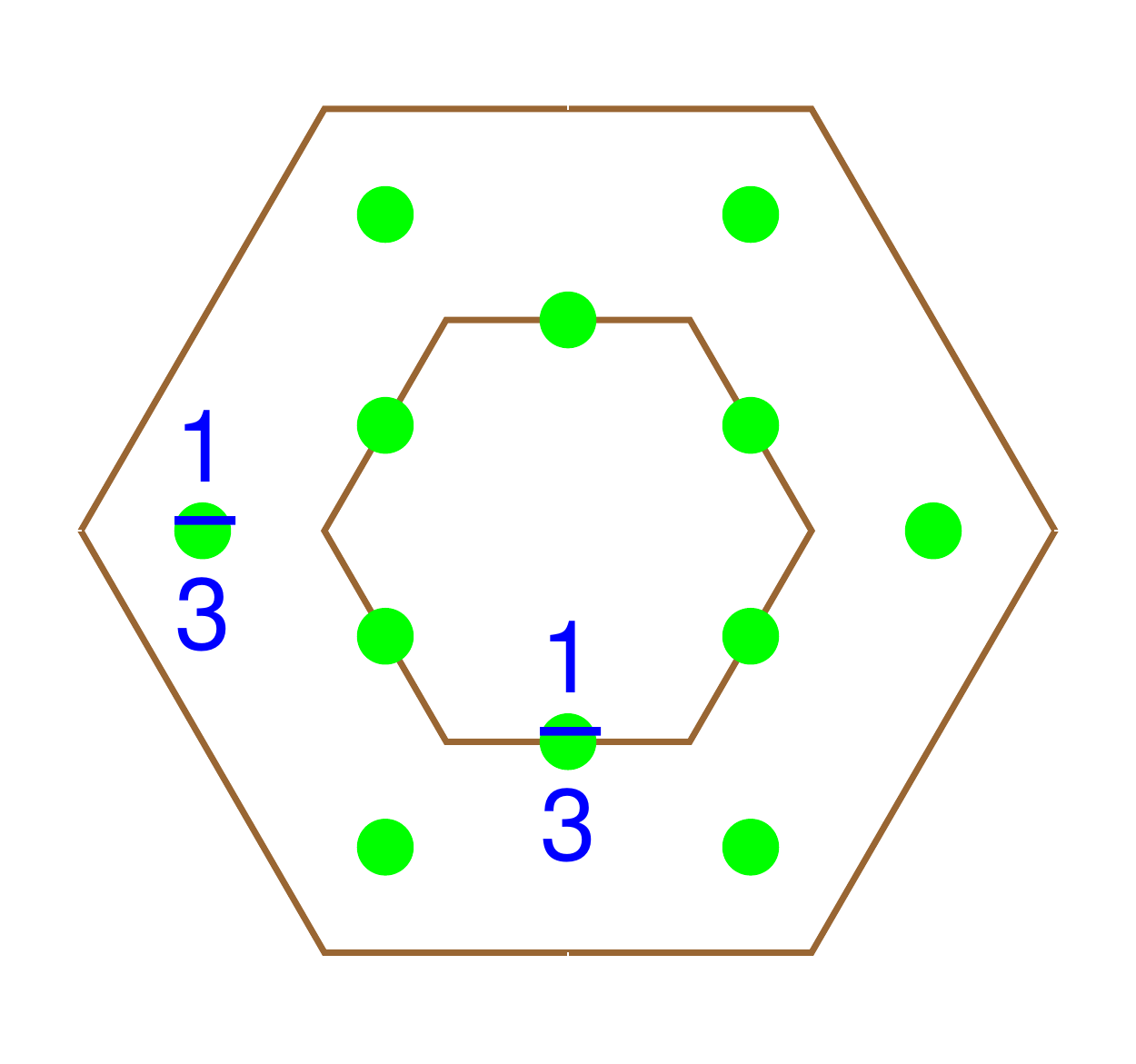}
\caption{Octahedral state}
\label{Ch4:fig:Octahedral-kagome-state}
\end{center}
\end{figure}

\vspace{0.2cm}
\noindent \textbf{(vi) Octahedral(O) state: } The octahedral state has six sub-lattices, and the spins are pointing towards the corner of an octahedron. The unit cell contains twelve sites, as shown in Fig.~\ref{Ch4:fig:Octahedral-kagome-state}. The energy per site is given by $E = - 4 J_3 + 2 J_{3h}$.

\begin{figure}[ht!]
\begin{center}
\includegraphics[height=2.75cm,width=4.25cm]{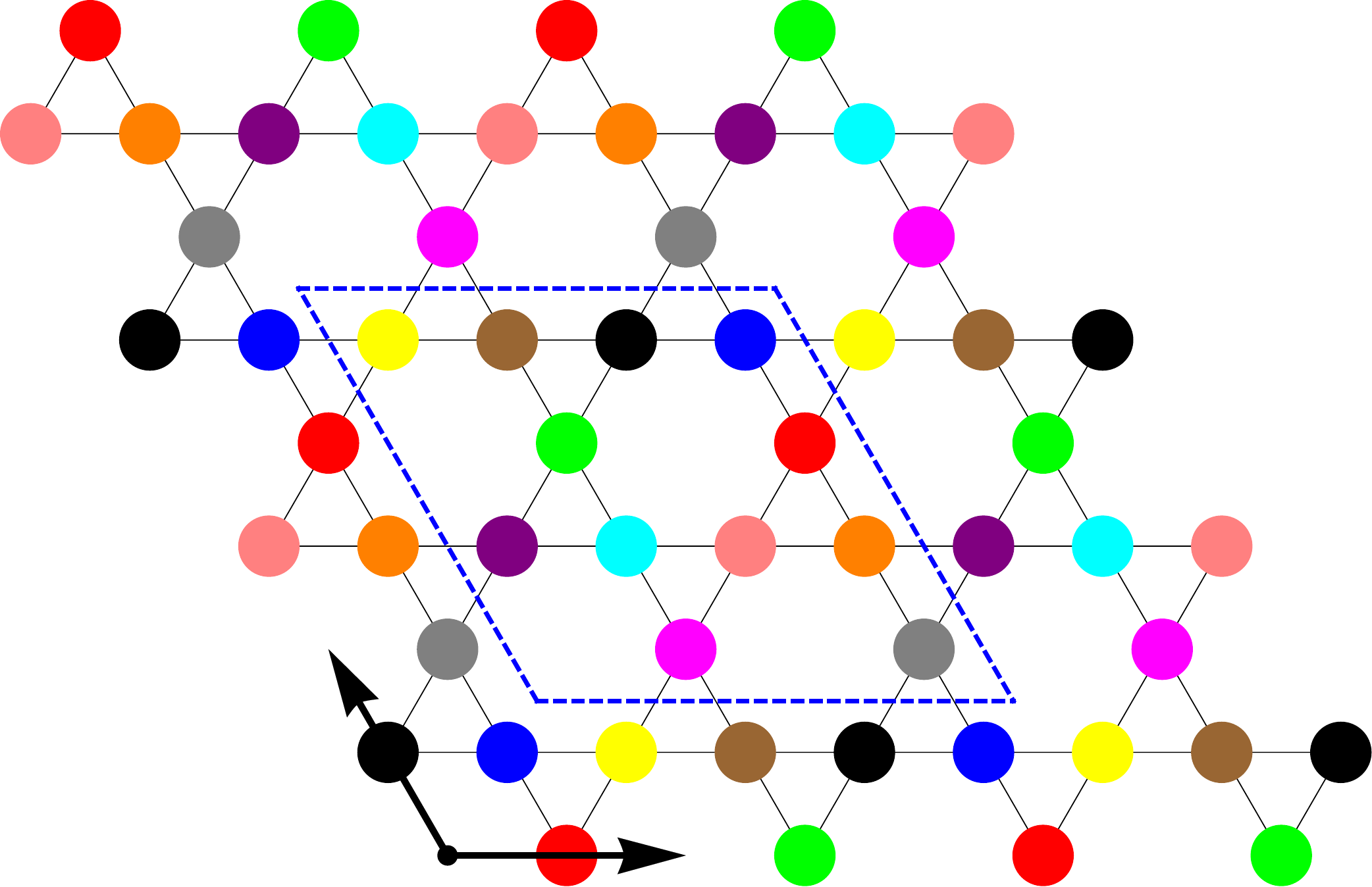}
\includegraphics[height=1.5cm,width=1.5cm]{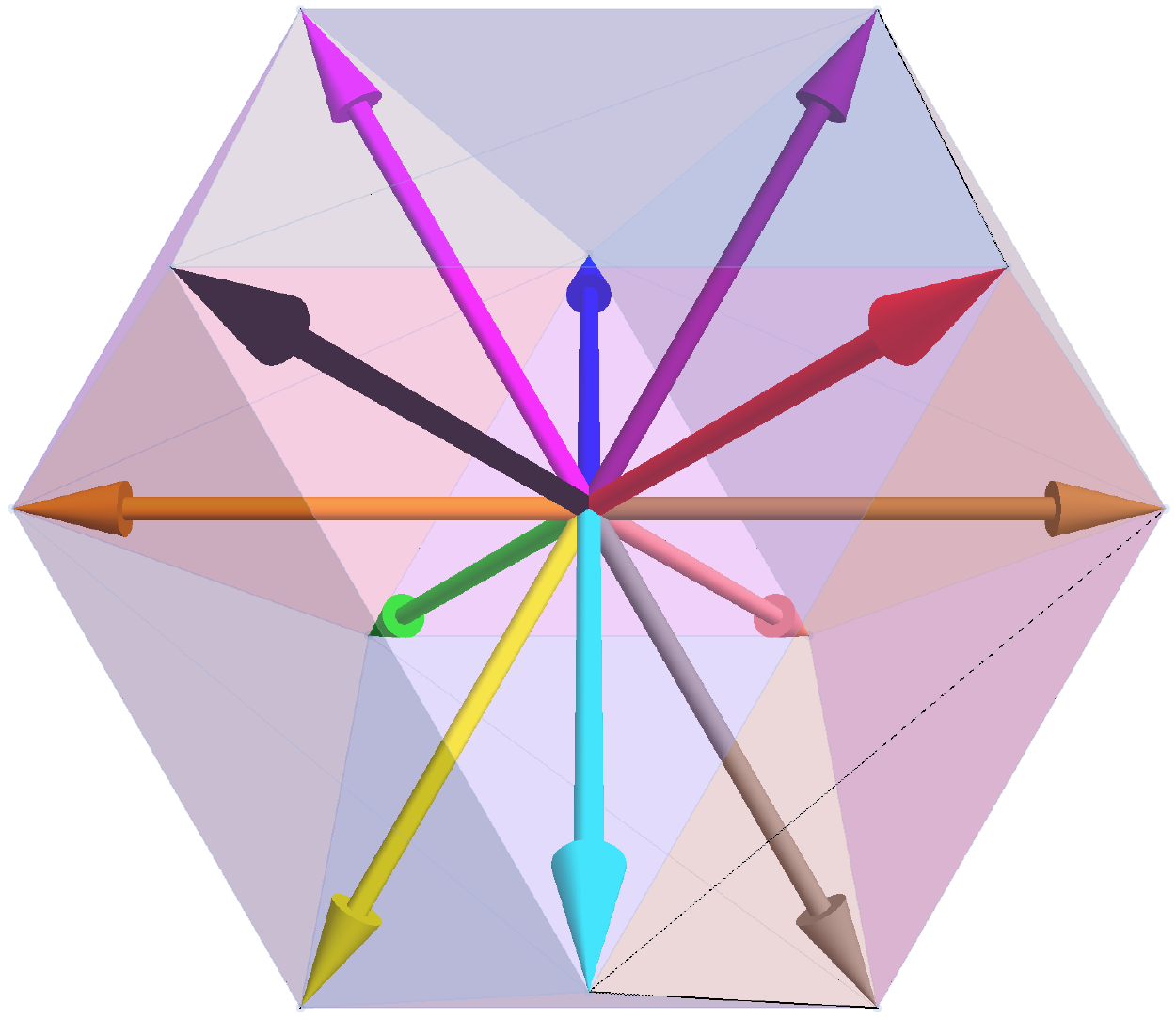}
\includegraphics[height=2.65cm,width=2.5cm]{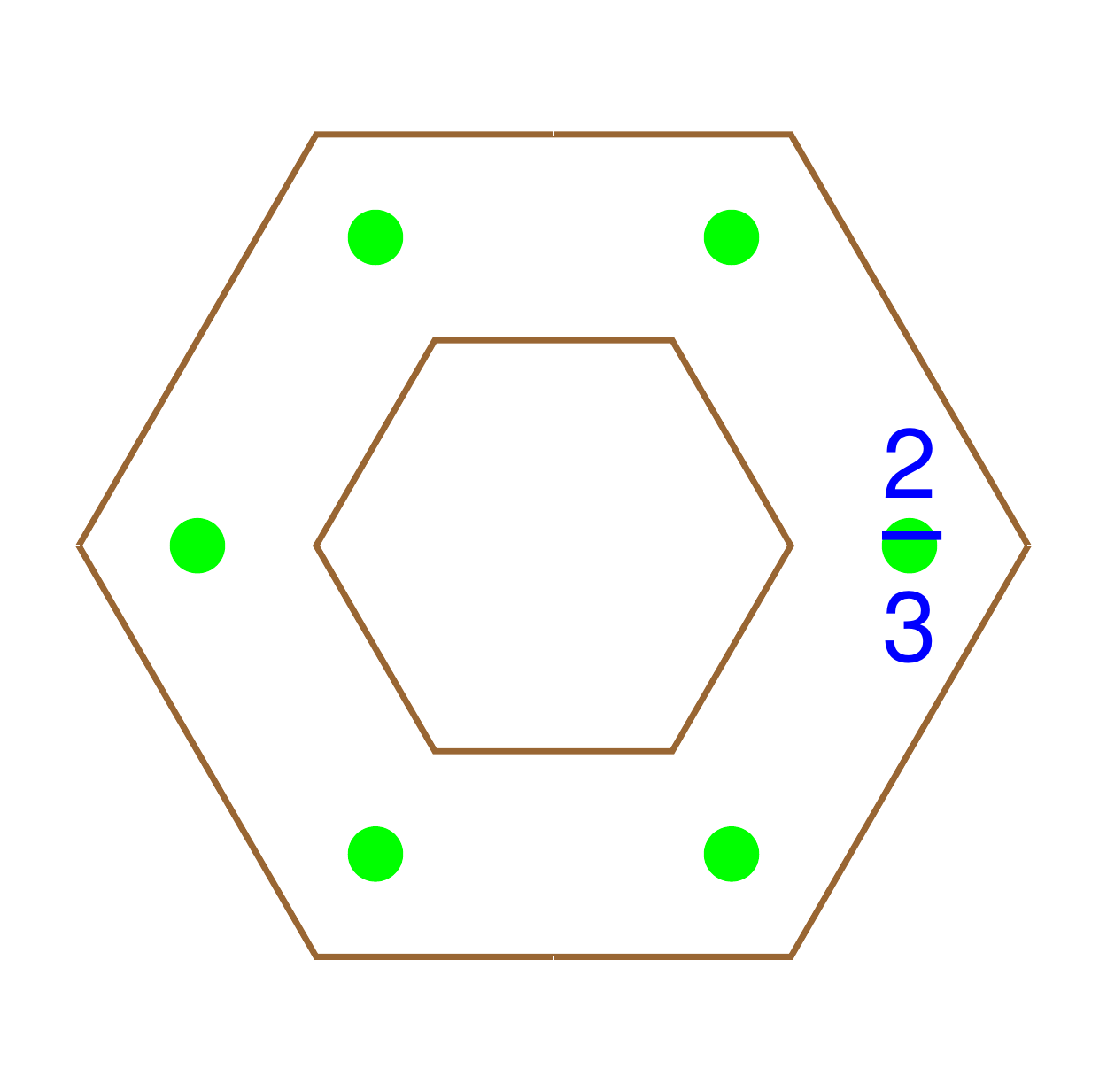}
\caption{Cuboc1 state}
\label{Ch4:fig:cuboc1-kagome-state}
\end{center}
\end{figure}

\vspace{0.2cm}
\noindent \textbf{(vii) Cuboc1(C$_1$) state: } Cuboc1 state has 12 sub-lattices, and the spins are pointing towards the corner of a cuboctahedron as shown in Fig.~\ref{Ch4:fig:cuboc1-kagome-state}. In this case, the relative angle between the neighboring spins is $120^o$, and the magnetic unit cell contains twelve sites. The energy per site is given by $E = -2 J_1 + 2 J_2 - 2 J_{3h}$.

\begin{figure}[ht!]
\begin{center}
\includegraphics[height=2.75cm,width=4.25cm]{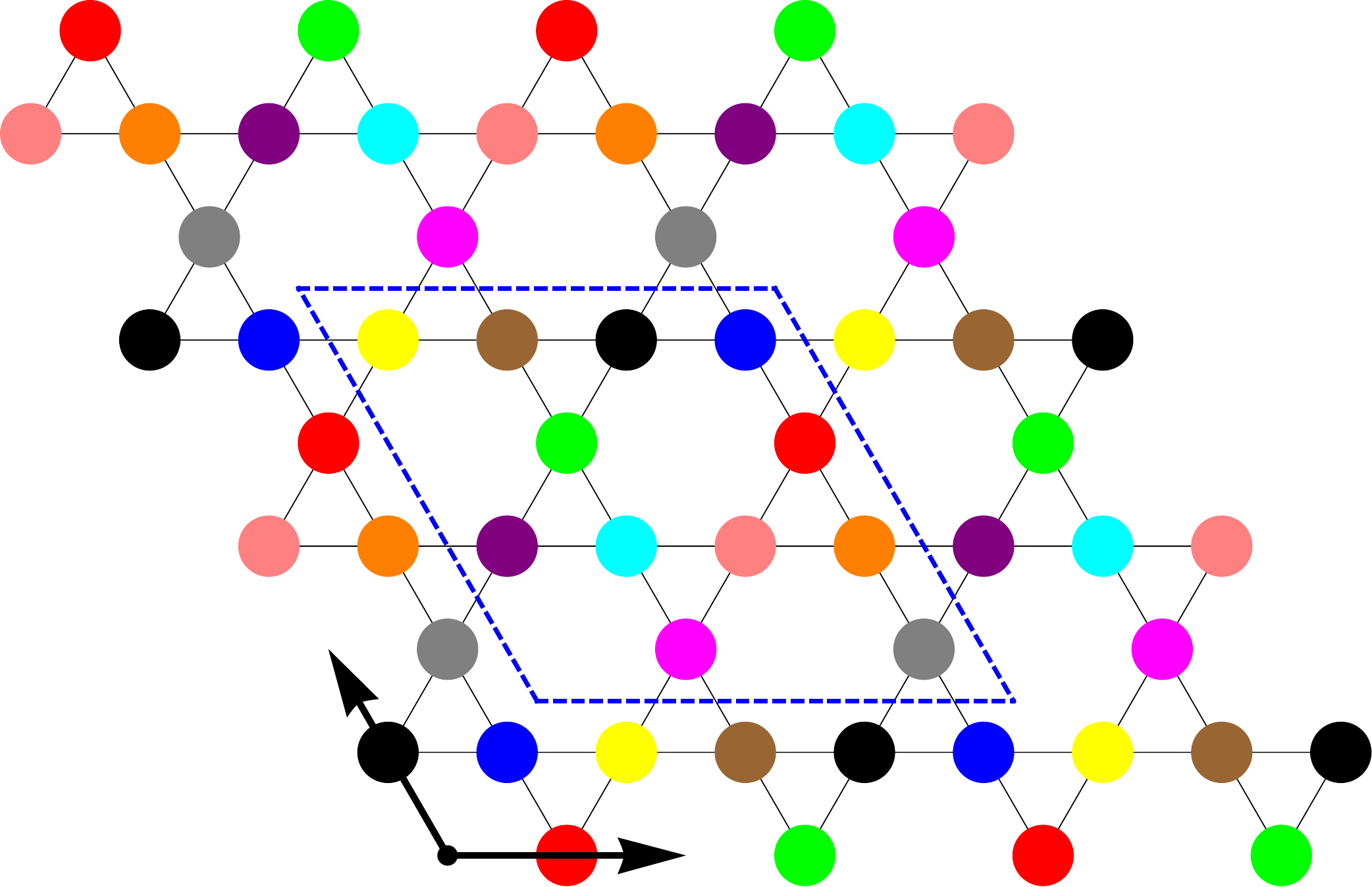}
\includegraphics[height=1.5cm,width=1.5cm]{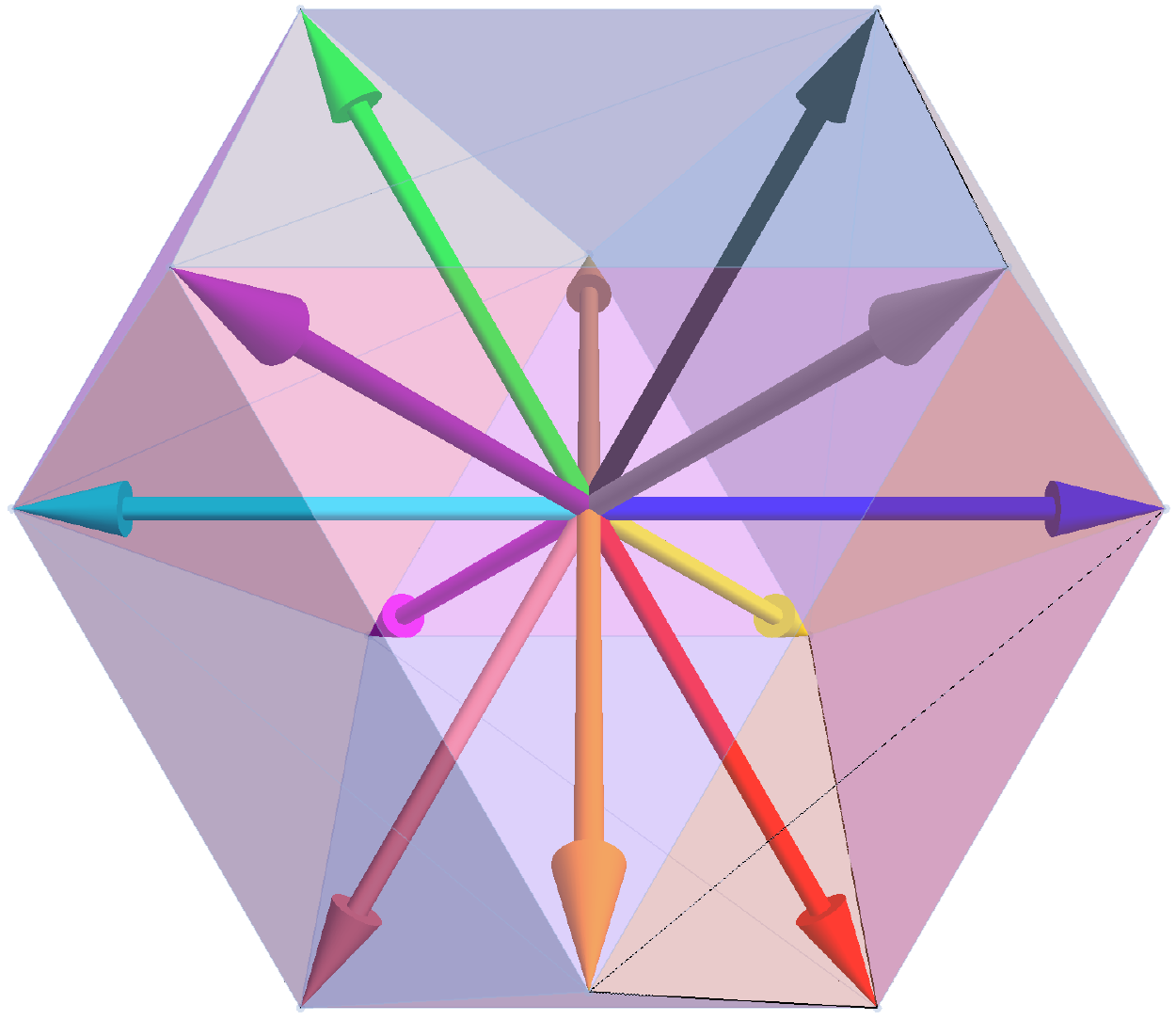}
\includegraphics[height=2.5cm,width=2.5cm]{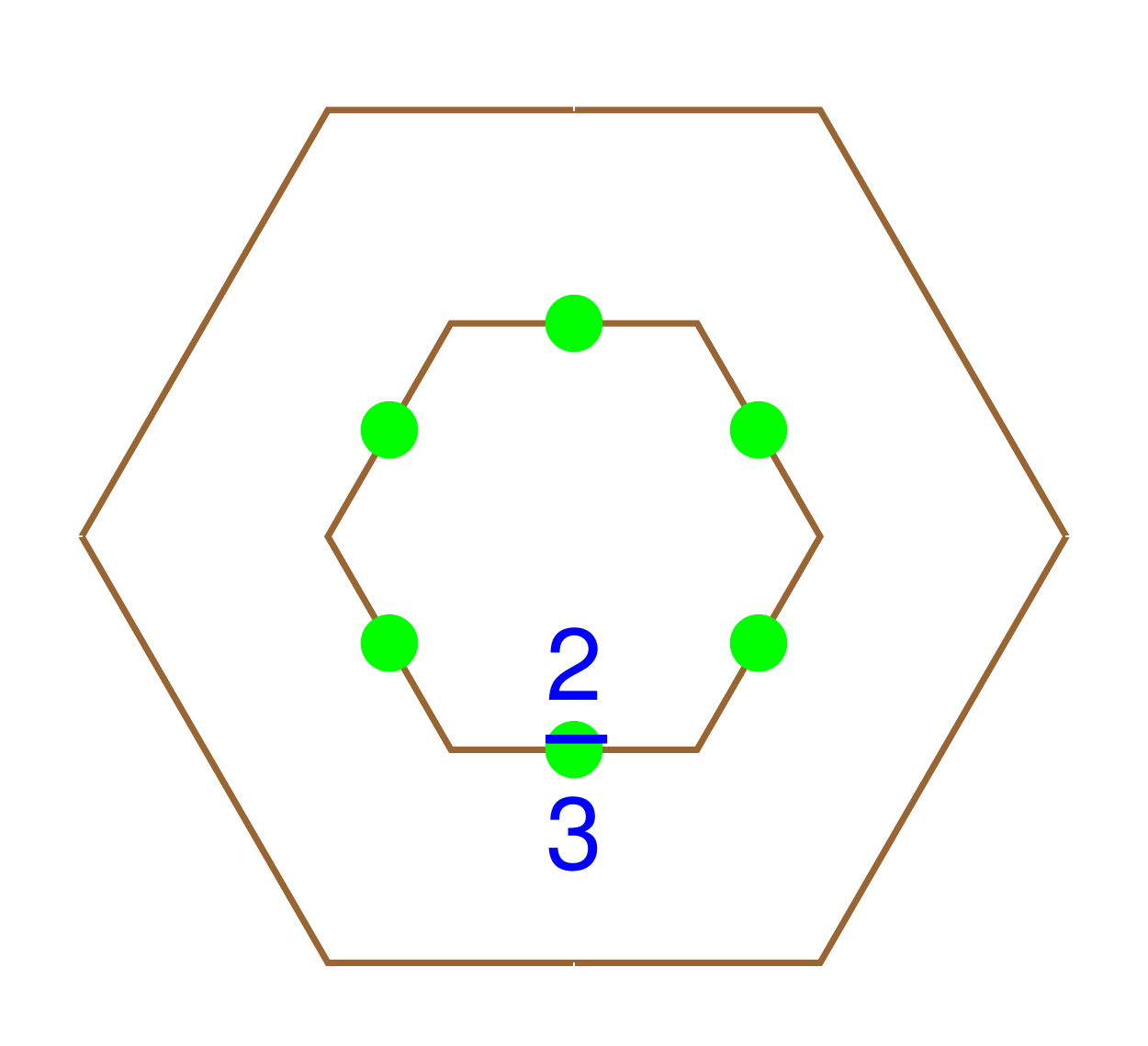}
\caption{Cuboc2 state}
\label{Ch4:fig:cuboc2-kagome-state}
\end{center}
\end{figure}

\vspace{0.2cm}
\noindent \textbf{(viii) Cuboc2(C$_2$) state: } Cuboc2 state has 12 sub-lattices, and the spins are pointing towards the corner of a cuboctahedron as shown in Fig.~\ref{Ch4:fig:cuboc2-kagome-state}. In this case, the relative angle between the neighboring spins is $60^o$ in contrast to the cuboc1 state. The magnetic unit cell contains twelve sites. The energy per site is given by $E = 2 J_1 - 2 J_2 - 2 J_{3h}$.

\begin{figure}[ht!]
\begin{center}
\includegraphics[height=2.75cm,width=4.25cm]{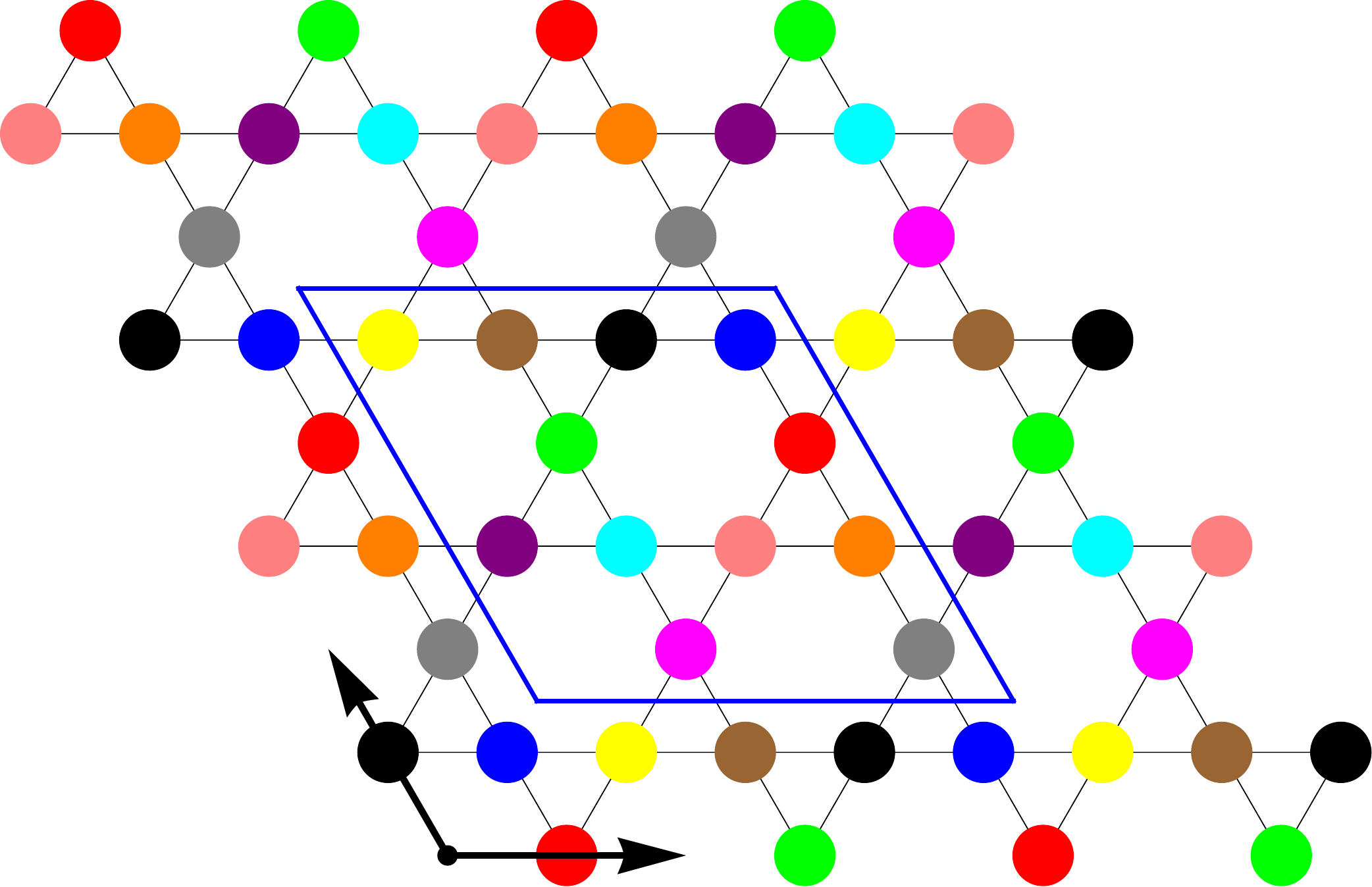}
\includegraphics[height=1.5cm,width=1.5cm]{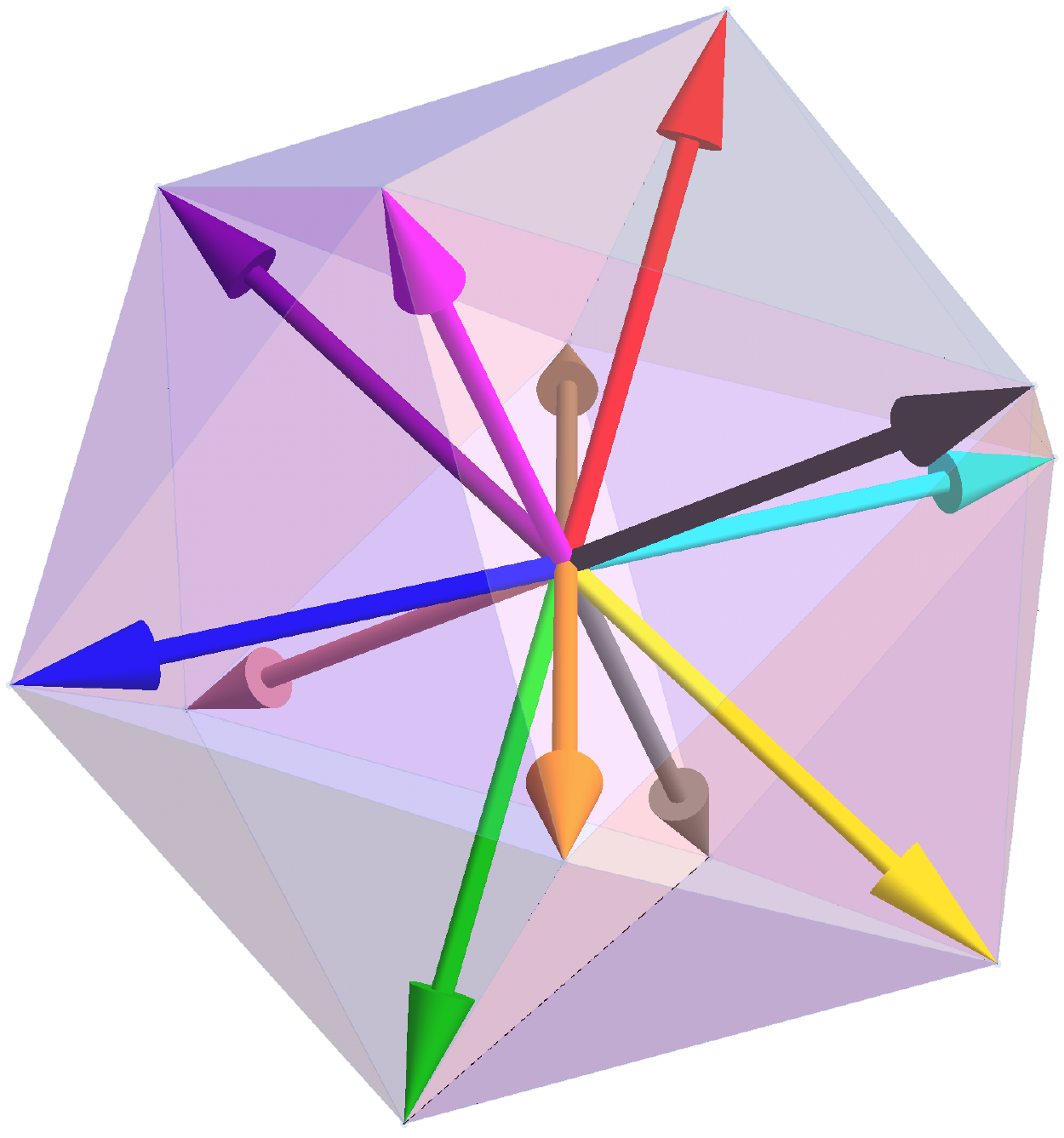}
\includegraphics[height=2.5cm,width=2.5cm]{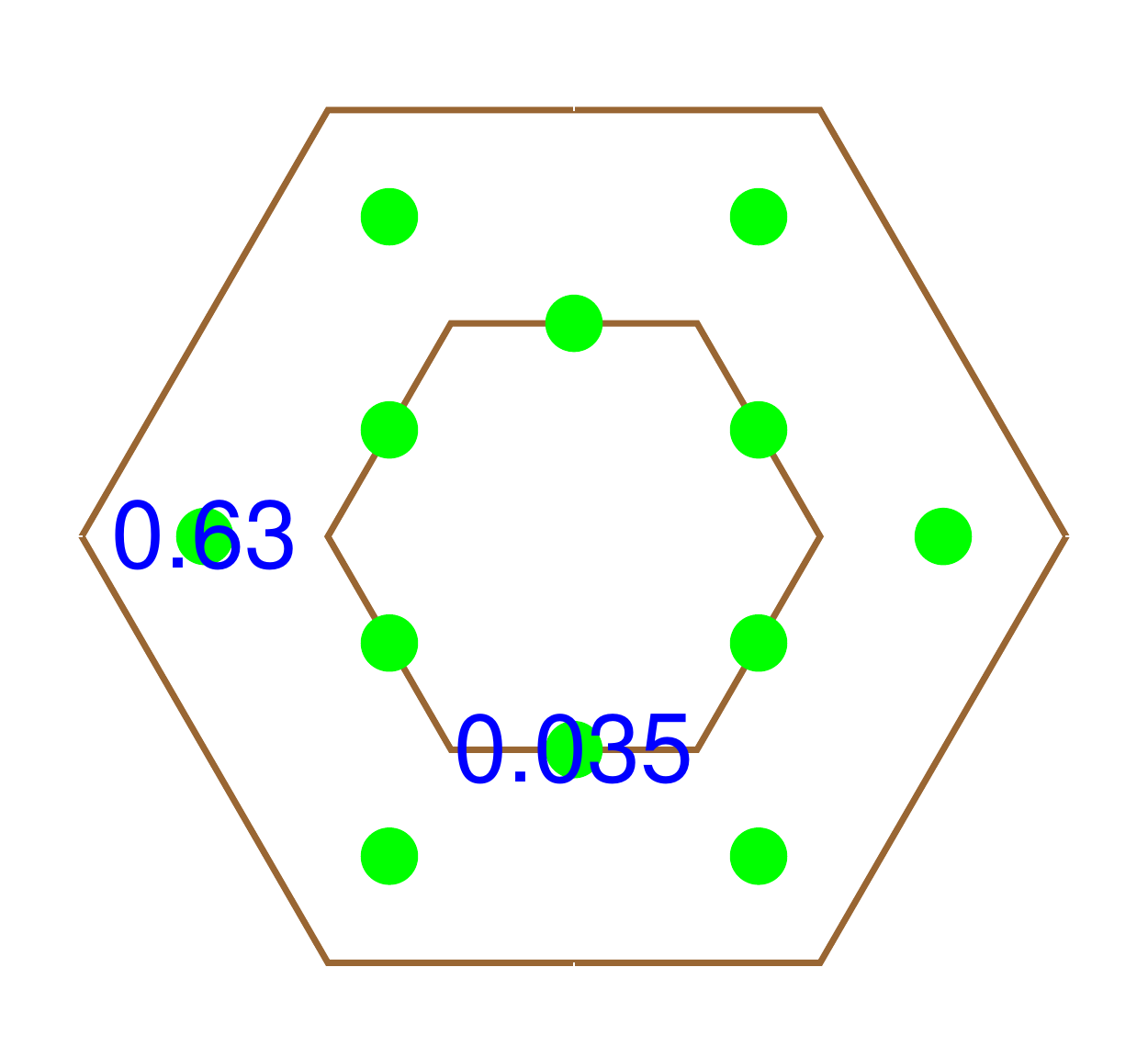}
\caption{Icosahedron1 state}
\label{Ch4:fig:Icosahedron1-kagome-state}
\end{center}
\end{figure}

\vspace{0.2cm}
\noindent \textbf{(ix) Regular icosahedron1(I$_1$) state: } The icosahedron1 state has 12 sub-lattices, and the spins are pointing towards the corner of an icosahedron. The unit cell also contains 12 sites. The neighboring spins make an angle of $~116.565^o$ to each other. It has 30 edges, and 20 equilateral triangle faces with five triangles sharing a common vertex, as shown in Fig.~\ref{Ch4:fig:Icosahedron1-kagome-state}. Here, the energy per site is given by $E = -\frac{4 }{\sqrt{5}} J_1 +\frac{4 }{\sqrt{5}} J_2 - 2 J_{3h}$.

\begin{figure}[ht!]
\begin{center}
\includegraphics[height=2.75cm,width=4.25cm]{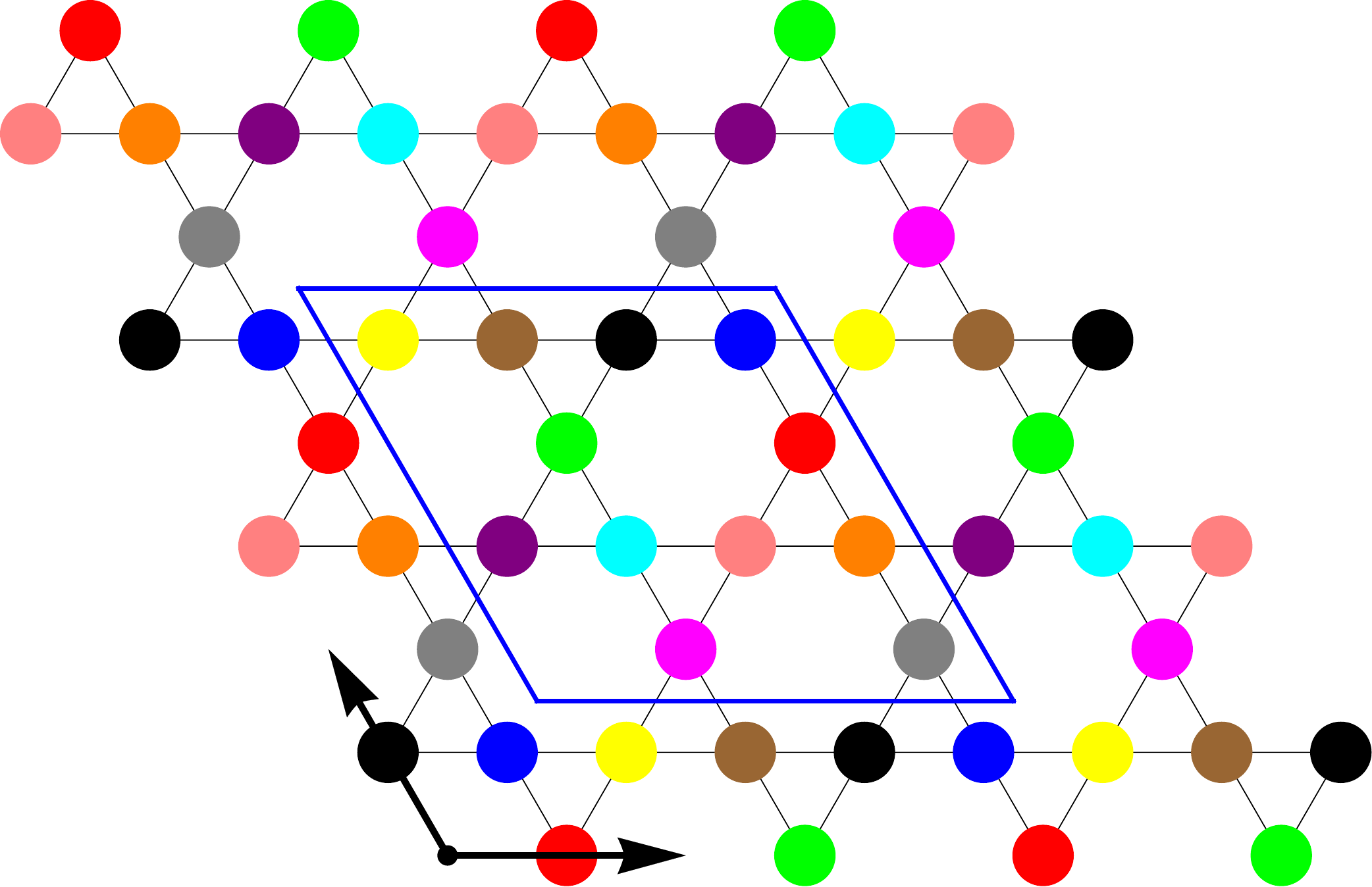}
\includegraphics[height=1.5cm,width=1.5cm]{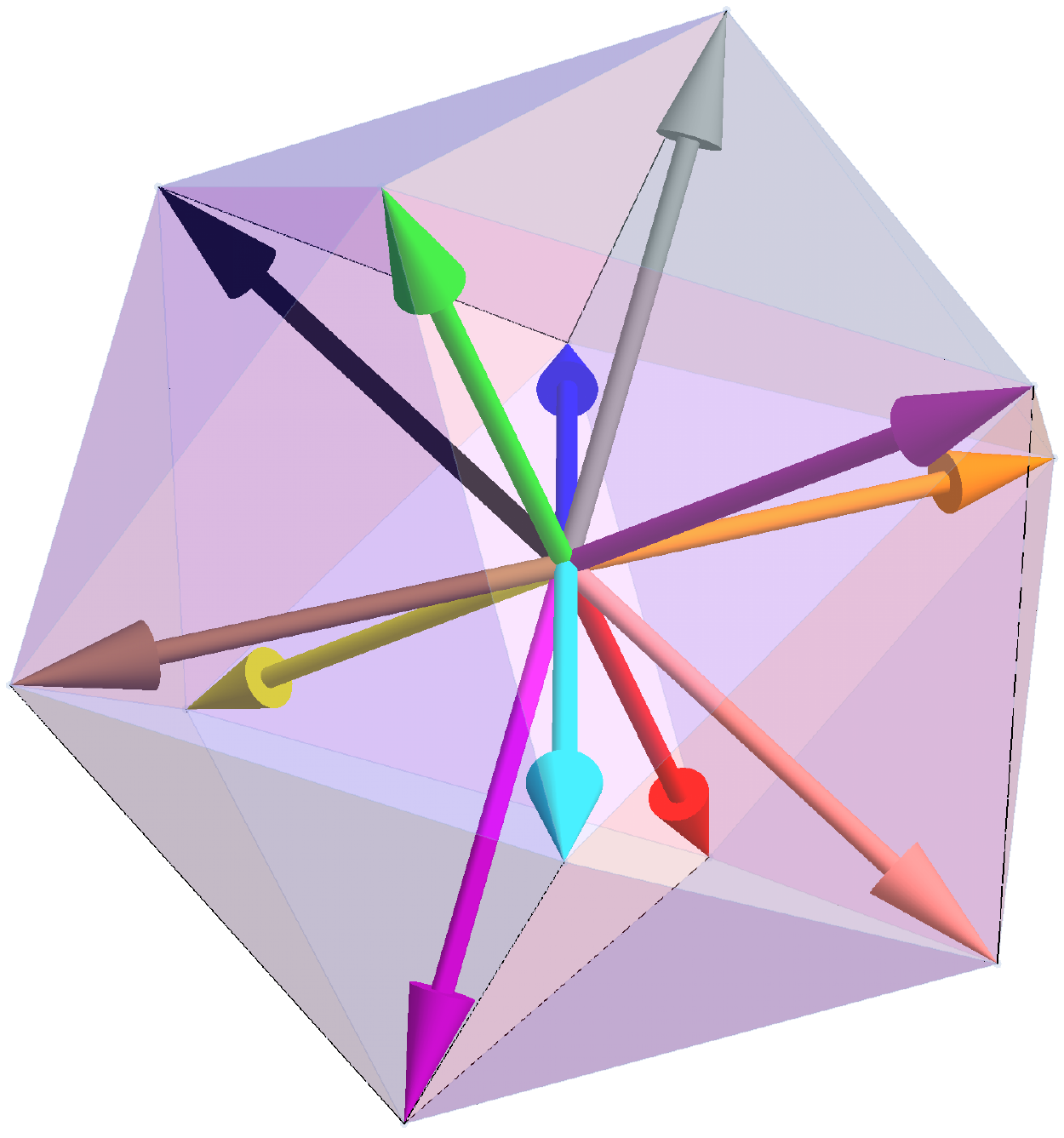}
\includegraphics[height=2.5cm,width=2.5cm]{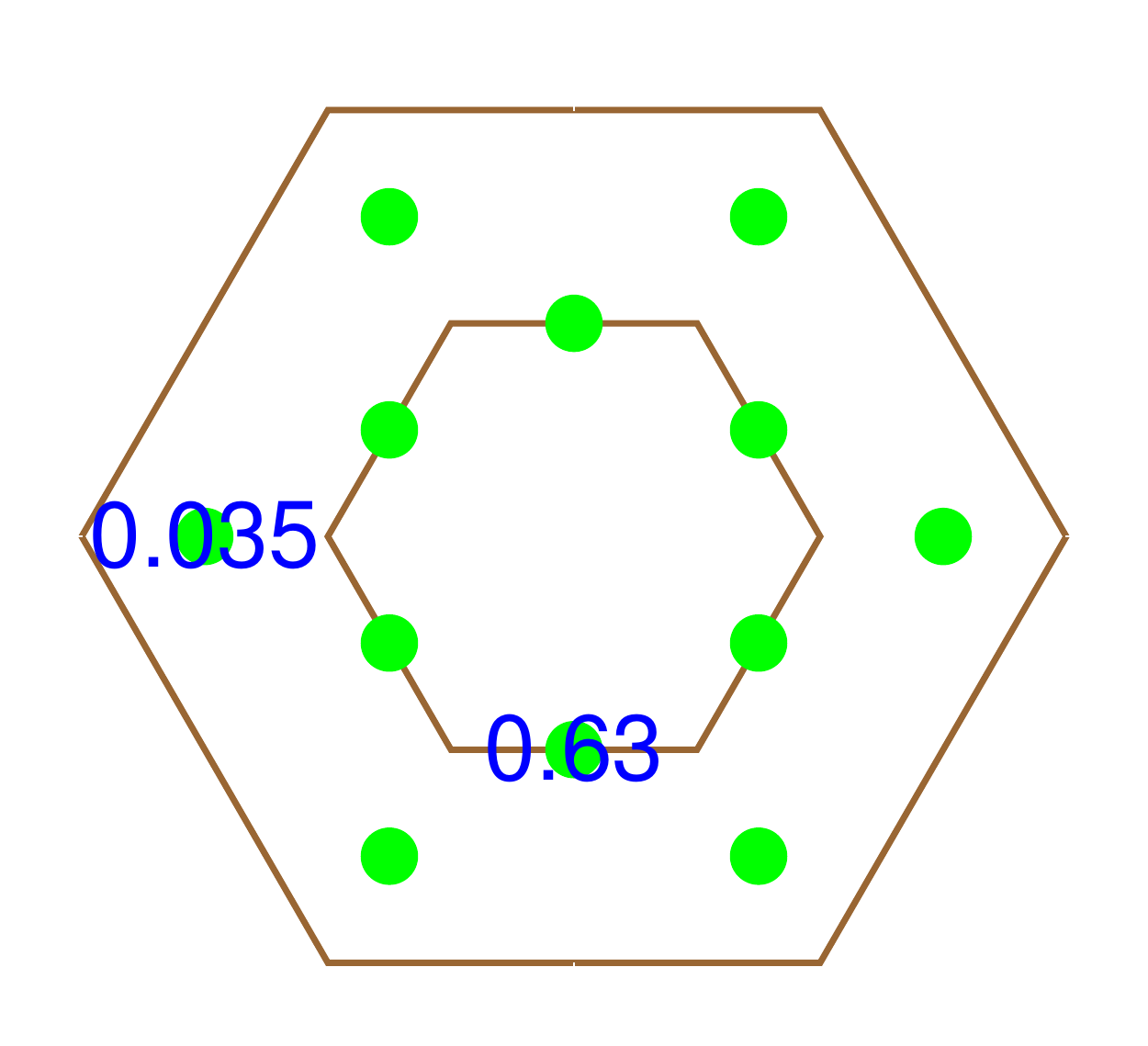}
\caption{Icosahedron2 state}
\label{Ch4:fig:Icosahedron2-kagome-state}
\end{center}
\end{figure}

\noindent \textbf{(x) Regular Icosahedron2(I$_2$) state: }In icosahedron2 structure, the neighboring spins are at $63.435^o$ to each other. Here, too the unit cell contains 12 sites, and all the spins are pointing towards the corner of an icosahedron, as shown in Fig.~\ref{Ch4:fig:Icosahedron2-kagome-state}. The energy per site is given by $E = \frac{4 }{\sqrt{5}} J_1 - \frac{4 }{\sqrt{5}} J_2 - 2 J_{3h}$.

\begin{figure}[ht!]
\begin{center}
\includegraphics[height=2.75cm,width=4.25cm]{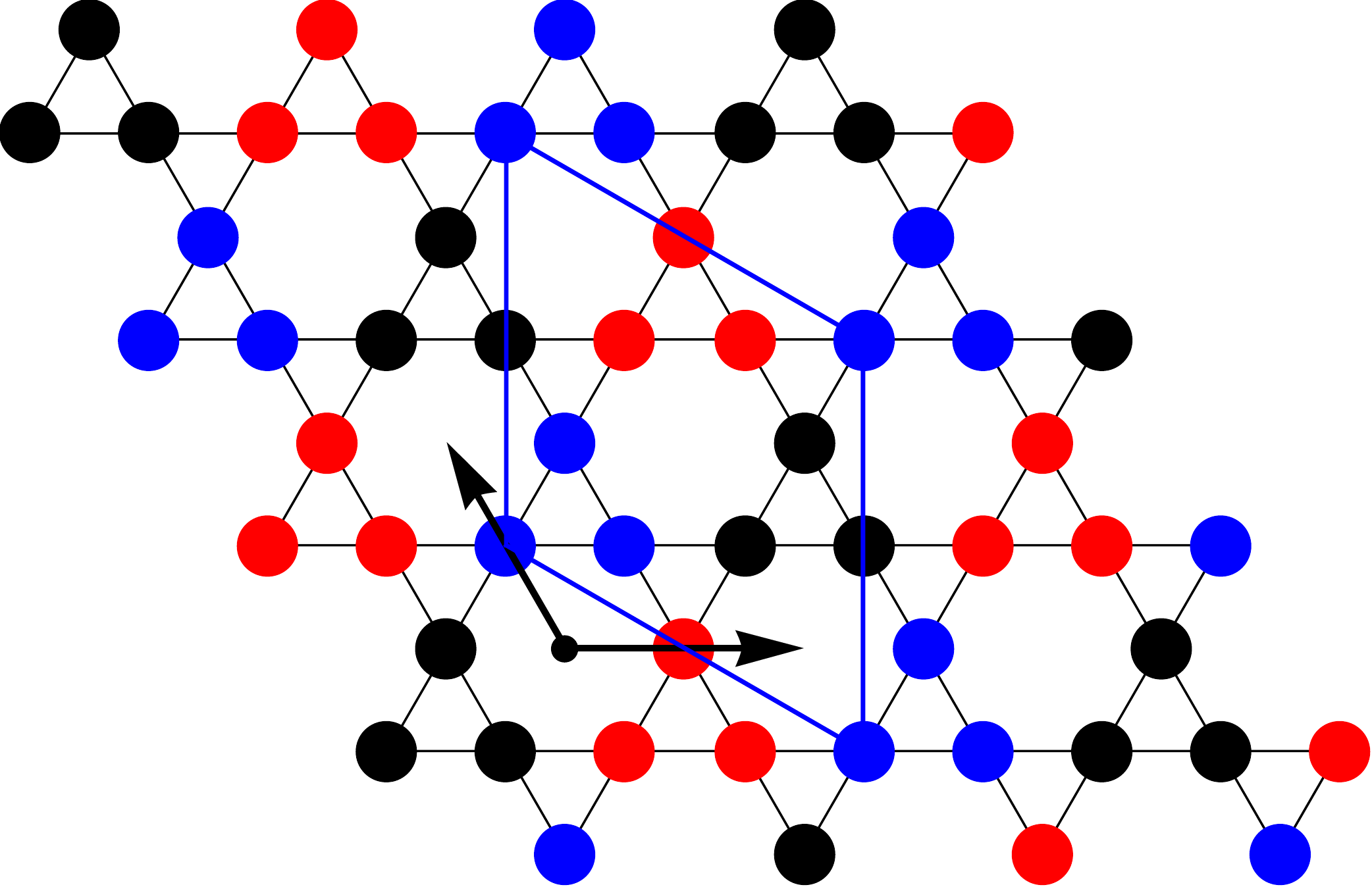}
\includegraphics[height=1.5cm,width=1.5cm]{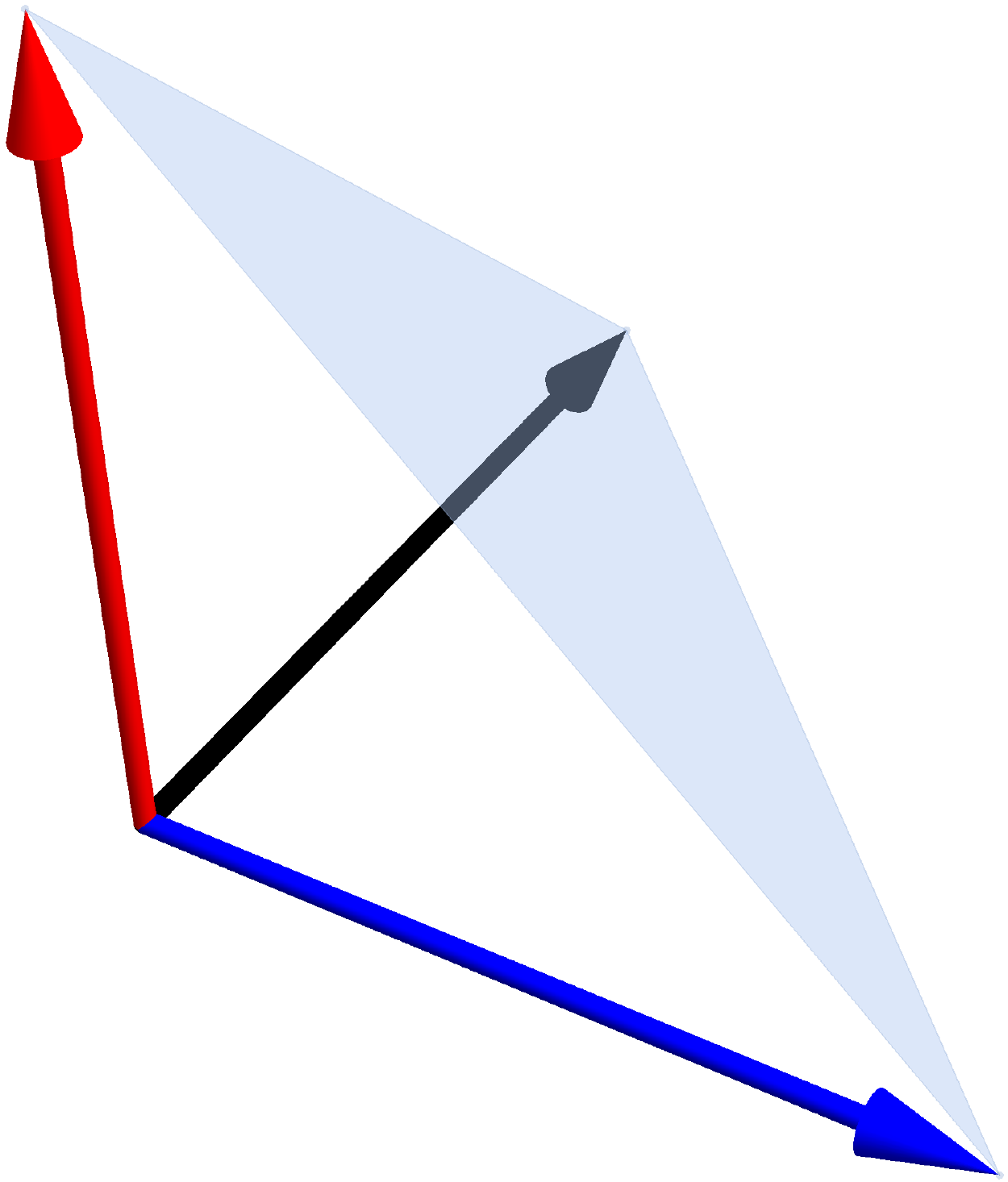}
\includegraphics[height=2.5cm,width=2.5cm]{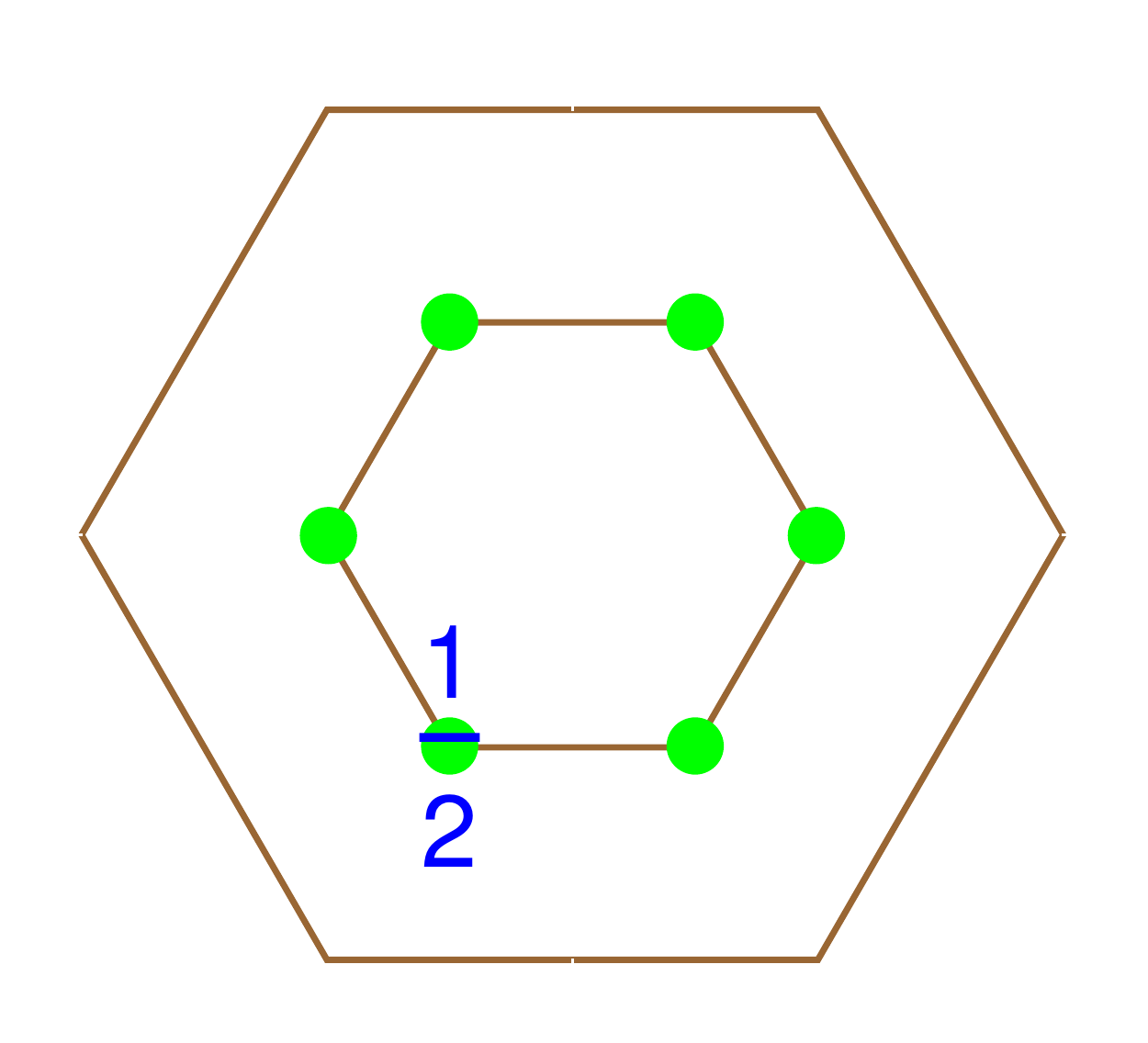}
\caption{ Umbrella1$(U_1)$ state}
\label{Ch4:fig:extendedQ=0-state}
\end{center}
\end{figure}

\noindent \textbf{(xi) Umbrella1$(U_1)$ state: } This state has three sub-lattices and an umbrella kind of structure which includes planar structure and ferromagnetic state. The relative angle between spins is identical. The angle varies from $0$ to $2 \pi/3$. Each of the sub-lattices occupies an up triangle, as shown by a different color in Fig.~\ref{Ch4:fig:extendedQ=0-state}. The unit cell contains nine sites, and the energy per site is given by $E = J_1 - 2 J_2 - 2 J_3 - J_{3h}$.

\begin{figure}[ht!]
\begin{center}
\includegraphics[height=2.75cm,width=4.25cm]{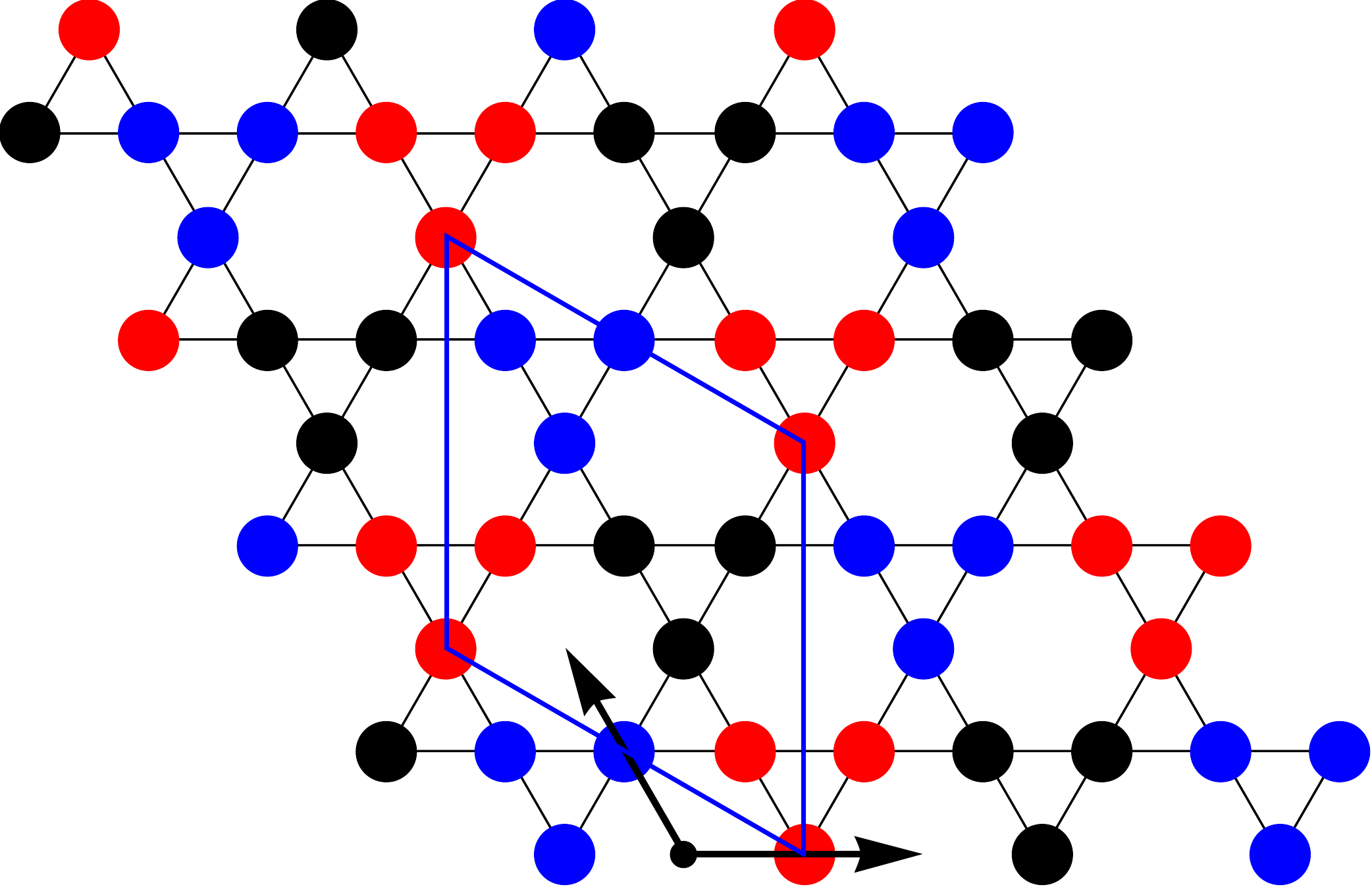}
\includegraphics[height=1.5cm,width=1.5cm]{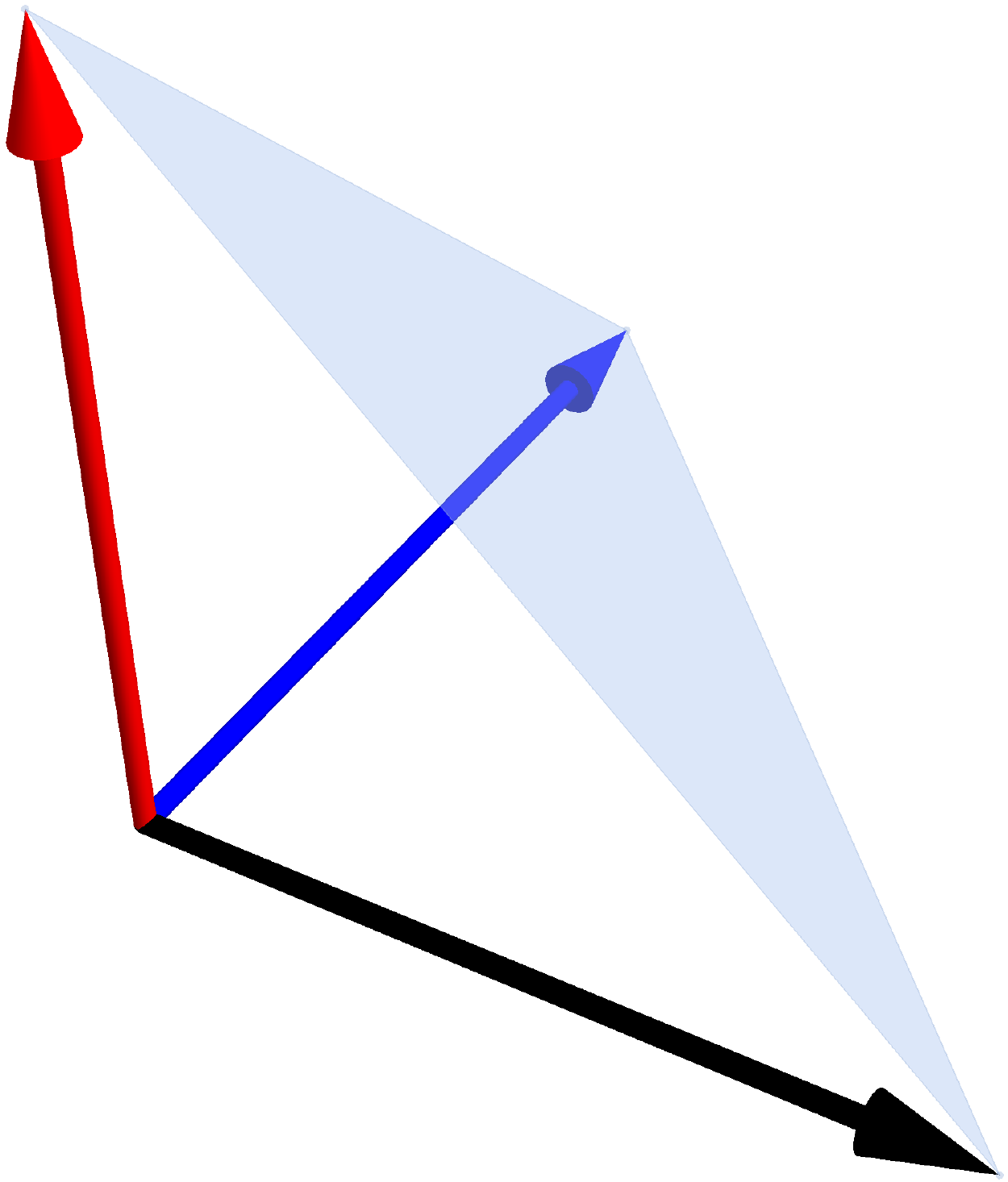}
\includegraphics[height=2.5cm,width=2.5cm]{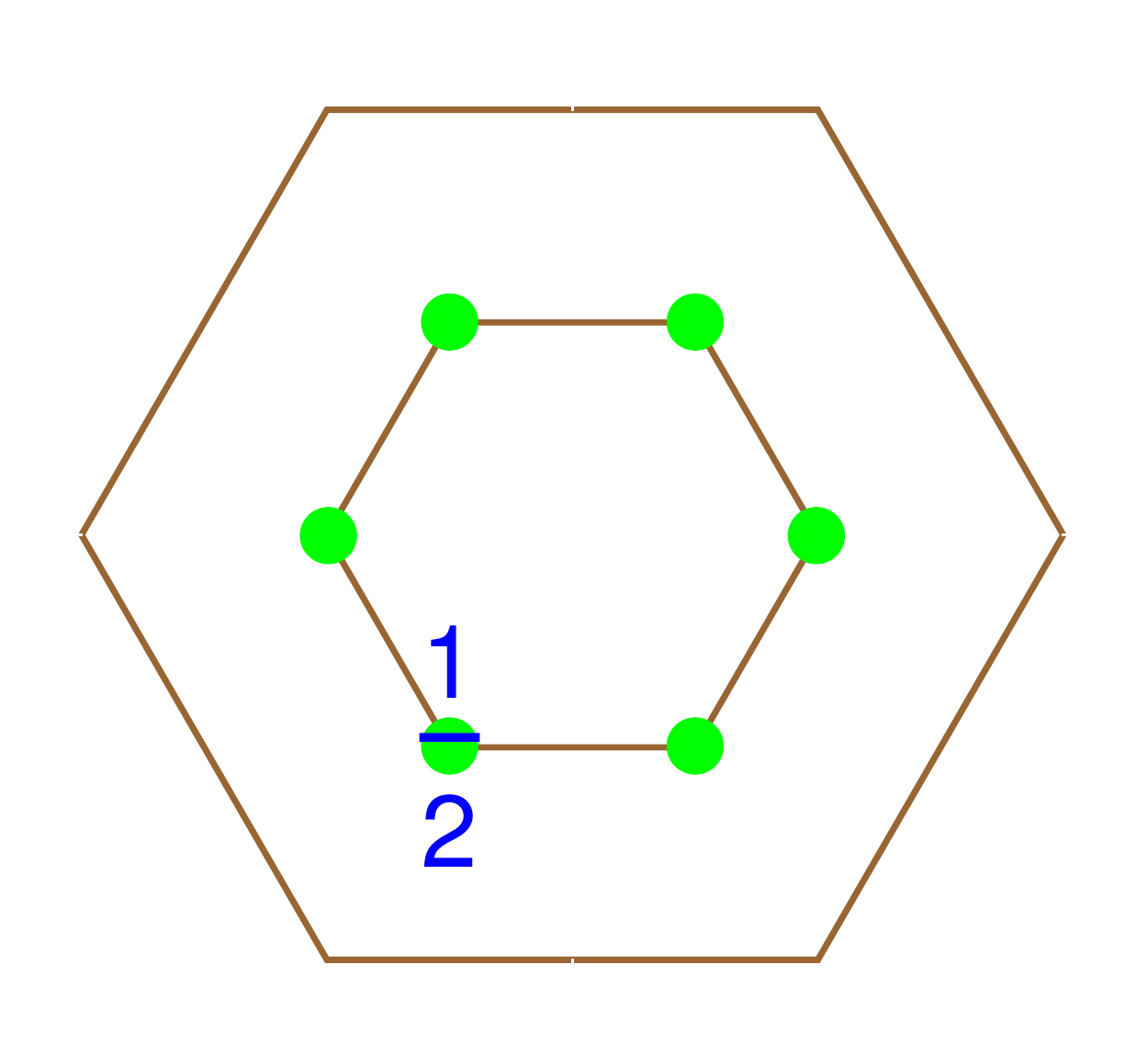}
\caption{ Umbrella2$(U_2)$ state}
\label{Ch4:fig:extendedQ=0-state2}
\end{center}
\end{figure}

\noindent \textbf{(xii) Umbrella2$(U_2)$ state: }This state is similar to the Umbrella1 state. Here each of the sub-lattices occupies a down triangle as shown by a different color in Fig.~\ref{Ch4:fig:extendedQ=0-state2}. The energy per site is given by $E = J_1 - 2 J_2 - 2 J_3 - J_{3h}$.

\begin{figure}[ht!]
\begin{center}
\includegraphics[height=2.75cm,width=4.25cm]{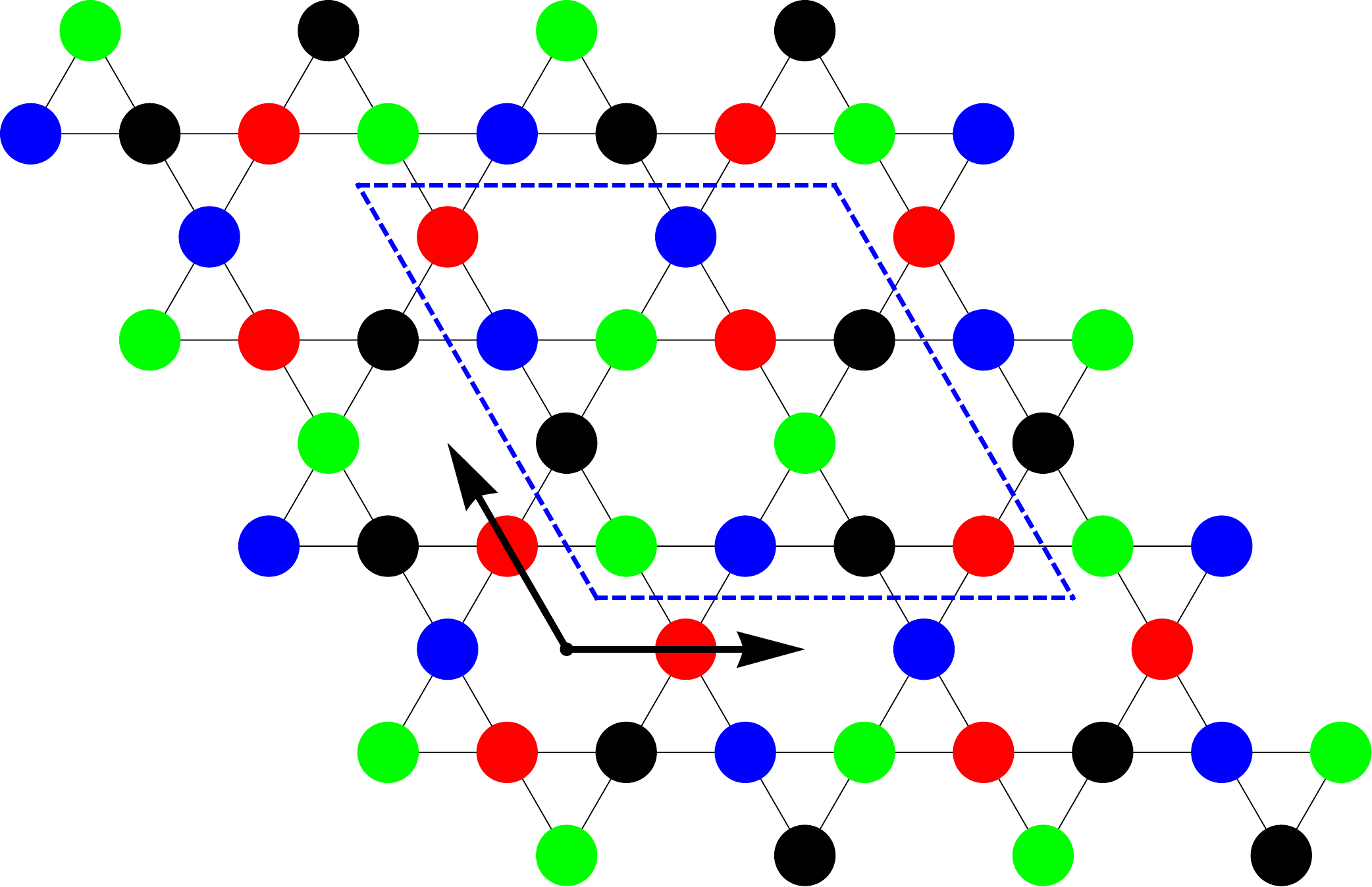}
\includegraphics[height=1.5cm,width=1.5cm]{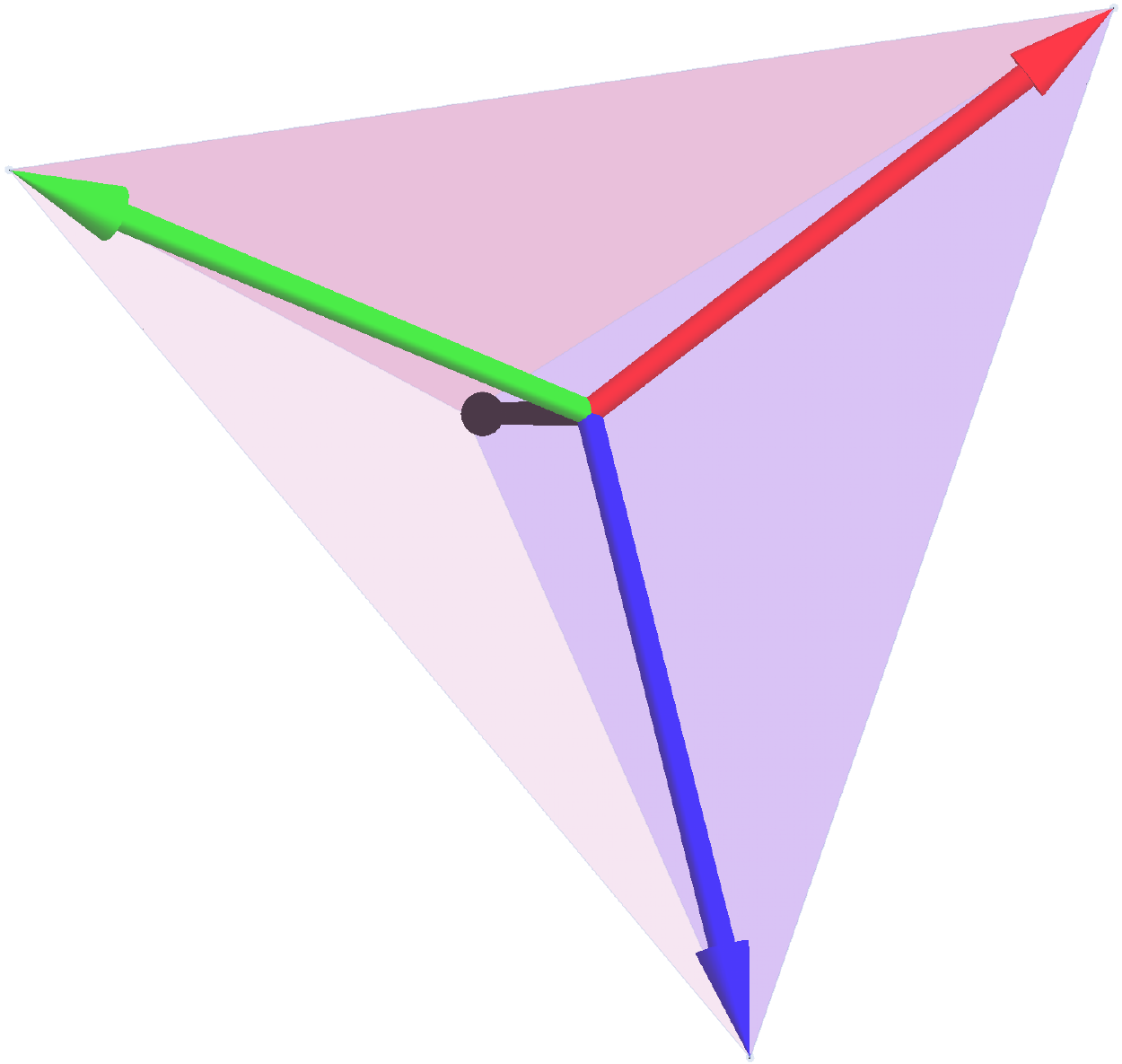}
\includegraphics[height=2.5cm,width=2.5cm]{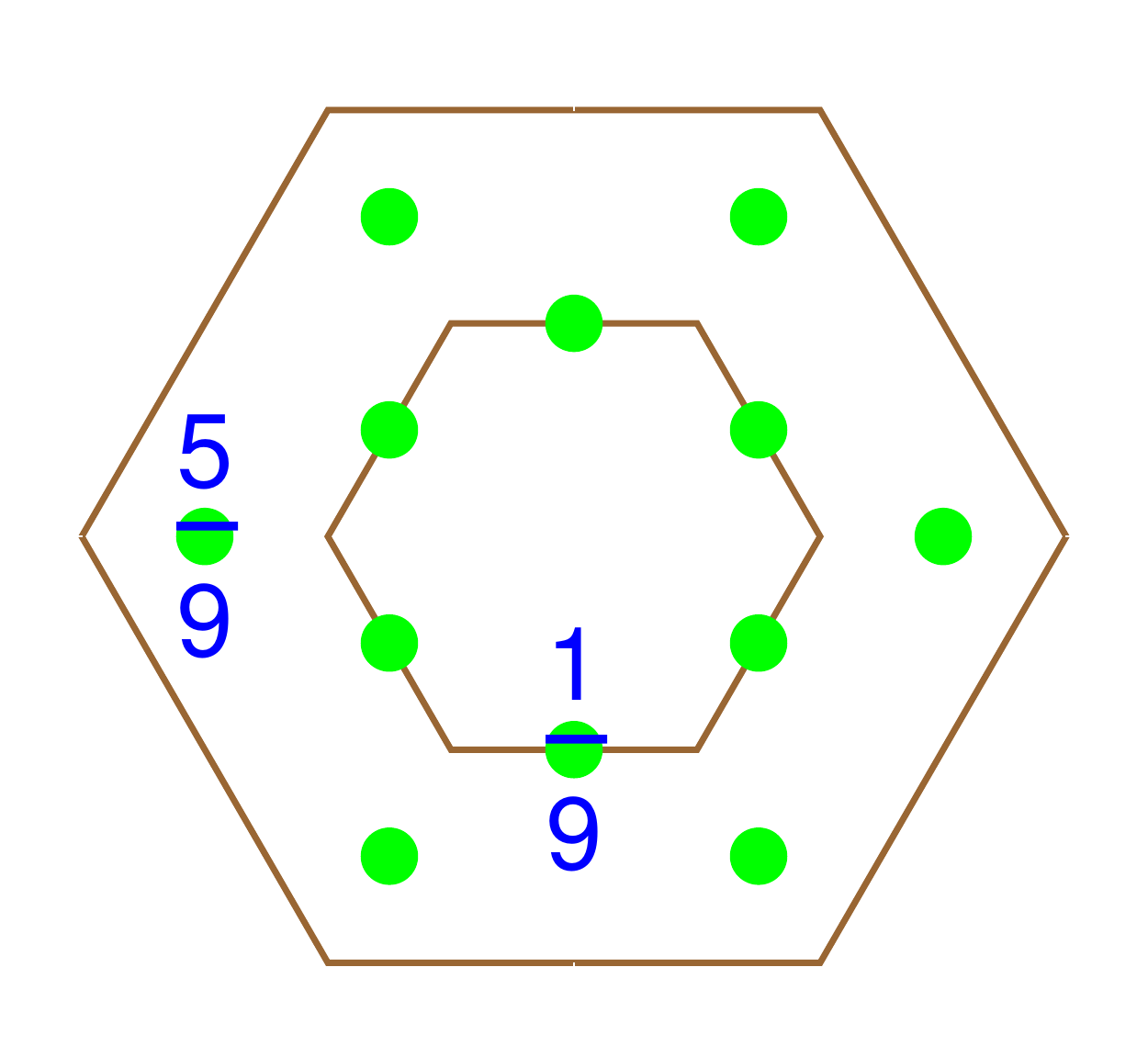}
\caption{Tetrahedral state}
\label{Ch4:fig:tetrahedral.}
\end{center}
\end{figure}

\noindent \textbf{(xiii) Tetrahedral(T) state : } This state has four sub-lattices, and the spins are pointing towards the corner of a tetrahedron. The unit cell contains 12 sub-lattices as shown in Fig.~\ref{Ch4:fig:tetrahedral.}
The energy per site is given by $E = -4 J_1 + 4 J_2 - 4 J_3 - 2 J_{3h}$.

\begin{figure}[ht!]
\begin{center}
\includegraphics[height=2.75cm,width=4.25cm]{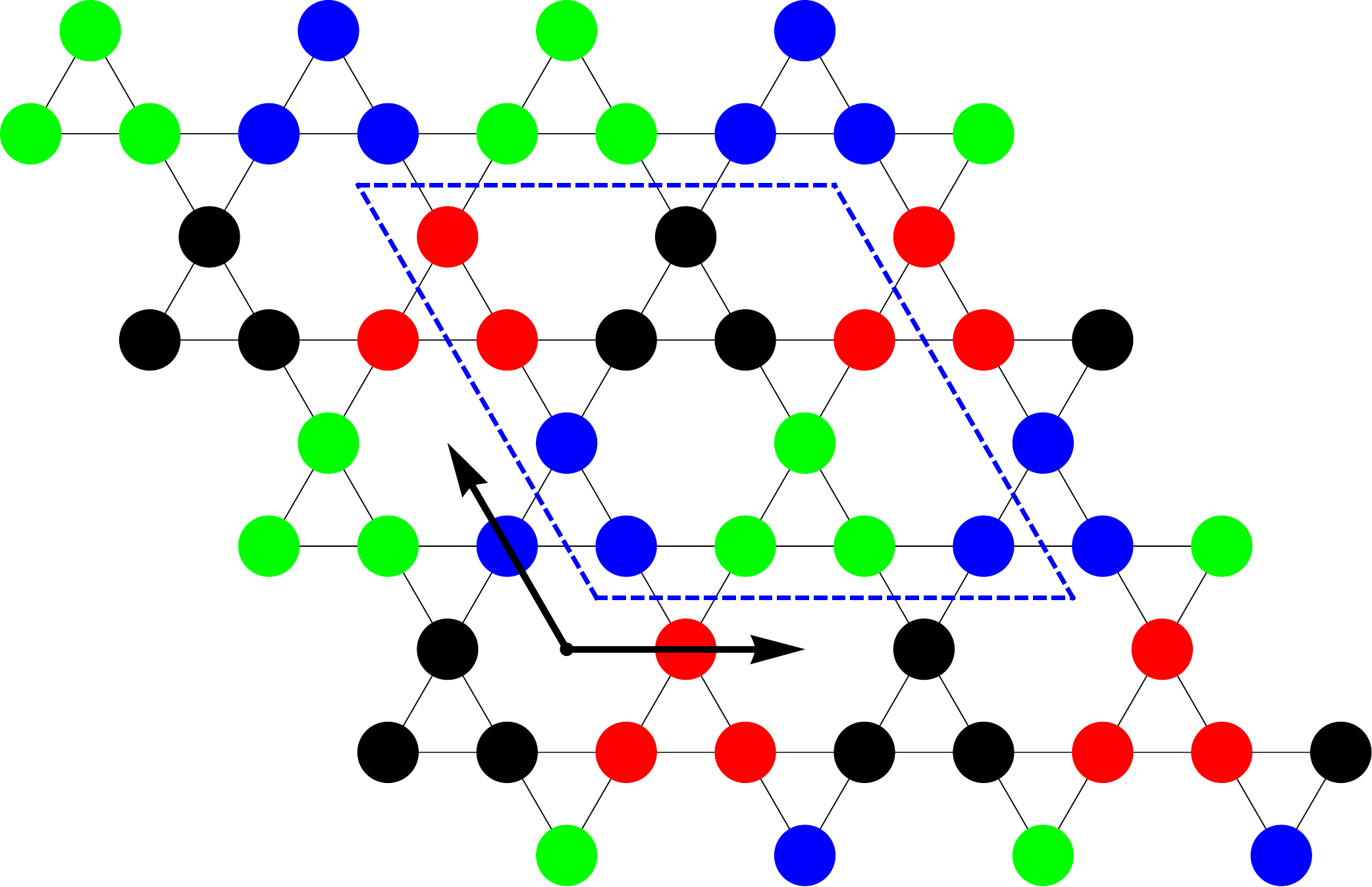}
\includegraphics[height=1.5cm,width=1.5cm]{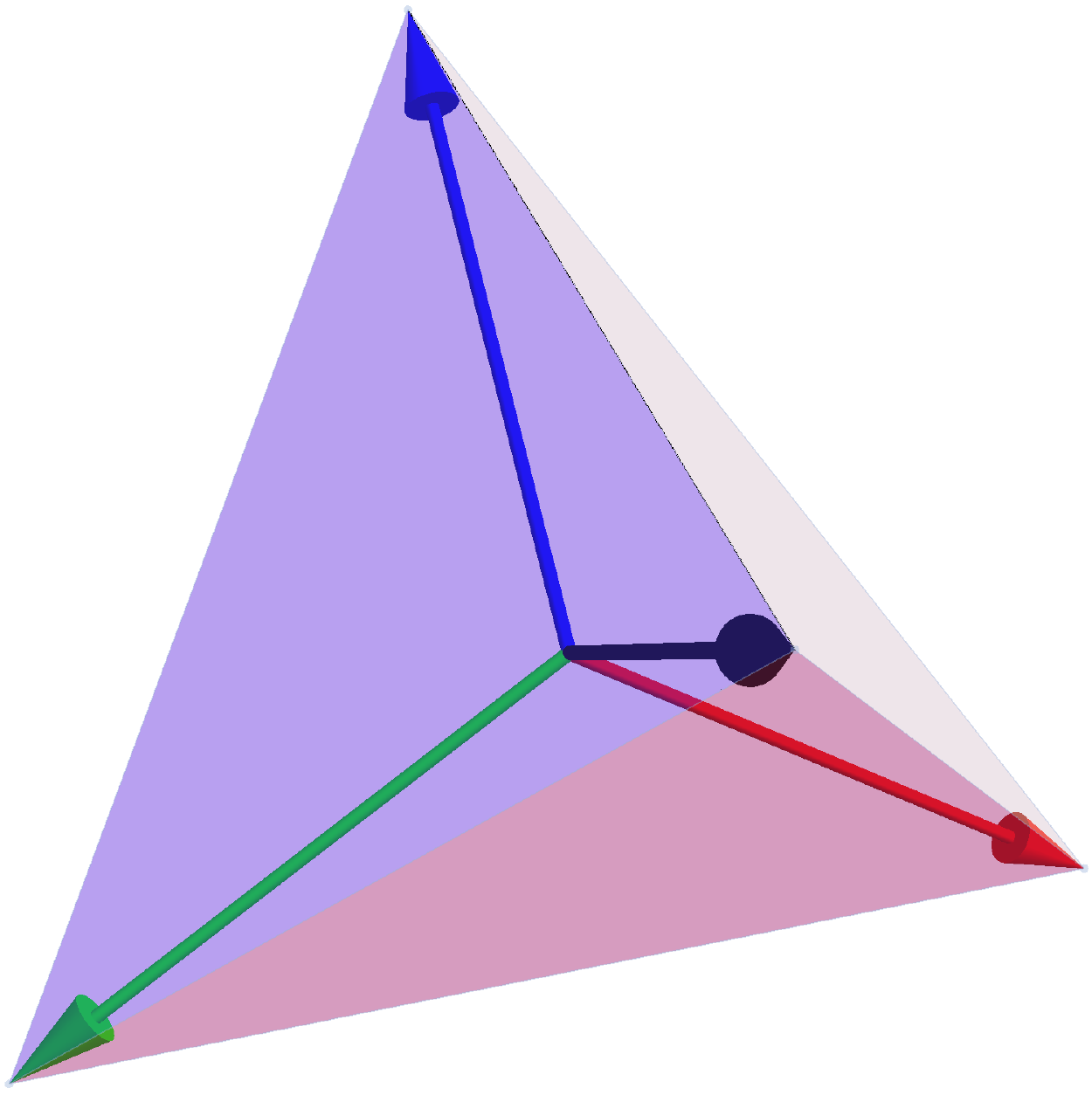}
\includegraphics[height=2.5cm,width=2.5cm]{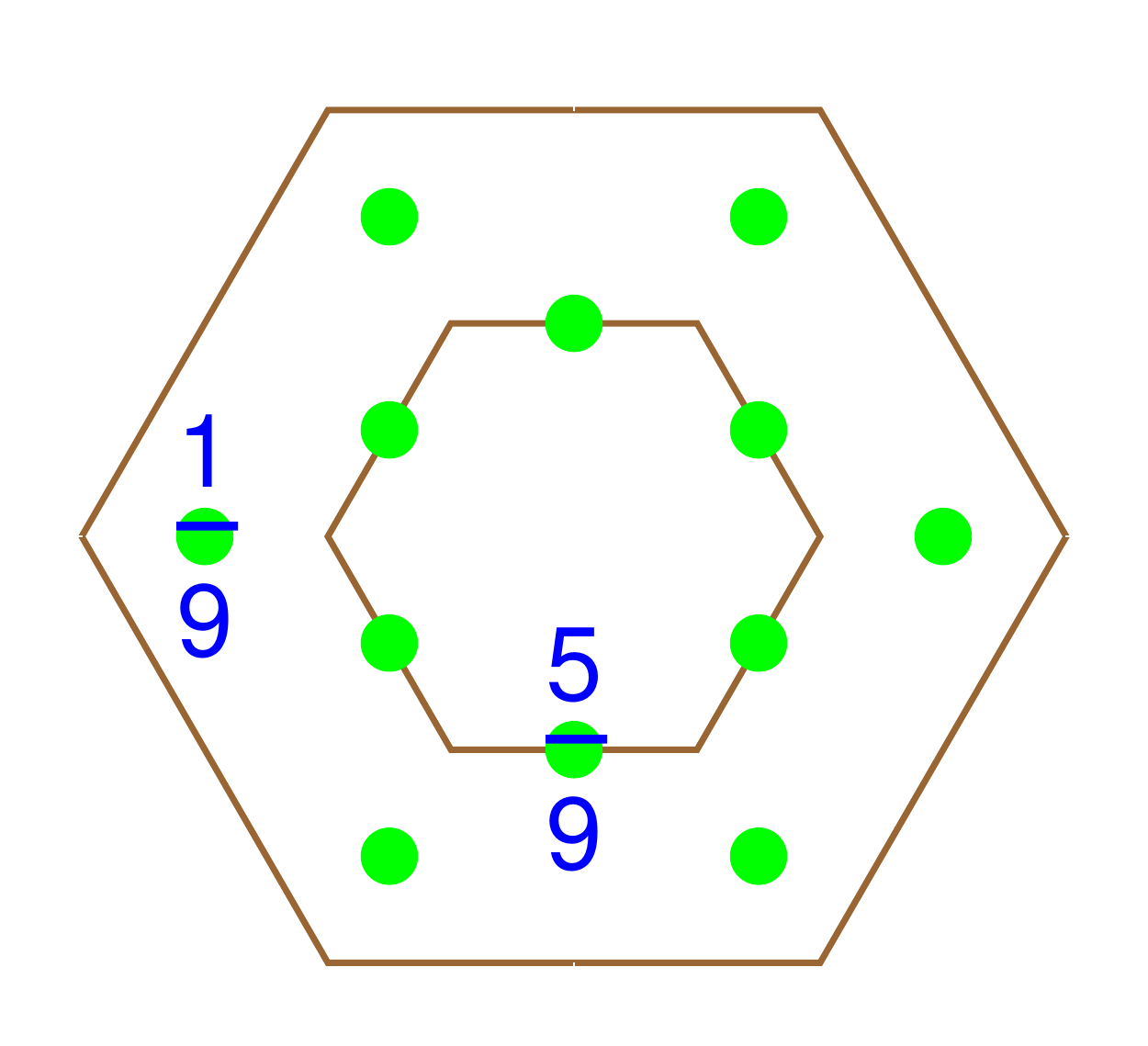}
\caption{Tetrahedral1(T$_1$) state}
\label{Ch4:fig:extendedTetrahedral2-state}
\end{center}
\end{figure}

\noindent \textbf{(xiv) Tetrahedral1(T$_1$) state : } This state has four sub-lattices, and the relative angle between the spins is $109.47$ degree. Each of the sub-lattices occupies an up triangle, as shown by a different color in Fig.~\ref{Ch4:fig:extendedTetrahedral2-state}. The unit cell is quite large, which contains 12 sites. Each of the sublattices is pointing to the corner of the tetrahedron. Here, the energy per site is given by $E = \frac{2}{3} (2 J_1 - 2 J_2 - 2 J_3 - J_{3h})$.

\begin{figure}[ht!]
\begin{center}
\includegraphics[height=2.75cm,width=4.25cm]{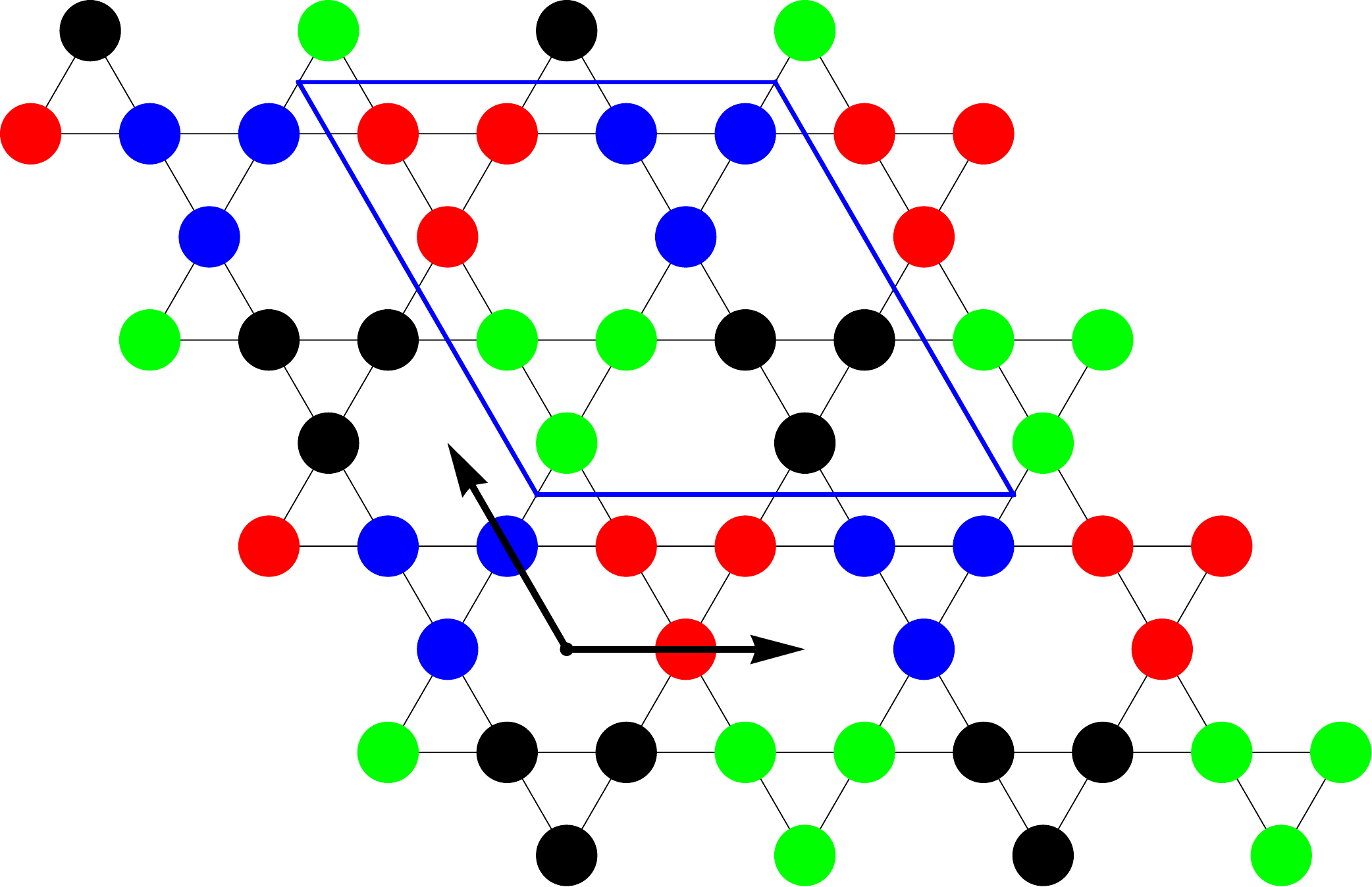}
\includegraphics[height=1.5cm,width=1.5cm]{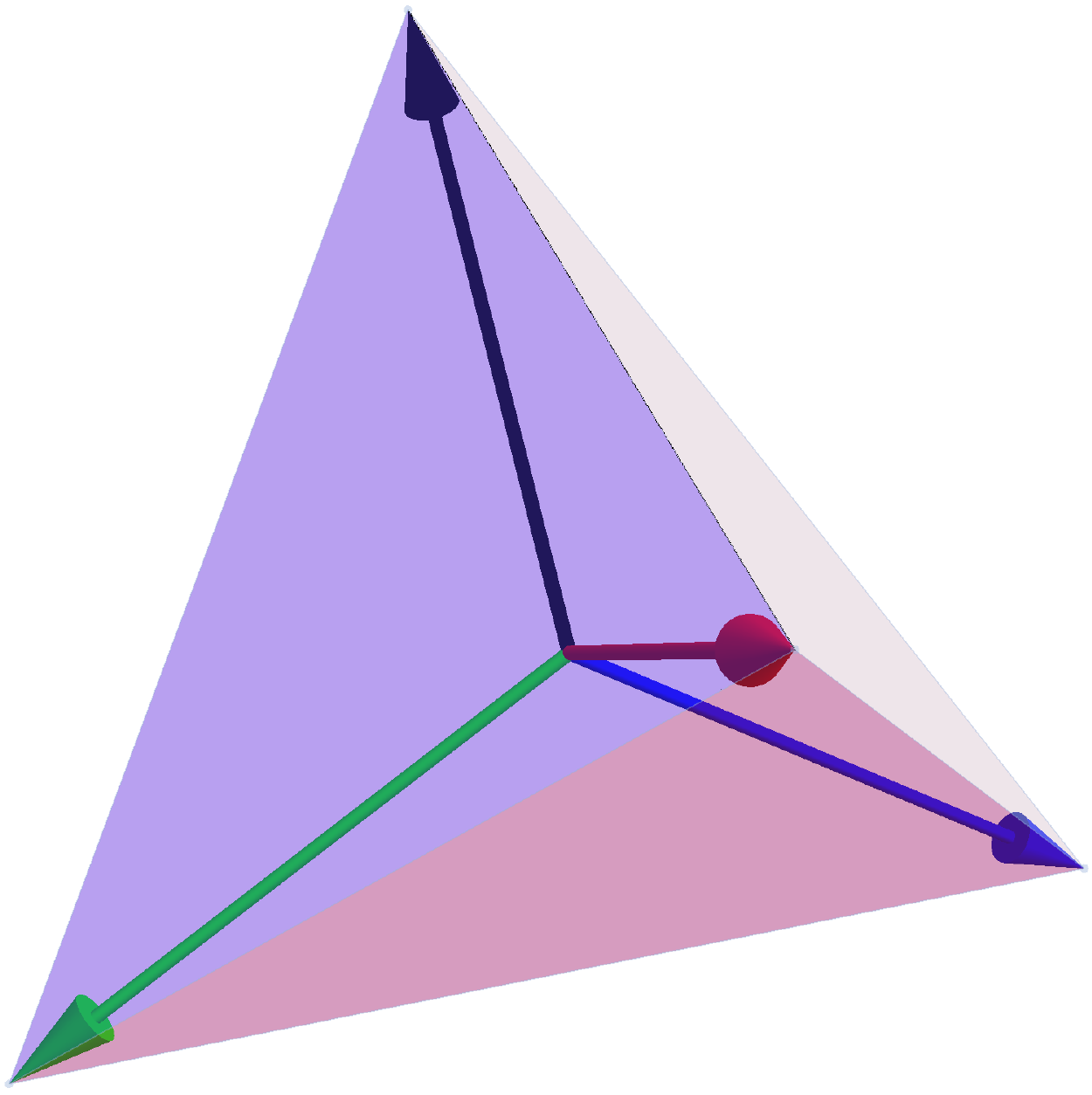}
\includegraphics[height=2.5cm,width=2.5cm]{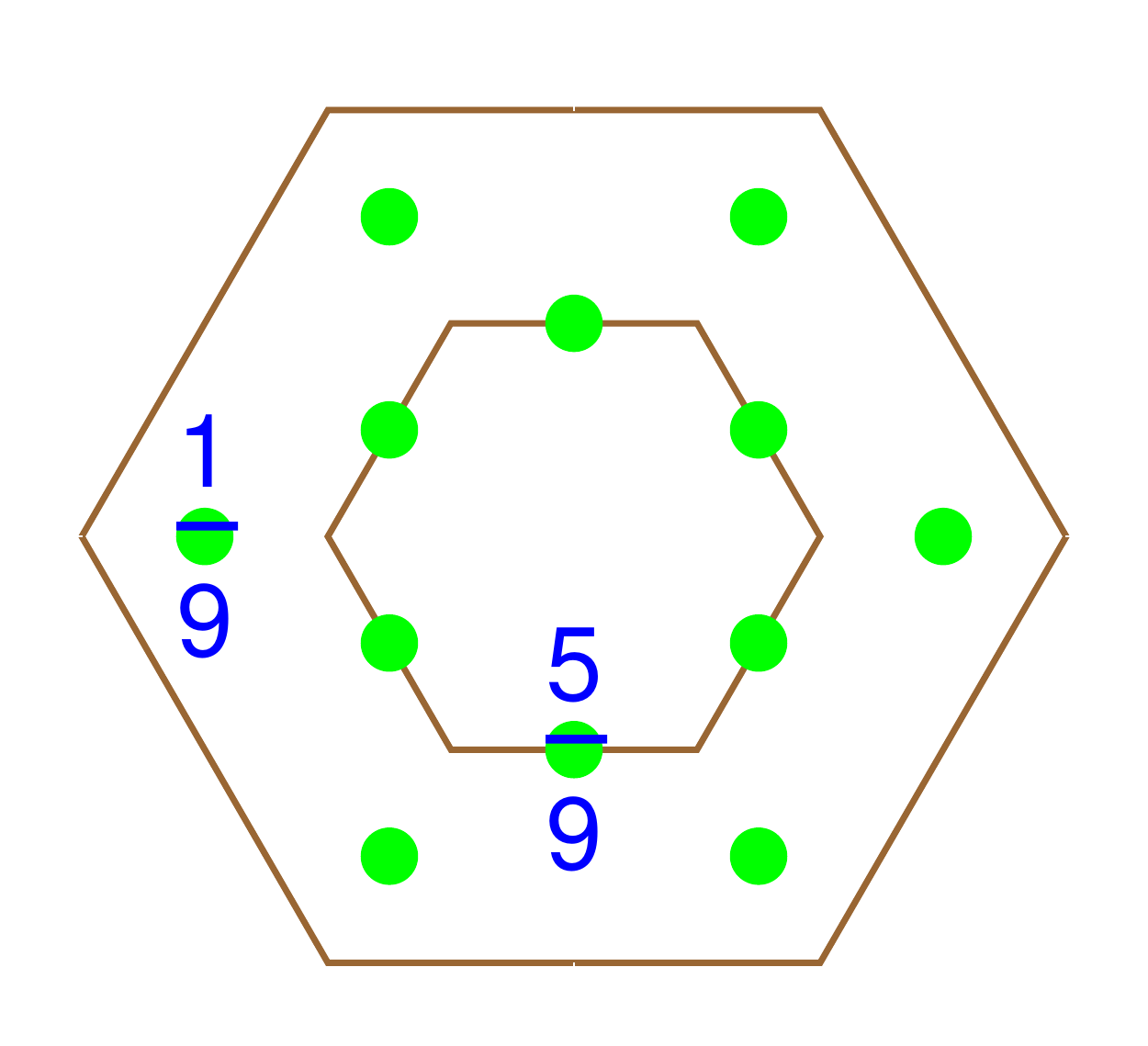}
\caption{Tetrahedral2(T$_2$) state}
\label{Ch4:fig:extendedTetrahedral1-state}
\end{center}
\end{figure}

\noindent \textbf{(xv) Tetrahedral2(T$_2$) state : }This state is similar to the tetrahedral1 state. Here, each of the sub-lattices occupies a down triangle as shown by a different color in Fig.~\ref{Ch4:fig:extendedTetrahedral1-state}. In this case, the energy per site is given by $E = \frac{2}{3} (2 J_1 - 2 J_2 - 2 J_3 - J_{3h})$

In this work, we have started with the unbroken symmetry group ISG as $\{ I \}$ but other possibilities like $\mathbb{Z}_2$ or $O(2)$ which leads to the co-planar state and ferromagnetic state, is also included. One can easily do the exercise for other choices of ISG, but that would not lead to any other new RMOs. In our study, we have considered the manifold $\mathcal{A} = S_2$, i.e., and the spins are treated as three-dimensional unit vectors. However, on the same footing, one can study other manifolds. For example nematic orders can be studied using the manifold $\mathcal{A} = S_2/ \mathbb{Z}_2$ with spin symmetry group $O(3)$.

\section{\label{sec:con}Conclusions}
Based on the group theoretical approach, we have constructed a family of classical magnetic orders, termed as regular magnetic order in kagome and triangular lattice. This approach is introduced by Messio et al.~\cite{messio2011lattice} and is analogous to Wen's classification of quantum spin liquids based on projective symmetry group~\cite{wen2002quantum}. Such RMOs can be constructed for any models on any lattice and hence a general method to construct classical magnetic orders. It turns out that these states are a good candidate as a variational state to study the ground-state phase diagram for many spin systems~\citep{PhysRevB.95.134404,mondal2021q, messio2012kagome, wang2006spin}. It also provides useful insights into the couplings present in a magnetic material by comparing the magnetic correlation where the range or strength of the interactions is not known.

In this study, we have extended the work of Messio et al.~\cite{wen2002quantum} for other wallpaper groups, which includes kagome and triangular geometry. We have obtained a few new classical orders which are not reported earlier. For example, in the case of $p6$ group, depending upon the initial vectors, we get icosahedron states with a series of regular structures like  icosahedron1$(I_1)$,  icosahedron2$(I_2)$, octahedral(O), cuboc1$(C_1)$, and cuboc2$(C_2)$. The realization of these states as a ground state of specific spin models is the subject of future study.
\section{Appendix}
\noindent\textbf{Theorem} : \textit{If a spin configuration $\phi$ is regular then the group $~\mathcal{G}_\phi/ \mathcal{G}_\phi^S~$ is isomorphic to  $~\mathcal{G}_{\mathcal{L}}$}. 

\vspace{0.25cm}
\noindent\textbf{Proof} : There are several steps in the proof.

\vspace{0.25cm}
\noindent (i) \textit{Define mapping : } Let $\sigma \in \mathcal{G}_{\mathcal{L}}$ then, $\exists ~g ~\in O(3)$ such that $g \sigma \in G_\phi$. Define a mapping $\xi : \mathcal{G}_{\mathcal{L}} \rightarrow G_\phi/G_\phi^S$ such that 
\begin{equation}
\xi : \sigma \mapsto (g\sigma) ~G_\phi^S
\end{equation}
This is well defined  mapping since if there are another $g_1 \in O(3)$ such that $g_1 \sigma \in G_\phi$ then, $g_1 \sigma \in (g \sigma)G_\phi^S$. To show this, note that $(g_1 \sigma)^{-1} \in G_\phi$ and hence $(g_1 \sigma)^{-1} g_1 \sigma = g_1^{-1}g_{1} \in G_\phi^S$. Thus, $g_1 \sigma = \sigma g~ (g_1 g^{-1}) $. Thus $g_1 \sigma \in (g \sigma) ~G_\phi^S$

\vspace{0.25cm}
\noindent (ii) \textit{To show that $\xi$ is a homomorphism :} If $\sigma_1$ and $\sigma_2$ are two elements of $\mathcal{G}_{\mathcal{L}}$. Thus if $g_1 \sigma_1 \in G_\phi$ and $g_2 \sigma_2 \in G_\phi$, then clearly $(g_1 \sigma_1) ~(g_2 \sigma_2)\in G_\phi$. Thus clearly, $\xi$ preserves the multiplication and hence it a homomorphism.

\vspace{0.25cm}
\noindent (iii) \textit{To show that $\xi$ is one to one :} If $\sigma_1$ and $\sigma_2$ are two elements of $\mathcal{G}_{\mathcal{L}}$ and  if 
\begin{eqnarray}
(g_1 \sigma_1) G_\phi^S & = & (g_2 \sigma_2) G_\phi^S \nonumber \\
g_1 \sigma_1 & = & g_2 \sigma_2 g \nonumber \\
(g_1 g g_2^{-1}) (\sigma_1 \sigma_2^{-1})  & = & e \nonumber
\end{eqnarray}
This can only happen when $\sigma_1 = \sigma_2$ and then $g_2^{-1}g$ must be in $G_\phi^S$.
\bibliography{refs}
\bibliographystyle{apsrev4-1}
\end{document}